\documentclass[12pt]{elsart}
\usepackage{graphicx}
\def\be{\begin{eqnarray}}
\def\ee{\end{eqnarray}}
\def\bc{\begin{center}}
\def\ec{\end{center}}
\def\om{\omega}

\def\prt{\partial}

\def\lsim{\stackrel{\scriptstyle <}{\phantom{}_{\sim}}}
\def\gsim{\stackrel{\scriptstyle >}{\phantom{}_{\sim}}}

\def\rmd{{\rm d}}
\begin{document}
\begin{frontmatter}
\title{ Relativistic Mean-Field Models
 with Effective Hadron Masses and
Coupling Constants,\\ and $\rho^-$ Condensation\\
%%{\rm\normalsize [kv-Wal-9]}
}
\author[NBI]{E.E.Kolomeitsev} \and \author[MEPHI,GSI]{D.N. Voskresensky}
\address[NBI]{Niels Bohr Institute, Blegdamsvej 17, Copenhagen, Denmark}
\address[MEPHI]{Moscow Engineering Physical Institute,\\ Kashirskoe
  Avenue 31, RU-11549 Moscow, Russia}
\address[GSI]{GSI, Plankstra\ss{}e 1, D-64291 Darmstadt, Germany}
\begin{abstract}
We study relativistic mean-field models with hadron masses and
coupling constants depending self-consistently on a scalar meson
field. We demonstrate that by the field redefinition some models
can be equivalently transformed into each other. Thereby the
large variety of scaling functions for masses and couplings can be
reduced to a restricted set of functions with a constrained
dependence on a scalar field. We show how by choosing
properly the latter scaling functions one may stiffen or soften
the equation of state at high densities and simultaneously
increase the threshold density for the direct Urca process without
any change of the description of nuclear matter close to the
saturation density. The stiffening of the equation of state might
be motivated by recent neutron star mass measurements, whereas the
increase of the threshold density for the direct Urca process is
required by the analysis of neutron star cooling data. We
demonstrate that if a rho meson is included in a mean-field model
as a non-Abelian gauge boson, then there is a possibility for a
charged rho-meson condensation in dense nuclear matter. We show
that such a novel phase can be realized in neutron star interiors
already for sufficiently low densities, typically $\sim 3\div
4~n_0$, where $n_0$ is the nuclear saturation density. In the
framework of the relativistic mean field model the new phase
arises in a second-order phase transition. The appearance of a
$\rho^-$
condensate significantly alters the proton fraction in a neutron
star but changes moderately the equation of state. The neutrino
emissivity of the processes involving a $\rho^-$ meson condensate
is estimated.
\end{abstract}\end{frontmatter}
%\tableofcontents
%------------------------------------------------------------%
\section{Introduction}

The baryon matter at densities relevant for neutron star (NS)
interiors is convenient to describe in terms of relativistic
mean-field (RMF) models, cf. ~\cite{G00} and references therein.
In these models the nucleon mass decreases with an increase of the
nucleon density due to coupling to a scalar field, whereas the
effective masses of scalar ($\sigma$), vector ($\omega$) and
isovector ($\rho$)  mesons, and their coupling constants, are usually
assumed to be density and field independent quantities. The RMF
models provide a simple and economical tool for construction of
the equation of state (EoS) for dense baryonic matter. By tuning
parameters of a RMF model at a saturation density one may get
an EoS similar to the one that follows from more involved
microscopic calculations, e.g., as that of Urbana--Argonne group
(A18+$\delta v$+UIX*)~\cite{APR98}.  The
latter calculation provides probably the most realistic EoS for
nuclear densities up to $4\,n_0$, but it uses a non-relativistic
potential and violates causality at higher
densities. As an extrapolation to higher densities RMF models are
practical. However all RMF models have the very same an unpleasant
feature. They produce  a  large fraction of protons in the NS matter, cf. ~\cite{BB77,Boguta81}.
This permits a very efficient cooling process, direct
Urca (DU) process $n\rightarrow pe\bar{\nu}$, for $n>n_{\rm crit}^{\rm
DU}$, with a low threshold density $n_{\rm crit}^{\rm DU}\lsim 3~n_0$.
Ref.~\cite{MSTV} emphasized this fact as an obvious shortcoming of
RMF models. Oppositely, ref.~\cite{LP} and some subsequent works~\cite{YAKDU}
tried to use this
fact to develop a so-called DU-based scenario of a NS cooling
(based on $n_{\rm crit}^{\rm DU}\sim (2\div 3)n_0$).

The NS cooling "data" (surface temperatures of NSs as a function of the NS age) can be
subdivided in three groups, corresponding to "slow cooling",
"intermediate cooling" and "rapid cooling" of neutron stars. If the density in the NS
central region exceeds $n_{\rm crit}^{\rm DU}$, which corresponds to the mass
$M_{\rm crit}^{\rm DU}$, the star cools down very fast, radiating mostly from
its central region. Thus, the NSs with
$M<M_{\rm crit}^{\rm DU}$ cool slowly, whereas the stars with
the mass $M$ only slightly above $M_{\rm crit}^{\rm DU}$ would be already
characterized by a very rapid cooling.
Such a scenario was criticized in~\cite{TTTTT02,T04,BGV04}.
It is doubtful that many NSs, belonging to an intermediate cooling group,
have the very similar masses.
It has been already experimentally established that NSs may have essentially
different masses. E.g., masses of two NSs in double NS systems are
measured with the very high precision, $M_{\rm B1913+16}=1.4408\pm 0.0003
M_{\odot}$, and $M_{\rm J0737-3039B}=1.250\pm 0.005 M_{\odot}$~\cite{S04}.

Authors of many RMF models do not care about the
DU threshold at all and obtain so low values of $n_{\rm crit}^{\rm
DU}$ that $M_{\rm crit}^{\rm DU}<1\,M_{\odot}$.
However according to the standard mechanisms only the NSs with $M\gsim
1\,M_{\odot}$ can be formed in supernova explosions. The mentioned
above microscopic Urbana-Argonne model produces
$n_{\rm crit}^{\rm DU}\simeq 5~\,n_0$, that corresponds to a heavy
NS with $M_{\rm crit}^{\rm DU}\simeq 2\,M_{\odot}$. 
%%With the same
%%EoS the central density of $3.3\,n_0$ ($2\,n_0$) relates to
%%the NS mass  $M\simeq 1.3\,M_{\odot}$ ($0.75\,M_{\odot}$).
If one used a standard RMF model  fitted in \cite{KV} to reproduce the
Urbana-Argonne EoS for $n<4n_0$,
but,  as well as other RMF models, producing a low DU threshold ($n_{\rm crit}^{\rm DU}\simeq (2.6\div 2.7)\,n_0$),
one would conclude that the majority of the NSs seen in soft $X$ rays are low
mass objects with  the mass $M< 1.3\,M_{\odot}$.
The latter value is however below the averaged value of the NS mass
($M\simeq 1.35\pm 0.04\,M_{\odot}$) measured in NS binaries \cite{TC99}.

Another problem is that there might exist heavy NSs. Recent
measurements of NS masses in binary compact systems yielded
$M_{\rm J0751+1807}=2.2\pm 0.2 M_{\odot}$ \cite{S04} and
$M_{\rm 4U1700-37}=2.44\pm 0.27~ M_{\odot}$ \cite{Q03}.
The latter object might be a black hole due to a lack of
pulsation but the former object, J0751+1807, is
definitely a NS. The confidence level of 95\% still allows to
have a smaller limiting NS mass than
$2~M_{\odot}$ (however the  limiting NS mass is surely
higher than $1.6~M_{\odot}$). Additional information on heavy
NSs may be obtained from quasi-periodic oscillation (QPO) sources.
These are NSs that are $X$-ray sources at frequencies of the
orbiting accreting matter. If QPOs  originate from the innermost stable
orbit~\cite{Miller} of the accreting matter, their
observed values imply that the accreting NS has the mass $\simeq
2.2~M_{\odot}$ in the case of 4U1820-30. The value
$2.2~M_{\odot}$ is the limiting mass of a NS following
the Urbana--Argonne EoS. Unfortunately this EoS is acausal at high
densities. Ref.~\cite{HHJ} suggested an analytic fit of the
Urbana--Argonne EoS for $n\lsim 4 n_0$, which respects causality
for larger densities. We will denote this parameterization as the HHJ
EoS. As a price, the limiting mass decreased down to $\simeq
2M_{\odot}$. Moreover, all phase transitions, like pion and kaon
condensations, and quark matter,  which are assumed to be possible
in NS interiors, may essentially  soften EoS decreasing further the
limiting mass of a NS\footnote{The Urbana--Argonne calculation
\cite{APR98} already includes neutral pion condensate for
$n>1.3~n_0$ that only slightly softens EoS in their case.}. Thus
modern microscopically motivated EoS and possible phase
transitions in NS interiors depend crucially on one precision
measurement of a heavy NS  with $M>(2\div 2.2)~M_{\odot}$.

Within  a standard RMF model it would be hard to describe heavy NSs with
$M>2.2~M_{\odot}$ without either an essential increase of the
compressibility parameter $K$ or a decrease of the effective
nucleon mass at the saturation density. Both values are
rather constrained by the atomic nucleus data. The problem
originates from the fact that the RMF model parameters are fixed
at the saturation density and no free parameters remain to
control the pressure and the symmetry energy at higher densities.
The problem could be in principle solved by introducing new terms
with new fitting parameters.

There are strong indications that meson masses as well as the
nucleon mass depend on the nucleon density. A
general trend for such a dependence is
governed by the property that  masses of all hadrons except
Goldstone bosons, like pions and kaons,\footnote{The kaon
effective mass may also decrease with the density, whereas the
nature  of  this decrease is different \cite{NK,G00,KV}.} should
decrease, when the nucleon density $n$ increases towards a
critical point of the chiral symmetry restoration~\cite{chiral}.
According to the conjecture of Brown and Rho~\cite{BR} the nucleon
mass and the masses of vector $\om\,, \rho$ and scalar $\sigma$
mesons may obey an approximately universal scaling law (a so-called
"Brown--Rho scaling")
\be
\frac{m_\rho^* }{m_\rho}\simeq  \frac{m_\om^*}{m_\om}\simeq
\frac{m_\sigma^*}{m_\sigma}\simeq \frac{m_N^* }{m_N}=\Phi_{BR}\,,
\label{BRm}\ee
where asterisks indicate in-medium masses. According to ~\cite{BR}
$\Phi_{BR}$ decreases with the increase of the density. In
ref.~\cite{SBMR97} an attempt was made to include density
dependent masses in the RMF model in order to incorporate the
Brown--Rho scaling, explicitly. A theoretical disadvantage of this
model is that the  density dependence, $\Phi_{BR} \simeq (1+yn/n_0
)^{-1}$, $y\simeq 0.28$, was inserted by hands into the model
Lagrangian rather than it followed from the corresponding
equations of motion. Insertion of a density dependence  into
parameters of a mean-field Lagrangian violates conservation of the
energy-momentum and spoils the thermodynamic  consistency.
Therefore, such models require a special treatment~\cite{consist}.
Ref. \cite{SBMR97} also realized that the scaling of meson masses
(\ref{BRm}) would result in a dramatic stiffening of the EoS, if
one does not simultaneously scale the corresponding coupling
constants in a similar way. As an argument for such a procedure,
authors relied on the description of the flow in heavy ion
collisions. Below we will show that this prescription relies on
equivalence of some RMF models that allows to use different
interpolating mean fields without changing physical quantities on
a mean-field level in homogeneous matter.

In ref.~\cite{manka} the density dependence of  effective hadronic
masses  was incorporated into the mean-field Lagrangian via extra
terms describing a hadron interaction with a scalar field. It
resulted, however, in \emph{an increase} of the $\rho$ and
$\omega$  meson masses with increase of the nucleon density in
disagreement  with (\ref{BRm}). At the same time the $\sigma$ mass
was assumed to be constant. Thus, the inclusion of the hadron mass
scaling into RMF models deserves a further consideration. Some
other works, cf. \cite{THIM}, played with a field-dependent
parameters of generalized mean field models, permitting changes of
both  masses and couplings.  Equivalence between different parameter
choices  was not realized in these papers.

Appropriate knowledge of the  EoS of hot, dense, and isospin asymmetrical
nuclear matter is essential for understanding
the results of heavy-ion collisions.
Flow, particle yields, fragmentation may give some constraints
on the EoS and the symmetry energy, cf.~\cite{dan,Gai}.
Often these constraints disagree with those obtained
from NS physics, e.g., yielding very low DU threshold density.
The proper choice of the EoS is particularly important for the CBM (compressed
baryon matter) experimental program at the future facility at GSI.
Precision measurements of heavy ion collisions are planned
in the whole energy range  from $\sim 1$~GeV/A
up to $\sim 40$~GeV/A.

To  describe correctly the symmetry energy of the nucleon matter
one needs to include the $\rho$-meson field into a mean-field
Lagrangian. The $\rho$-meson field is traditionally assumed to
have only one mean-field component,
$(\vec{\rho\,}_\mu)^a=\delta_{\mu 0}\, \delta^{a 3}\,
\rho_0^{(3)}$, where $\mu=0,1,2,3$ is the Lorentz index, and
$a=1,2,3$ is the isospin index. This field produces then the
expected contribution to the energy density $\propto
(n_p-n_n)^2$\,, where $n_p$ and $n_n$ are the proton and neutron
densities. In ref.~\cite{v97} it was realized that, if the
$\rho$-meson is introduced as a non-Abelian gauge boson,  e.g., in
the framework of the hidden local symmetry model~\cite{bando83},
then at a density $n>n_{\rm c}^{\rho}$ the charged $\rho$-meson
condensation occurs. In a new phase the symmetry energy  grows as
$\propto |n_p-n_n|$ only. The critical density was found to be
$n_{\rm c}^{\rho}\sim m_\rho^{*3}$, being sensitively dependent on
the value of the in-medium mass of the $\rho$-meson
($m_{\rho}^{*}$). The phenomenon of the charged $\rho$-meson
condensation is similar to the condensation of gauge bosons in
QCD~\cite{migdal} (gluon condensate) and to
electro-weak~\cite{linde} ($W$-boson condensate) sectors of the
Standard model.

In this paper we first construct a generalized RMF model with all
parameters, being field dependent
(secs.~\ref{sec:grmfm},\ref{sec:mass}). We derive the energy
density functional, and show how various standard cases can be
reproduced. We discuss the equivalence relations among some cases.
Then we demonstrate shortcomings  of the models with a hadron mass
scaling  without a scaling of coupling constants. We show  that
the RMF models with the universal scaling for masses and coupling
constants can be reduced to the models without any scaling. Such
models do not solve, however, the problem of the low
$n_{\rm crit}^{\rm DU}$. 
%%In sec.~\ref{sec:absence} 
Then we allow for a
different scaling of hadron masses and coupling constants in order
to generate a model that produces a stiffer EoS at high densities.
This model supports a large limiting mass of the NS and rather
high critical density for the DU process. On the other hand, it
uses  appropriate values of compressibility and effective nucleon mass
at the saturation nuclear density. Then we demonstrate another example
of the RMF model with a non-universal scaling of masses and
coupling constants that fits well the HHJ EoS including the
correct density dependence of the proton concentration. In
sec.~\ref{sec:rho} we investigate a possibility for a charged
$\rho$-meson condensation. We demonstrate that in the framework of
our generalized RMF models the charged $\rho$-meson condensation
occurs as a second-order phase transition. Emissivity of the NS
cooling processes involving the charged rho condensate is
estimated in sec.~\ref{sec:rhocool}. Results are summarized in the
concluding section~\ref{sec:concl}. In Appendix we show how one can
introduce inhomogeneous electric potential. Throughout the paper
we use units $\hbar =c=1$.

\section{Generalized relativistic mean-field model} \label{sec:grmfm}
\subsection{Mean field Lagrangian}
We start with a generalized RMF model,
where  effective hadron  masses and coupling constants
are assumed to be field-dependent from the very beginning.
The Lagrangian density is given by
\be
&&\mathcal{L}=\mathcal{L}_N+\mathcal{L}_M+\mathcal{L}_l\,,
\label{lag}\\
&&\mathcal{L}_N =  a_N\, \bar \Psi_N\, \Big(i\,D\cdot
\gamma\Big)\, \Psi_N -m_N\,\phi_N\, \bar\Psi_N\,\Psi_N\,,
\nonumber \\
&&D_\mu=\prt_\mu+i\,g_{\om}\,\widetilde\chi_\om \om_\mu+
\frac{i}{2}g_{\rho}\,
 \widetilde\chi_\rho\, \vec{\rho}_\mu\,\vec{\tau}\,,
\label{lagN}\\
&&\mathcal{L}_M =
a_\sigma\frac{\prt^\mu \sigma \prt_\mu
  \sigma}{2}-\phi^2_\sigma\frac{m_\sigma^{2}\, \sigma^2}{2}-\widetilde U(\sigma)
  \nonumber\\
&&-a_\om\,\frac{\omega_{\mu\nu}\, \omega^{\mu\nu}}{4} +
  \phi^2_\om\,\frac{m_\om^{2}\, \om_\mu\om^\mu}{2}
-a_\rho\,\frac{\vec{\rho\,}_{\mu\nu}\,\vec{\rho\,}^{\mu\nu}}{4} +
  \phi^2_\rho\,\frac{m_\rho^{2}\, \vec{\rho\,}_\mu\vec{\rho\,}^\mu}{2}\,,
\label{lagM}\\
&&\om_{\mu\nu}=\prt_\nu\om_\mu-\prt_\mu\om_\nu\,,\quad
\vec{\rho}_{\mu\nu}=\prt_\nu\vec{\rho}_\mu-\prt_\mu\vec{\rho}_\nu
+ g_\rho^{'}\,\widetilde{\chi}'_\rho\,[\vec{\rho}_\mu\times \vec{\rho}_\nu]\,,\\
&&\mathcal{L}_l=\sum_{l}
\bar \Psi_l [i(\gamma\cdot\prt)-m_l]\Psi_l\,.
\label{lagOR} \ee
Here $\Psi_N=(\Psi_n,\Psi_p)^{\rm T}$ is  the isospin doublet of the
nucleon bispinors,
$\sigma$, $\omega_\mu$, $\vec{\rho}_\mu$ are the $\sigma$, $\omega$ and $\rho$ meson
fields, $\vec{\tau}$ is the isospin Pauli matrix. The part $\mathcal{L}_l$
describes the contribution of the light leptons (electrons and
muons, $l=e,\mu$). For the sake of simplicity we will neglect hyperons in
the Lagrangian (\ref{lag}), which can contribute, when the hyperon Fermi
seas are filled  in the NS matter.
An extension for  hyperons is straightforward, but is associated with
large uncertainties due to poorly-known coupling constants of hyperons to
mean-fields, cf. \cite{G00,KV}.
The meson sector in (\ref{lag}) is treated  on a mean-field level
as described in Ref.~\cite{G00}.

The scaling parameters of the model, the ``dielectric'' constants
$a_{i}$, renormalizations of the coupling constants and masses,
$\widetilde\chi_{i}$ and $\phi_{i}$, are dimensionless
functions of the fields and effective couplings. The potential $\widetilde U(\sigma)$
represents the possible self-interaction of the $\sigma$ field,
which was suggested in ref.~\cite{BB77}. In the general case the
scaling parameters can depend also on $(\om_\mu\om^\mu)$ and
$(\vec{\rho\,}_\mu\vec{\rho\, }^\mu)$\, yielding corresponding terms in the potential. For instance, writing
$\phi_\om^2=\phi_\om^2(\sigma)+\zeta\, \om_\mu\, \om^\mu/4$ we recover
the $(\om_\mu\om^\mu)^2$ terms proposed in ref.~\cite{bodmer} to
improve the description of finite nuclei within the RMF model approach. A
systematic construction of mean-field Lagrangians with
high-order terms based on constraints from the chiral symmetry of
QCD can be found in ref.~\cite{furnstahl}.
In our treatment we will neglect
a possible dependence of the potential and the parameters on the $\omega$ and
$\rho$ meson fields, elaborating only the $\sigma$-field
dependence.  We assume that the density dependent
scaling functions $a_N$ and $\phi_N$
are basically due to a nucleon--sigma-meson interaction. Thus, we
can write
\be
a_N= a_N ( g_{\sigma}\widetilde\chi_\sigma \sigma )\,,\quad
\phi_N =\phi_N (g_{\sigma}\widetilde\chi_\sigma \sigma)\,.
\ee
Simplifying we will also assume that all other scaling functions,
$\widetilde{\chi}_{\om,\rho}$, $a_{\om,\rho,\sigma}$, and the potential
$\widetilde U$ are due to the same
nucleon--sigma-meson interaction, i.e. are the functions of
$g_{\sigma}\widetilde\chi_\sigma \sigma$. For the rho meson included according to
the hidden local symmetry principles~\cite{bando83}
one has  $g_{\rho}=g_{\rho}^{'}$. Despite this
the scaling of $g_{\rho}$ and $g_{\rho}^{'}$
may be quite different in dense nuclear matter,
$\widetilde\chi_{\rho}\neq\widetilde\chi_{\rho}^{'}$.
The scaling factor $\widetilde\chi_{\rho}$ arises due to the renormalization of the
nucleon-sigma interaction in medium, whereas the factor $\widetilde\chi_{\rho}^{'}$
is due to a non-Abelian interaction between $\rho_i^{\alpha}$ fields and does not depend
directly on the nucleon-field source. Thus within above assumptions we may take
$\widetilde\chi_{\rho}^{'}\simeq 1$.

The nucleon density is given by the zero component of the Noether current
\be
n_N = a_N <\Psi_N^{\dagger} \Psi_N >\,.
\ee

\subsection{Field redefinition in RMF models}

Describing an  extended homogeneous system like a NS we may
assume that mean fields of mesons do not depend on  coordinates.
Then the physical results obtained on a mean-field level would not change if
all particle fields are rescaled by constant factors.
For instance we may change meson and nucleon fields  as
\be\label{sc}
\Psi_N\to \Psi_N/\sqrt{a_N}\,, \quad \sigma \to \sigma/\sqrt{a_\sigma}\,,
\quad \om_\mu\to \om_\mu/\sqrt{a_\om}\,,\quad
\vec{\rho}_\mu\to \vec{\rho}_\mu/\sqrt{a_\rho}
\ee
and obtain the new nucleon and meson Lagrangians equivalent to those
in (\ref{lagN},\ref{lagM}) in a mean-field approximation
\be
\mathcal{L}_N
&=& \bar \Psi_N\, \Big(i\,D\cdot
\gamma\Big)\, \Psi_N -m_N^*\, \bar\Psi_N\,\Psi_N,
\nonumber \\
&& D_\mu=\prt_\mu+i\,g_{\om }\,{\chi}_\om \om_\mu+ \frac{i}{2}g_{\rho
}\, {\chi}_\rho\, \vec{\rho}_\mu\,\vec{\tau}\,,
\label{lagNn}\\
\mathcal{L}_M
&=&\frac{\prt^\mu \sigma \prt_\mu
\sigma}{2}-\frac{m_\sigma^{*2}\, \sigma^2}{2}-{U}(\sigma)
  \nonumber\\
&-&\frac{\omega_{\mu\nu}\, \omega^{\mu\nu}}{4}
+\frac{m_\om^{*2}\, \om_\mu\om^\mu}{2}
-\frac{\vec{\rho\,}_{\mu\nu}\,\vec{\rho\,}^{\mu\nu}}{4}
+\frac{m_\rho^{*2}\, \vec{\rho\,}_\mu\vec{\rho\,}^\mu}{2}\,,
\label{lagMn}\\
&&\om_{\mu\nu}=\prt_\nu\om_\mu-\prt_\mu\om_\nu\,,\quad
\vec{\rho}_{\mu\nu}=\prt_\nu\vec{\rho}_\mu-\prt_\mu\vec{\rho}_\nu
+g_\rho\,{\chi'}_\rho\,[\vec{\rho}_\mu\times \vec{\rho}_\nu]\,.
\label{lagORn}
\ee
We denoted $U(\sigma)=\widetilde U(\chi_\sigma\,\sigma)$.
The effective masses of particles introduced in this Lagrangian
are equal to
\be
{m_N^*}/{m_N}&=&{\phi_N(\chi_\sigma \sigma)}
/{{a_N(\chi_\sigma\sigma)}}=\Phi_N(\chi_\sigma\sigma)\,,\nonumber \\
{m_i^*}/{m_i}&=&{\phi_i(\chi_\sigma\sigma)}/{\sqrt{a_i(\chi_\sigma\sigma)}}=\Phi_i(\chi_\sigma\sigma)\,,
\quad i=\sigma,\om,\rho\,,
\label{effmass}
\ee
where the dimensionless functions $\Phi_N$ and $\Phi_i$ depend on the
scalar field in the combination $\chi_\sigma\sigma$.
The scaling functions of
coupling constants $g_\sigma , g_\om , g_\rho ,g_\rho^{'}$ are given by
\be
{\chi}_i={\widetilde\chi_i(\chi_\sigma\sigma)}/{\sqrt{a_i(\chi_\sigma\sigma)}}\,,\quad
i=\sigma,\om,\rho\,,\quad
{\chi'}_\rho={\widetilde\chi'_\rho(\chi_\sigma\sigma)}/{\sqrt{a_\rho(\chi_\sigma\sigma)}}\,.
\label{chibar}
\ee
All quantities are now expressed in terms of the new renormalized $\sigma$
field.
%The combination $\chi_\sigma \sigma$ did not change under the rescaling.

The thermodynamic potential density $\Omega =-P$, where $P$ is the
  pressure, is related to the energy density by the standard  equation
  \be\label{presmu}
  E[n_n,n_p , n_l ;f,\vec{\rho}_\mu] =\sum_{i} \mu_i n_i +\Omega\,,\quad
  P=\sum_{i}\mu_i n_i
  -E,\quad \mu_i=\frac{\prt E}{\prt n_i}\,.
  \ee

Chemical potentials enter  the Green functions in the standard gauge
  combinations $\varepsilon_i +\mu_i$. Chemical potentials of the charged boson
  mean fields are also introduced
by the gauge replacements
  $\varepsilon_i\rightarrow \varepsilon_i +\mu_i$.

%Bearing in mind a
%possibility of production of a charged $\rho$-meson field, characterized by
%the chemical potential $\mu_{ch}^{\rho}$, with the help of the gauge replacement
%we should accordingly rewrite the contribution of the
%term $\frac{\vec{\rho\,}_{\mu\nu}\,\vec{\rho\,}^{\mu\nu}}{4}$.  Thus we arrive
%at the shift
For charged rho-meson fields above replacement implies
\be
&&\rho_{\mu\nu}\to \rho_{\mu\nu}+ \Delta\rho_{\mu\nu}\,,
\qquad \Delta\rho_{\mu\nu}=
\mu_{\rm ch}^{\rho}\delta_{\nu 0}[\vec{n_3}\times
\vec{\rho}_\mu]-\mu_{\rm ch}^{\rho}\delta_{\mu 0}
[\vec{n_3}\times \vec{\rho}_\nu]\,,
\ee
where $(\vec{n}_3)^a=\delta^{a3}$ is the unit vector in the isospin
space, $a=1,2,3$\,,
and $\mu_{\rm ch}^{\rho}$ is the chemical potential of rho
mesons.
The explicit expression for $\Omega$ is given in Appendix.

After calculation of
the nucleon contribution
the energy-density functional takes the form
\be
&&E[n_n,n_p , n_l ;f,\om_0 ,\vec{\rho}_\mu ] = E_N[n_n,n_p;f]+E_l[n_e,n_\mu]\nonumber\\
&&\qquad\qquad\qquad\qquad\qquad
+E_\om[n_n,n_p;f,\om_0 ]+E_\rho[n_n,n_p;f,\vec{\rho}_\mu]\,,
\label{Efun}\\
&&E_N[n_n,n_p;f] =\frac{m_N^4\,f^2}{2\, C_\sigma^2}\, \eta_{\sigma}(f)
+{U}(f)
\nonumber\\
&&\qquad\qquad\qquad+
\left(\intop_0^{p_{{\rm F},n}}+\intop_0^{p_{{\rm F},p}}\right)
\frac{\rmd p p^2}{\pi^2}\,\sqrt{m_N^2\, \Phi_N^2(f)+p^2}\,,
\label{ENfun}\\
&&E_l[n_e,n_\mu]=\sum_{i=e\,, \mu}\intop_0^{p_{{\rm F},i}}\frac{\rmd p
  p^2}{\pi^2}\sqrt{m_i^2+p^2}\,,
\label{elept}\\
&&E_\om [n_n,n_p;f,\om_0 ] =
\frac{C_\om^2\, (n_n+n_p)^2}{2\, m_N^2\, \eta_{\rm \omega}(f)}
\nonumber\\
&&\qquad\qquad\qquad\qquad\quad
-\frac{m_N^2\eta_\om(f)}{2\,C_\om^2}\left[g_\om\,{\chi}_\om\,\om_0 -
\frac{C_\om^2(n_p + n_n)}{\,m_N^2\,\eta_\om (f)}\right]^2\,,
\label{ome}\\
&&E_\rho[n_n,n_p;f,\vec{\rho}_\mu] =E_{\rho,\rm neut}[n_n,n_p;f,\rho_0^{(3)}]
+E_{\rho,\rm ch}[n_n,n_p;f,\vec\rho_\mu]\,,
\nonumber\\
&&E_{\rho,\rm neut}[n_n,n_p;f,\rho_0^{(3)}]=
\frac{C_\rho^2 (n_n-n_p)^2}{8\,m^2_N\, \eta_\rho(f)}
\nonumber\\
&&\qquad\qquad\qquad\qquad\quad
-\frac{m_N^2\eta_\rho(f)}{2\,C_\rho^2}\left[g_\rho\,{\chi}_\rho\rho_0^{(3)}-
\frac{C_\rho^2(n_p-n_n)}{2\,m_N^2\,\eta_\rho (f)}\right]^2 ,
\nonumber\\
&&E_{\rho,\rm ch}[n_n,n_p;f,\vec\rho_\mu]=
-\frac{m_\rho^2}{2}\, \Phi_\rho^2(f)\, \Big( \vec{\rho\,}_\mu\,
\vec{\rho\,}^\mu-\rho_0^{(3)}\,\rho_0^{(3)}\Big)
\nonumber\\
&&\qquad\qquad\qquad \qquad\quad+\frac14\, (\Delta\rho_{\mu\nu}\big)^2
-\frac14\,\mu_{\rm ch}^\rho\frac{\prt}{\prt\mu_{\rm ch}^\rho}\,
\big(\Delta\rho_{\mu\nu}\big)^2\,.
\label{ERfun}
\ee
We have introduced here the dimensionless field
\be\label{f}
f=g_\sigma \chi_\sigma\sigma/m_N\,,
\ee
and the dimensionless
coupling constants $C_i={g_i\, m_N}/{m_i}$\,, for $i=\sigma,\,\om\,\rho$\,.
The nucleon and lepton Fermi momenta  $p_{{\rm F},i}$ are related to the
corresponding partial density as
$n_i={p_{{\rm F},i}^3}/(3\, \pi^2)$ for $i=n,p,e,\mu$\,.
The   contributions of mesonic fields entered with the scaling factors
\be
\eta_{i}(f)={\Phi_i^2(f)}/{{\chi}_i^2(f)}\,,\quad i=\sigma\,,\om\,,\rho.
\label{eta}\ee

Then one may
use equations of motion for $\om_0$ and
 $\rho$-meson fields
$\vec{\rho}_\mu$:
\be
&&\frac{\prt}{\prt \om_0 }\,\Omega [n_n,n_p;f,\om_0]=\frac{\prt}{\prt \om_0 }\,E [n_n,n_p;f,\om_0]=0\,,\nonumber\\
%\quad
&&\frac{\prt}{\prt \vec{\rho}_\mu}\,\Omega [n_n,n_p;f,\vec{\rho}_\mu]=\frac{\prt}{\prt \vec{\rho}_\mu}\,E [n_n,n_p;f,\vec{\rho}_\mu]=0\,.
\label{extreme}
\ee
The first line in (\ref{extreme}) yields
\be
E_\om [n_n,n_p;f] =\frac{C_\om^2\, (n_n+n_p)^2}{2\, m_N^2\, \eta_{\rm \omega}(f)}\,.
\label{omf}
\ee
Traditionally one considers only a neutral $\rho$-meson field $\rho_0^{(3)}$
disregarding a possibility for the presence of charged $\rho$-meson mean fields.
Then the second line in (\ref{extreme}) yields
\be
\rho_0^{(3)}= g_\rho \chi_\rho\, (n_p-n_n)/(2m_\rho^2
\Phi_\rho^2)\,.
\label{solt}
\ee
The corresponding contribution to the energy density is
\be\label{Er}
E_{\rho} [n_n,n_p;f]=\frac{C_\rho^2 (n_n-n_p)^2}{8\,m^2_N\,
\eta_\rho(f)}\,.
\label{enrhot}
\ee
As one can see, the source of the isovector $\rho$-meson field $\rho_0^{(3)}$ is the
isovector nucleon density $(n_p-n_n)$\,. Therefore for the isospin
symmetrical matter ($n_p=n_n$) we always have  $E_\rho=0$\,.

The main merit of this section is in the following.
Eqs.~(\ref{Efun}--\ref{Er}) demonstrate that instead of \emph{twelve} scaling
functions of a scalar field, $a_{N,\sigma,\om,\rho}$, $\widetilde\chi_{\rho,\om,\sigma}$,
$\phi_{N,\sigma\,\om,\rho}$ and $\widetilde U(\sigma)$, which enter the
initial Lagrangian (\ref{lag}), the energy density
functional  depends actually only on \emph{four} particular combinations of
these functions, $\eta_{\sigma,\rho,\om}(f)$ and $U(f)$. Note that the dependence
on the scaling function $\eta_{\sigma}$ can be always presented as a part
of the potential $U$ and absorbed in it by the replacement
$U\to U+\frac{m_N^4\,f^2}{2\, C_\sigma^2}\,
(1-\eta_{\sigma}(f))$\,, and vise versa the potential $U$ can be absorbed in
$\eta_{\sigma}$. 
Thus only \emph{three} independent functions
enter the energy-density functional.
The expression (\ref{Efun}) demonstrates explicitly equivalence
of mean-field Lagrangians  for constant fields with various $a$, $\widetilde\chi$ and $\phi$
parameters if they correspond to the same functions
$\eta_{\rho,\om}(f)$ and $\eta_{\sigma}$ (either $U(f)$). In sec.~\ref{sec:mass} we discuss
possible constraints on the $f$ dependence of the functions.
We would like to stress that the field $f$ relates now to a scalar
field $\sigma$ in a non-trivial way determined by eq.~(\ref{f}).

In order to obtain final expression for the energy of the system, one should
still use equation of motion for the scalar field $f$,
\be
\frac{\prt}{\prt f}\,\Omega [n_n,n_p;f]=\frac{\prt}{\prt f}\,E [n_n,n_p;f]=0\,.
\label{extremef}
\ee

\subsection{Determination of parameters of a RMF model}

Parameters of the RMF model, $C_\sigma$, $C_\om$, $C_\rho$, and
parameters of the potential $U$ are to be adjusted to reproduce
the nuclear matter properties at the saturation and have rather
broad uncertainties, cf.~\cite{ASK,asym}. One usually takes the
saturation density $n_0\simeq 0.16\pm 0.015$~fm$^{-3}$\,, the
binding energy per particle, $e_{\rm B}(n_0)\simeq -15.6\pm
0.6$~MeV, the compressibility modulus $K(n_0)\simeq 240\pm
40$~MeV, the effective  nucleon mass $m_{N}^* (n_0)/m_N \simeq
0.75\pm 0.1$\,, and the symmetry energy coefficient $a_{\rm
sym}(n_0)=32\pm 4$~MeV. Some models exploit parameters beyond
even these large error bars. E.g., the original Walecka
model~\cite{walecka} used $n_0 \simeq 0.191$~fm$^{-3}$ producing
$K\simeq 540$~MeV and $m^*_N /m_N \simeq 0.56$.

To facilitate the comparison among different RMF models which we
study below we will use the same basic input parameters
\be
&&n_0 =0.16~\mbox{fm}^{-3}, \quad e_{\rm B}=-16~\mbox{MeV},  \quad
a_{\rm
sym}(n_0)=32~\mbox{MeV},\nonumber \\
&&m_N=938~\mbox{MeV}\,.
\label{param}
\ee

The saturation density and the binding energy are related as
\be
\frac{\prt E[n;f]}{\prt n}\Bigg|_{n_0,f(n_0)}=
\frac{1}{n_0}\,E[n_0;f(n_0)]=m_N+e_{\rm B}\,,
\label{satur}
\ee
and the compressibility modulus is given by
\be
K=9 \,n_0\left[\frac{\prt^2 E}{\prt n^2}\Bigg|_{n_0,f(n_0)}-
\left( \frac{\prt^2 E}{\prt n\, \prt f}\Bigg|_{n_0,f(n_0)}\right)^2\,
\left[\frac{\prt^2 E}{\prt
f^2}\Bigg|_{n_0,f(n_0)}\right]^{-1}\right]\,.
\label{kmod}
\ee
Here $f(n_0)$ is a solution of eq.~(\ref{extremef}) at the density $n_p=n_n=n_0/2$\,.
The coupling constant $C_\rho$ is determined from the symmetry
energy coefficient of the nuclear matter
\be
a_{\rm sym}(n)=\frac{n}{8}\frac{\prt^2}{\prt n_p^2} E[n-n_p,n_p]\Big|_{n_p=n/2}=
\frac{C_\rho^2\, n}{8 m_N^2\, \eta_\rho}+\frac{\pi^2\, n}{4\, p_F\,
 \sqrt{m_N^{*2}+p_{\rm FN}^2}}\,,
\label{symen}
\ee
$p_{\rm FN}$ is the nucleon Fermi momentum in isospin-symmetric matter.

\begin{figure}
\bc
\includegraphics[height=6cm,clip=true]{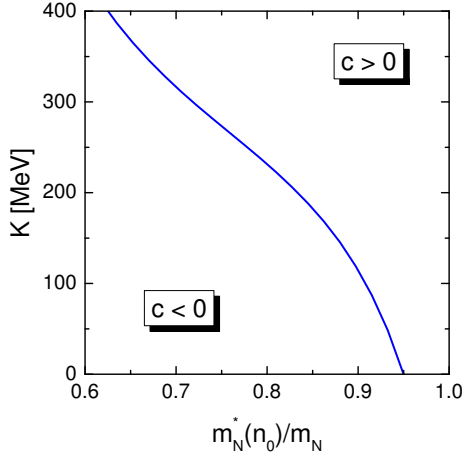}
\ec
\caption{The line shows the compressibility   $K$ of the isospin symmetrical
  matter vs.
the effective nucleon mass $m_N^* (n_0)$ corresponding to $c=0$ for the
MW model, input set (\ref{param}).}
\label{fig:cbord}
\end{figure}

Generic Lagrangians, $\mathcal{L}_N$ and $\mathcal{L}_M$ in (\ref{lag})
and the energy density functional (\ref{Efun}) reproduce various
types of mean-field models. Consider several examples.
The choice
\be
a_{N,\, \sigma,\, \om,\, \rho}&=&\widetilde\chi_{N,\, \sigma,\, \om,\, \rho}
=\phi_{\sigma,\, \om,\, \rho}=1\,,
\quad \phi_N=1-{g_\sigma\, \sigma}/{m_N}\,, \\
\widetilde\chi_\rho^{'} &=&\widetilde{U}=0,\nonumber
\label{w}
\ee
yielding the energy-density functional (\ref{Efun},\ref{ENfun},\ref{omf},\ref{Er}) with
$\eta_{\sigma,\om,\rho}=1$, and $\Phi_N=1-f$,
corresponds to the standard Walecka model (W)\footnote{Inclusion of the
neutral $\rho_0^{(3)}$ field generalizes the model to
describe asymmetrical nuclear matter. We call the models, that include
$\rho$-meson fields, by the same names as those without $\rho$-meson fields.}
with the minimal number of free
parameters (since $U=0$)~\cite{walecka}.
For (\ref{param}) we find the parameter set
\be
\mbox{W}: \quad C_\sigma^2&=&329.70 \,,\,\, C_\om^2=249.40\,, \quad C_\rho^2 =68.09\,,
\label{par-W}\ee
which produces
\be
K&\simeq& 553~\mbox{MeV},\quad m_N^*(n_0)/ m_N \simeq 0.54.
\label{W-Km}
\ee
Thus this
minimal model does not allow to fit appropriately the values of the
nucleon  effective mass  and  compressibility modulus.
In order to cure these problems, it was suggested in ref.~\cite{BB77}
to  introduce a  scalar-field self-interaction
\be\label{nonlinpot}
U(f) =m_N^4 (b\,f^3/3+c\,f^4/4 )=\frac{b}{3}\,(m_N\,g_{\sigma}\,\sigma)^3
+\frac{c}{4}\,(g_{\sigma}\,\sigma)^4\,.
\ee
Two additional parameters, $b$ and
$c$, make the model (called the modified Walecka model (MW))
able to  accommodate realistic values of
the nuclear  compressibility and  effective nucleon mass.
An extra  attention
should be paid to the fact that the coefficient ``$c$'' should be positive, otherwise
there is no stable vacuum for the Lagrangian
(\ref{lagM},\ref{lagMn})\footnote{Some parameter sets used, e.g.,
in~\cite{G00}, do not respect this constraint dealing thus with a metastable
state. One then assumes that a mean field energy is valid  only near a
local minimum.}.
The solid line in Fig.~\ref{fig:cbord} shows values of $K$
and $m_N^*(n_0)$, for  the boundary case $c=0$. Above this line $c>0$, and below
$c<0$. With a positive $c$, low values of $m_N^*(n_0)$ correspond to
unrealistically high values of the compressibility $K$.

 Using as input (\ref{param}) and
\be
\quad K=270~\mbox{MeV}, \quad m^*_N(n_0)/m_N =0.8
\label{MW-Km}
\ee
we obtain
\be
\mbox{MW}: \quad &&C_\sigma^2=189.94\,,\,\,
C_\om^2=90.768\,, \quad C_\rho^2=100.18\,,
\,\, \nonumber\\
&&b=6.3714\times 10^{-3}\,,\,\,
c=1.6288\times 10^{-2} \,.
\label{parMW}
\ee

The choice, cf. (\ref{lag}),
\be
a_N&=&1+{g_\sigma \sigma}/{m_N}\,,\quad
a_{\sigma,\,\om,\,\rho}=\phi_{N,\,\sigma,\,\om,\,\rho}=
\widetilde\chi_{N,\,\sigma,\,\om,\,\rho}=1, \\
\widetilde\chi_\rho^{'} &=&\widetilde{U}=0\nonumber
\label{zm}
\ee
corresponds to the original Zimanyi--Moszkowski model with the
derivative coupling (ZM)~\cite{zm}. In
eqs.~(\ref{Efun},\ref{ENfun},\ref{omf},\ref{Er})
we have then $U=0$, $\eta_{\sigma,\om,\rho}=1$,
$\Phi_{\sigma,\om,\rho}=1$ and
$\Phi_N=1/(1+f)$\,. For the ZM model with the input parameters (\ref{param})
we obtain
\be\label{parZM}
\mbox{ZM}: \quad &&C_\sigma^2=169.36\,, \quad C_\om^2=59.055,\quad
C_\rho^2=104.56\,.
\ee
This model reproduces realistic values of the compressibility modulus and the effective
nucleon mass at $n_0$,
\be
K\simeq 224.5~\mbox{MeV}, \quad m_N^* (n_0)/m_N\simeq 0.85\,,
\label{ZM-Km}
\ee
without introduction of extra free parameters.
As the model, which eliminates successfully the deficiencies of the
original Walecka model keeping the minimal number of parameters, the
ZM model has gained a lot of attention and has been widely used.
Notice that, the ZM model can be also considered as a
generalization of the W model done with the introduction of  a scaling of the $\sigma N$ coupling
\be
g_\sigma^*=g_\sigma\chi_{\sigma}=
{g_\sigma}/{(1+g_\sigma \sigma/m_N)}\,.
\label{gs}
\ee

To close this section we note that
eqs.~(\ref{effmass}) and (\ref{chibar}) demonstrate that for constant mean
fields under consideration {\em{
"dielectric constants'' $a_i$ can be always absorbed into
effective masses  and effective coupling constants.}}
Therefore, below we will put $a_i=1$ and deal with the functions $\Phi_i$
and $\chi_i$ only.

\section{RMF models with dropping hadron masses}\label{sec:mass}

In this section we construct RMF models which support a decrease
of hadron masses with an increase of the baryon density.
Simplifying we follow the original
Brown-Rho conjecture~\cite{BR} about the hadron mass scaling, which
in our case imposes
\be
\Phi_N=\Phi_\sigma=\Phi_\om=\Phi_\rho=\Phi (f)\,.
\label{BR}\ee
Thus for the generalized W and
MW models we would have $\Phi=\Phi_{\rm W}=\Phi_{\rm MW}(f)=1-f$ and for the generalized
ZM model, $\Phi=\Phi_{\rm ZM}(f)=1/(1+f)$\,.
We recall that $f$ is related to  the $\sigma$ field via
(\ref{f}).

We have to note that due to non-linear interactions effective masses obtained on a mean-field level
should not be in general identified with positions of the poles in propagators
of corresponding excitations.\footnote{In our case a difference appears for the
  $\sigma$-meson and nucleon masses.} Following  Brown and Rho~\cite{BR} we
apply a mass scaling  on a mean-field level.

\subsection{RMF models with decreasing hadron masses and fixed coupling constants}
\label{ssec:nogscale}

Assuming that coupling constants do not change in the medium,
i.e. ${\chi}_{i}=1$, we obtain the following scaling factors in the
energy-density functional
\be
\eta_{i}=\Phi^2(f)\,, \quad i=\sigma\,,\omega\, , \rho .
\label{efm}
\ee
For the generalized W and MW models we have
$\Phi=\Phi_{\rm W}=\Phi_{\rm MW}(f)=1-f$
and for the generalized ZM model, $\Phi=\Phi_{\rm ZM}(f)=1/(1+f)$.
In terms of a $\sigma$ field these functions have familiar forms
$\Phi_{\rm MW}(\sigma)=1-g_\sigma\,\sigma/m_N$ and
$\Phi_{\rm ZM}(\sigma)=1/(1+g_\sigma\,\sigma/m_N)$\,.

The resulting functional possesses several unpleasant features.
First, effective hadron  masses do not decrease monotonously
with the density increase. Decreasing at small densities, they start to
increase at higher densities. Such a non-linear behavior has been observed, e.g.,
in ref.~\cite{manka}. To show that this is a generic feature of such a
type of models we calculate $\prt f(n)/\prt n$, where $f(n)$ is a
solution of eq.~(\ref{extremef}) for isospin symmetrical
nuclear matter with the total density  $n=n_p+n_n$\,. From
(\ref{extremef}) we have
\be
\frac{\prt f}{\prt n}=-\frac{\prt^2 E_N[n;f]}{\prt n \prt f}\Bigg|_{f(n)}\Bigg/
\frac{\prt^2 E_N[n;f]}{\prt f^2}\Bigg|_{f(n)}\,,
\ee
where
the denominator $\prt ^2 E_N/\prt f^2$ should be always positive to assure the
stability of the mean-field solution (\ref{extremef}).
The nominator,
\be
\frac{\prt^2 E[n;f]}{\prt n \prt f}\Bigg|_{f(n)}=
-\frac{C_\om^2\, n}{m_N^2\, \eta_{\rm V}^2}\,
\frac{\prt \eta_{\om}}{\prt f}+
\frac{\prt \Phi_N}{\prt f}\,\frac{m_N^2\,\Phi_N(f)}
{\sqrt{m_N^2\, \Phi_N^2(f)+p_{\rm F}^2}}\,,
\label{dEdndf}
\ee
changes the sign at some value of $n$, provided $\prt \eta_{\rm
\om}/\prt f$ and $\prt \Phi_N/\prt f$ have the same sign. Indeed
at small densities the second term  dominates, whereas with the
density increase the first term begins to win. Thus, if $f$
increases at small densities (hadron masses drop), at larger
densities $f$ begins to decrease, resulting in an increase of
hadronic masses. This effect can be enhanced in the isospin
asymmetric nuclear matter due to the rho-meson contribution, if
$\prt \eta_{\rho}/\prt f$ has the same sign as $\prt
\eta_{\om}/\prt f$ and $\prt \Phi_N/\prt f$\,. The second feature
of the energy-density functional (\ref{Efun}) with scaling
factors (\ref{efm}) is that it does not guarantee the continuity
of  the density dependence of the scalar field $f$. The equation
of motion (\ref{extremef}) for $f$ may have several solutions,
which can appear or disappear with a change of the density,
leading to jumps of the $f(n)$ function. Third, the scaling of the
mass of the scalar $\sigma$-meson field included in a
Dirac-Brueckner-Hartree-Fock calculation of nuclear matter leads
to over-binding of the nuclear matter, as has been  found in
ref.~\cite{rmdb99}. Fourth, in ref.~\cite{LBLK97} it was noticed
that increasing the vector repulsive potential, due to the factor
$1/\eta_{\om}$, would produce too strong nucleon flow in heavy-ion
collisions. Such a flow would be in odd with experimental data.
This observation stimulated authors of ref.~\cite{SBMR97} to
propose  the scaling of the vector-meson coupling constant
${\chi}_\om\sim \Phi_\om$, which leads to $\eta_{\om}\simeq 1$\,.
In the scalar sector the problems can be solved if we put
$\eta_{\sigma}\simeq 1$ and choose the potential $U$ so that there
is only one solution of the equation of motion for $f$.

Information about density dependence of the symmetry energy
$a_{\rm sym}(n)$, cf. (\ref{symen}), is rather controversial.
The microscopic calculation within the relativistic
Dirac-Brueckner-Hartree-Fock approach~\cite{lklb98} shows that
$a_{\rm sym}(n)$ increases almost linearly with the density. This
would require $\eta_\rho\simeq 1$ in (\ref{symen})\,.
Ref.~\cite{Gai} discussing heavy ion collision properties uses a
stiffer density dependence of $a_{\rm sym}(n)$ ($\eta_\rho$ decreases
with a density increase). On an other hand, the HHJ
EoS \cite{HHJ} fitting the microscopic model of Urbana-Argonne
(A18+$\delta v$ +UIX*)\cite{APR98} produces $a_{\rm
sym}(n)\propto n^{0.6}$ that corresponds to increasing
$\eta_{\rho}\simeq (n/n_0)^{0.4}$. Note that the latter case
allows to solve the problem of the low DU threshold density.

\subsection{RMF models with a universal scaling of
hadron masses and coupling constants}\label{brUn}

\subsubsection{Construction of the model}

Following the conjecture of ref.~\cite{SBMR97}
and argumentation in the previous subsection we study now the case
of a universal scaling of hadron masses and coupling constants,
\be
\Phi_N=\Phi_\sigma=\Phi_\om=\Phi_\rho= \chi_\sigma =\chi_\om =\chi_\rho =
\Phi(f)\,,
\label{BR-un}\ee
which implies
\be
\eta_\sigma=\eta_\om=\eta_\rho=1\,.
\label{eta-MWu}
\ee
Note that according to a hypothesis of a "vector manifestation"~\cite{Rho-vm} the vector meson mass must go to
zero in proportion to the corresponding coupling. At the saturation nuclear matter density
refs. \cite{SBMR97,Rho} give
$\Phi (n_0)\simeq 0.78$ according to their analysis  of the gyro-magnetic
ratio that agrees with our input value (\ref{MW-Km}), $m_N^* (n_0)/m_N \simeq
0.8$. 
%%%The latter ratio relates to the Landau mass,  entering nucleon
%%%quasiparticle distributions near Fermi surface, as $m_N^{\rm L}
%%%(n_0)\simeq 0.85 m_N$. This is only slightly less than the value  $m_N^{\rm
%%%  L}(n_0)\simeq 0.9m_N$ extracted from the analysis of the single-particle spectra in
%%%nuclei, see \cite{KS82}.

Note also that for the case $\chi_\sigma\neq 1$ we have a non-trivial relation between
$f$ field and an original $\sigma$ field (\ref{f}). The W and ZM models (in both of them $U=0$) with the universal
scaling (\ref{BR-un},\ref{eta-MWu}) we will denote as the W(u) and
ZM(u) models, respectively, and the MW model with the universal
scaling we will denote MW(u).
For the W(u) and MW(u) models we then have
\be
\Phi(f)=1-f\quad \longrightarrow\quad
f=\frac{g_\sigma\,\sigma/m_N}{1+\frac{g_\sigma\,\sigma}{m_N}}
\quad \longrightarrow\quad
\Phi(\sigma)=\frac{1}{1+\frac{g_\sigma\,\sigma}{m_N}}\,.
\label{MWf2s}\ee
For the ZM(u) model the correspondence is a more complex:
\be
\Phi(f)=(1+f)^{-1}\quad&\longrightarrow&\quad
f=\sqrt{\frac14+\frac{g_\sigma\,\sigma}{m_N}}-\frac12
\nonumber\\
\quad &\longrightarrow&\quad
\Phi(\sigma)=\left(\sqrt{\frac14+\frac{g_\sigma\,\sigma}{m_N}}+\frac12
\right)^{-1}\,.
\label{ZMf2s}
\ee

We note the obvious equivalence between the original Walecka model (W) and
the W(u) model, and also between the original 
ZM  and ZM(u) models for constant mean fields. We will denote models with a non-universal
scaling as (nu) models. From (\ref{MWf2s}) we see that W(u) model is in
turn equivalent to the ZM(nu) model with
$\Phi_\sigma(\sigma)=\Phi_N(\sigma)=1/(1+g_\sigma\,\sigma/m_N)$ but
$\chi_\sigma=1$ and $\Phi_{\om,\rho}=\chi_{\om,\rho}=1$.
Analogously we find equivalence of the  ZM(u) model and the W(nu) model with
$\Phi_{\sigma,\om,\rho}=\chi_{\om,\rho}=1$ and $\chi_{\sigma}(\sigma)=\Phi_N(\sigma)=
1 - g_\sigma\,\sigma/m_N$.

As it follows from (\ref{lagNn},\ref{lagMn},\ref{lagORn},\ref{effmass},\ref{chibar}),
and (\ref{BR-un}) in the absence of the charged $\rho$ meson mean fields and
the potential $U$, with the help of the replacements
\be\label{varr}
\Phi \xi \rightarrow \xi\,,\quad \mbox{where}\quad
\xi =\sigma,\om,\rho\,,
\ee
we can equivalently transform a RMF model
with a universal scaling of meson masses and coupling constants
($\eta_{\sigma,\om,\rho}=1$) to a model without any scaling of meson masses and coupling
constants.
The absence of the non-Abelian interaction of $\rho$ meson fields and the potential $U$
is very important in order to arrive at such a conclusion,
since the transition (\ref{varr}) would affect the non-linear sectors of the model.

For the MW model with $U$ given by (\ref{nonlinpot})
we can construct an equivalent MW(u) model with the universal
 scaling of meson masses and coupling constants
(\ref{BR-un}) for $\Phi =1-f$ and the same form of $U$ written in terms of an $f$ field.
In terms of the $\sigma$ field the potential of the new MW(u)
model looks more cumbersome:
\be
U&=&m_N^4 (\frac{b}{3}\,f^3 +\frac{c}{4}\,f^4 )
=\frac{b}{3}\,\frac{m_N\,(g_{\sigma}\,\sigma)^3}
{\left( 1+\frac{g_{\sigma}\,\sigma}{m_N}\right)^{3}}
+\frac{c}{4}\,\frac{(g_{\sigma}\,\sigma)^4}
{\left( 1+\frac{g_{\sigma}\,\sigma}{m_N}\right)^{4}}\,.
\label{Unew}
\ee
Note that potentials $U(\sigma)$ in the
MW and MW(u) models, cf. (\ref{nonlinpot},\ref{Unew}), look  quite  differently
in terms of the $\sigma$ field although in terms of the $f$
fields they are the same. We used the same letter $f$, since
the mean values of those $f$ fields,
determined from the very same equation of motion, are identical. This
illustrates that, if thermodynamic
characteristics look the same in terms of the variables found in
the minimization procedure, then the RMF models
are equivalent.

As an illustration we also exploit a simplified ansatz proposed in
ref.~\cite{SBMR97} (further the  SBMR model) for the isospin symmetrical matter, where the Brown--Rho
scaling was included into RMF EoS. Assume that the energy density
of the model  takes the form that we have used above with
\be
&&\eta_{\om}=\Phi^2(n;z)/\Phi^2(n;y)\,,\quad \eta_{\sigma}=\Phi^2(n;y)\,,\quad
\Phi_N(n,f)=\Phi(n;y)-f\,,
\nonumber\\
&&\Phi(n;y)={1}/{\Big(1+y\, p^3_{\rm F}(n)/(260~{\rm MeV})^3\Big)}\quad
p_{\rm F}^3(n)=(p_{{\rm F},i}^3+p_{{\rm F},i}^3)/2\,.
\label{cmbr}
\ee
Here $f=h\sigma /m_N$, $h$ is the coupling constant and $\sigma =\phi$ in
notations of ~\cite{SBMR97}, eq. (32).
Parameters of the model are  $m_N=939$~MeV, $y=0.28$, $z=0.31$,
$C_\om^2=332.27$ and $C_\sigma^2=50.546$, which correspond to
$n_0=0.151$~fm$^{-3}$, $e_B=-16.1$~MeV, $K=259.6$~MeV, and $m_N^*(n_0)=0.675\,m_N$.
Here $\eta_{\om}$ is very closed to unit and $\eta_{\sigma}$ depends on $n$
rather than on $f$. The pressure and chemical potentials of partilces
are related to the
energy following eqs. (\ref{presmu}).
Note that owing to an explicit dependence of the energy on the nucleon density
expressions for the neutron and proton chemical potentials deviate from those
given by eqs. (\ref{mun}) and (\ref{mup}).
%%As we have mentioned, such models need a special treatment
%%due to problems with thermodynamic consistency.

In order to consider asymmetrical matter within the same
framework we add the rho-meson contribution (\ref{Er}) with
$\eta_\rho=1$ and $C_\rho^2=100.08$.
The latter quantity follows from the value of $a_{\rm sym}(n_0)=32~$MeV in
(\ref{param}).

\begin{figure}
\begin{center}
\includegraphics[clip=true,height=6cm]{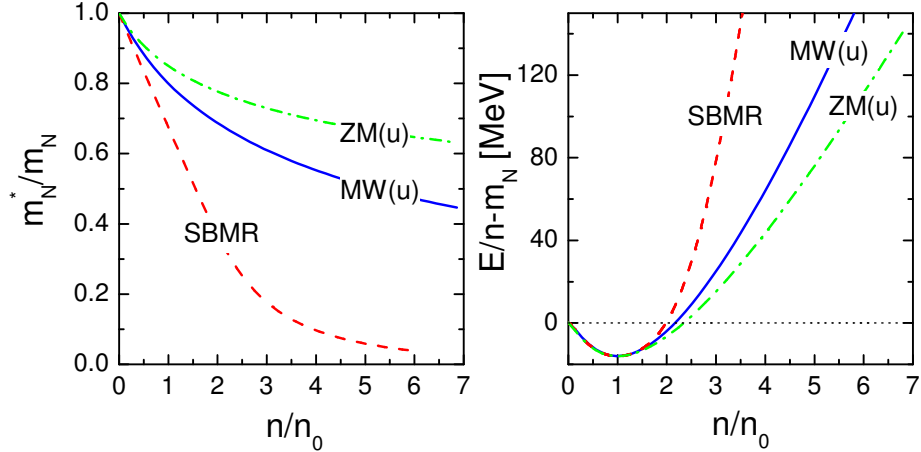}
\end{center}
\caption{The effective nucleon mass (left panel) and the energy per particle
  (right panel)
of the
isospin symmetrical nuclear matter within the MW(u)
(input-parameter set (\ref{param},\ref{MW-Km},\ref{parMW})), ZM(u)
(input-parameter set (\ref{param},\ref{parZM})) and
SBMR model (\ref{cmbr}).}
\label{fig:eos}
\end{figure}

In Fig.~\ref{fig:eos} we show the effective nucleon mass and the
EoS for the isospin symmetrical nuclear matter calculated for
three  models of EoS: MW(u) (input-parameter set (\ref{param},\ref{MW-Km},\ref{parMW})),
ZM(u) (input-parameter set (\ref{param},\ref{parZM}))
and SBMR, (\ref{cmbr}). We see that these
three models cover a broad range of possible density dependences
of the effective mass and the energy per particle. Compared to the RMF models the SBMR model produces
very rapid decrease of the effective nucleon mass and a stiff EoS (at 
a reasonable value of the compressibility for $n=n_0$).

\subsubsection{Neutron star composition}

The composition of the NS matter (in absence of mixed phases and a charged condensate)
is determined in the  standard manner by the
local charge-neutrality condition and the conditions of the equilibrium with respect
to weak reactions $e\leftrightarrow \mu$ and $n\leftrightarrow p+e$\,.
The latter ones impose chemical equilibrium conditions for  the corresponding chemical
potentials
\be
\mu_e=\mu_\mu=\mu_n-\mu_p\,,\qquad \mu_i=\frac{\prt E_{\rm
    tot}}{\prt n_i}\,, \quad i=e,\mu,n,p\,.
\label{weak}
\ee
Here ${E}_{\rm tot}$ is the full energy-density.
For the energy density (\ref{Efun})
conditions (\ref{weak}) and the local charge-neutrality condition result in
the system of equations
\be
&&\mu_e=\sqrt{m_N^2\,\Phi_N^2(\bar f)+p_{{\rm F},n}^2}-
\sqrt{m_N^2\,\Phi_N^2(\bar f)+p_{{\rm
F},p}^2}+\frac{C_\rho^2\,(n_n-n_p)}{2\,m_N^2\,\eta_\rho(\bar
f)}\,,
\nonumber\\
&&n_p=\frac{(\mu_e^2-m_e^2)^{\frac32}}{3\,\pi^2}\theta (\mu_e^2-m_e^2)
+ \frac{(\mu_e^2-m_\mu^2)^{\frac32}}{3\,\pi^2}\theta (\mu_e^2-m_\mu^2)\,,
\label{beta-eq}
\ee
with $\bar f=f(n)$ being a solution of the equation of motion
(\ref{extremef}) for the scalar field.
Eqs. (\ref{beta-eq}) are solved  with respect
to $n_p$ for a given total density $n=n_p+n_p$.
\begin{figure}
\begin{center}
\includegraphics[clip=true,height=6cm]{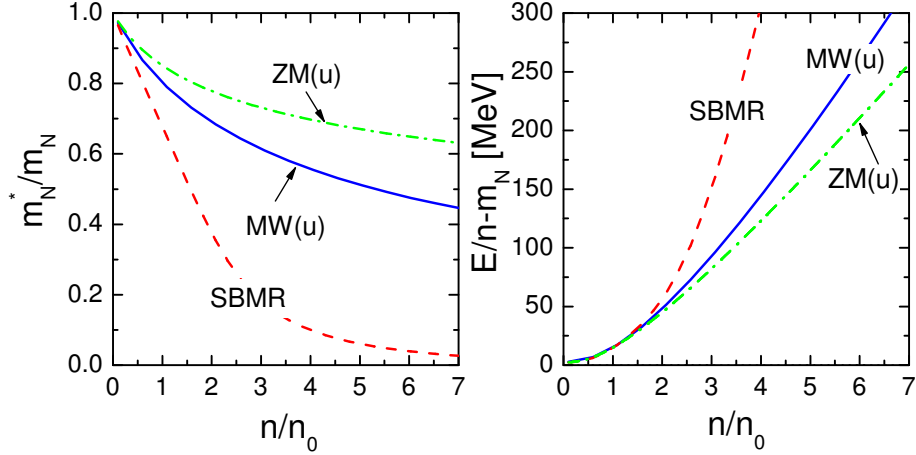}
\end{center}
\caption{The same as in Fig.~\ref{fig:eos} but for the neutron star matter.
}
\label{fig:eosN}
\end{figure}
In Fig.~\ref{fig:eosN}
we show the effective nucleon mass and the
EoS for the NS matter
calculated for MW(u) (input-parameter set (\ref{param},
\ref{MW-Km},\ref{parMW})), ZM(u) (input-parameter set (\ref{param},\ref{parZM}), and SBMR models.
Compared to the isospin symmetrical case (Fig.~\ref{fig:eos}) the EoS of the NS matter becomes
substantially stiffer for all three models.

\begin{figure}
\begin{center}
\includegraphics[clip=true,width=\textwidth]{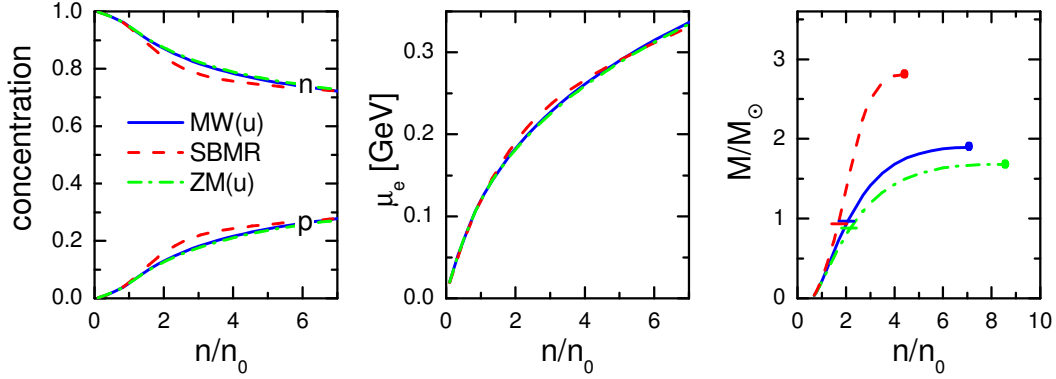}
\end{center}
\caption{The proton and neutron concentrations
(left panel), and the electron chemical potential (middle panel) in the
neutron star matter calculated for the same models as in Figs~\ref{fig:eos},
\ref{fig:eosN}. The right panel shows the neutron star mass as a
function of the central density. Horizontal cuts indicate the DU thresholds.
}
\label{fig:nspar}
\end{figure}

In Fig.~\ref{fig:nspar} (left and middle panels) we show the composition of
the NS
matter and the electron chemical potential, as they follow
from the solution of equations (\ref{weak}), (\ref{beta-eq})
for the three specified above  models (MW(u), ZM(u), and SBMR).
We see that  all three models produce
the very similar NS compositions and $\mu_e$\,. The proton
concentration $Y_p =n_p /n$ is rather high so that the DU process
becomes operative at $n=2.13\, n_0$, $2.07~n_0$, and $1.72~n_0$
for the ZM(u), MW(u), and SBMR models, respectively. The
corresponding NS masses are $M=0.88$, 0.99, and $0.97~M_\odot$.
{\em{Thus, yielding so low DU threshold densities these models
cannot adequately describe the NS cooling, cf. \cite{BGV04}.}} This
is a serious drawback. In the right panel of Fig.~\ref{fig:nspar}
we show the NS masses (in units of solar mass $M_\odot=2\times
10^{33}$~g) resulting from the integration of the differential
Oppenheimer-Volkoff equations~\cite{G00} for a given central
density of the star. We stopped the integration at the inner crust-core
boundary, 
because the EoS following (\ref{Efun}) cannot be used
any longer.  In different
models the value of this boundary density varies in the interval $n_{\rm crust}\simeq (0.5\div
0.8)\,n_0$. In our calculations we settle $n_{\rm crust}\simeq 0.8\,n_0$.
Since the crust contributes only a little to the NS
mass (yielding an additional mass $\leq 0.06~M_{\odot}$), we do not
supplement our calculations by subsequent calculation of the crust. We see that the 
ZM(u) RMF model has an extra shortcoming
that the limiting NS mass is rather low. Opposite, the SBMR model shows an example
of a very stiff EoS ($M_{\rm lim}\simeq 2.8~M_{\odot}$). 
%%However as we have mentioned the SBMR model suffers of a
%%serious shortcoming such as a thermodynamic inconsistency since the SBMR
%%energy depends not only on the field variables but also on an external
%%variable, the nucleon density.

\begin{figure}
\begin{center}
\includegraphics[clip=true,width=6cm]{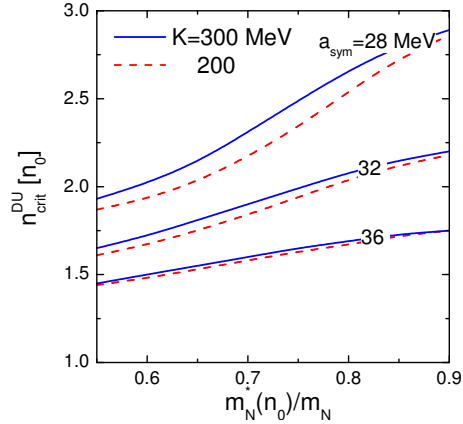}
\end{center}
\caption{
The threshold density for the DU process as a
function of the $m^{*}_N (n_0)$ for the neutron star matter, for the different
MW(u) models, the input set (\ref{param}),
at two values of  $K$ and three values $a_{\rm sym}$.
}
\label{fig:ndu}
\end{figure}

It is interesting to study the dependence of a DU threshold
density and a limiting NS mass on the input parameters of a RMF
model. We will use the MW(u) keeping the basic input parameters
(\ref{param}) and varying parameters $m_N^*(n_0)$ and $K$.
Fig.~\ref{fig:ndu} shows threshold densities for the DU process as
a function of the magnitude of the effective nucleon mass at $n_0$ for
two values of the compressibility $K=200$ and $300$~MeV and for
three values of the symmetry energy coefficient $a_{\rm
sym}(n_0)=28$, 32, $36~$MeV. We see that $n_{\rm crit}^{\rm DU}$
increases with increase of $K$, and $m^{*}_N (n_0)$ and decreases
with increase of $a_{\rm sym}$. In all cases $n_{\rm crit}^{\rm
DU}$ is more sensitive to the changes of $a_{\rm sym}$ and
$m^{*}_N (n_0)$, than to the change of the compressibility. We see
that all RMF models yield $n_{\rm crit}^{\rm DU}<2.7~n_0$ for
reasonable values of parameters. Hence all these models will meet
difficulties in applications to a NS cooling problem, cf. NS
masses in Fig.~\ref{fig:nspar}. As we have mentioned,  models with
$M (n_{\rm crit}^{\rm DU})<1.3~M_{\odot}$ meet serious problems,
since the latter value is smaller than the averaged value of the
NS mass, $M\simeq 1.35\pm 0.04~M_{\odot}$, and  models producing
$M (n_{\rm crit}^{\rm DU})<1~M_{\odot}$ should be  completely excluded.

\begin{figure}
\begin{center}
\includegraphics[clip=true,width=6.5cm]{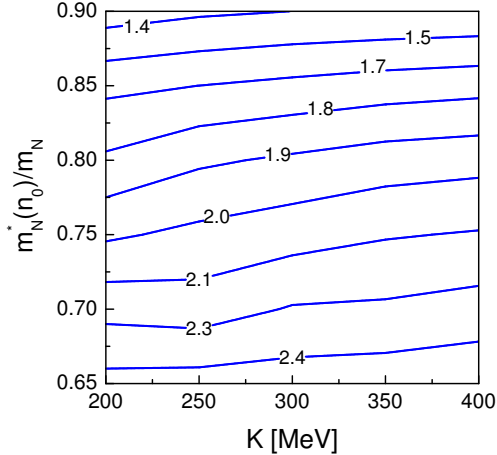}
\end{center}
\caption{
The counter plot of the limiting masses of neutron stars (counter labels
in $M_\odot$), for the MW(u) models, the input set (\ref{param}), for different values of the input
parameters $K$ and $m_{N}^{*}(n_0)$.
}
\label{fig:Mlim}
\end{figure}

The dependence of the limiting NS mass on the input parameters is
illustrated in Fig.~\ref{fig:Mlim}. The limiting mass is a
sharp function of $m_N^*(n_0)$ and a rather smooth function of
$K$. We see that the limiting mass $>(2.2\div 2.3) M_\odot$ can be
reached  either for models  exploiting unrealistic values of the
compressibility modulus $K>300$~MeV, or for $m_N^*(n_0)<0.7 m_N$.
Therefore if one observed  a NS of a mass $M\gsim 2.2~M_\odot$,
see \cite{S04}, this would mean that many RMF models without any
scaling or with a universal scaling of masses and
coupling constants would have a problem. One of the possibility to avoid
the problem is to use $m_N^*(n_0)<0.7 m_N$. 
However, in the latter case one may
meet a problem with description of the atomic nucleus data. Besides then
either the compressibility is unrealistically high or $c<0$, as it follows from Fig.~\ref{fig:cbord}.
%%%the single-particle spectra in nuclei \cite{KS82}.

\subsection{RMF models with a  non-universal scaling for masses and couplings}
\label{sec:absence}

\subsubsection{Solving problems with DU reactions and limiting NS mass}

Obviously the universal scaling law may hold only approximately.
In this section we demonstrate that allowing some departure from
the universal scaling of the hadron masses and the coupling
constants (\ref{BR-un}) we are able to obtain a somewhat stiffer
EoS and shift the threshold density for the DU process to higher
densities. We will consider the MW(nu) models with $\eta_{\sigma}=1$
but $\eta_\om (f)<1$, being decreasing with $n$ for $n>n_0$. This
behavior enhances the repulsion in the  potential at large
densities. In absence of negatively charged bosons to increase the critical density for the DU processes
we should reduce the proton concentration. This can be achieved by
choosing $\eta_\rho(f)>1$,  being increasing with $n$ for $n>n_0$,
that results in  a suppression of the rho-meson contribution to
the EoS.

As we have argued in
section~\ref{ssec:nogscale}, if $\eta_\om'(f)<0$, then a decrease of
a hadron mass will stop at some density and turn to an increase.
However we can choose the scaling to be weak enough to shift a
turning point  to a high density not reached  in the center of the heaviest
NS (of the limiting mass). In the case of the NS matter we can
exploit the fact that the first term in (\ref{dEdndf}) responsible for
the turn-over of the hadron mass scaling can be partially
compensated by the analogous term from the rho-meson contribution,
if $\eta_\rho'(f)>0$. For the pure neutron matter the exact cancellation
condition takes the form
$4\,C_\om^2\,\eta_\om'(f)\,\eta_\rho^2(f)=-C_\rho^2\,\eta_\rho'(f)\,\eta_\om^2(f)$\,.
An appropriate behavior of the scaling factors is obtained, e.g., with
\be\label{etnun}
\eta_\sigma=1\,,\quad
\eta_\om (f)=\frac{1+z\,f_0}{1+z\,f}\,,
\quad \eta_\rho (f)=\frac{\eta_\om (f)}{\eta_\om
(f)+4\,\frac{C_\om^2}{C_\rho^2}\,(\eta_\om (f)-1)},
\ee
where $f_0=1-m_N^*(n_0)/m_N$\,. Since $\eta_\om
(f_0)=\eta_\rho(f_0)=1$, the values of parameters $C_\om$ and
$C_\rho$ do not change.
Other parameters related to the scalar field  change just slightly.

\begin{figure}
\bc
\includegraphics[height=6cm,clip=true]{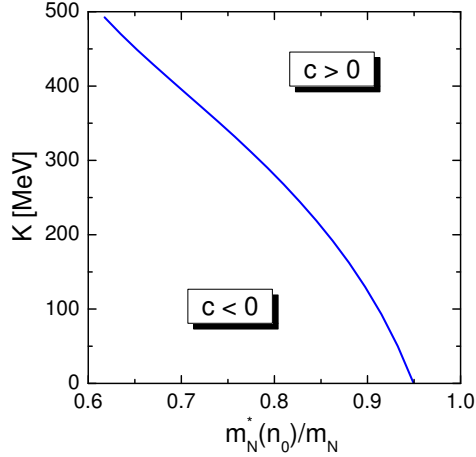}
\ec
\caption{The line shows the compressibility   $K$ of the isospin symmetrical
  nuclear matter vs.
the effective nucleon mass $m_N^* (n_0)$ corresponding to $c=0$ for the
MW(nu) model with the scaling~(\ref{etnun}) and $z=0.65$ for the input set
(\ref{param}).}
\label{fig:cbord_sc}
\end{figure}

We fit now the parameter $z$ in order to increase the limiting NS mass
$M_{\rm lim}$ and to push the threshold density  $n_{\rm crit}^{\rm DU}$
to higher densities. The optimal value is found to be $z=0.65$\,.
We use the same parameter set (\ref{param}), as before.
Had we taken the same values of the compressibility modulus
$K=270$~MeV  and $m_N^{*}(n_0)/m_N =0.8$, as we have used above,
we would obtain negative value of $c$.
The "$c=0$" border shown in Fig.~\ref{fig:cbord} shifts slightly up, if we
apply the scaling (\ref{etnun}).
In order to get positive $c$ we take
slightly different values for the effective mass
and the compressibility modulus compared to those in  (\ref{parMW}).
We use
\be
m_N^{*}(n_0)/m_N =0.805, \quad K=275~\mbox{MeV}.
\label{MWnu-inp}
\ee
Then we obtain
\be
\mbox{for \,\,MW(nu)} \quad &&z=0.65: \nonumber \\
&&C^2_\om = 87.600\,,\quad C_\rho^2 = 100.64\,,\quad C^2_\sigma = 179.56\,,\nonumber\\
&&b = 7.7346 \times 10^{-3}\,,\quad c = 3.4462\times 10^{-4}\,,
\label{param_sc_065}
\ee
and
\be
\mbox{for \,\,MW(u)} \quad &&z=0:\nonumber \\
&&C^2_\om = 87.600\,,\quad C_\rho^2 = 100.64\,,\quad C^2_\sigma = 184.36\,,
\nonumber\\
&&b = 5.5387 \times 10^{-3}\,,\quad c = 2.2976\times 10^{-2}\,.
\label{param_sc_0}\ee

\begin{figure}
\begin{center}
\includegraphics[clip=true,height=6cm]{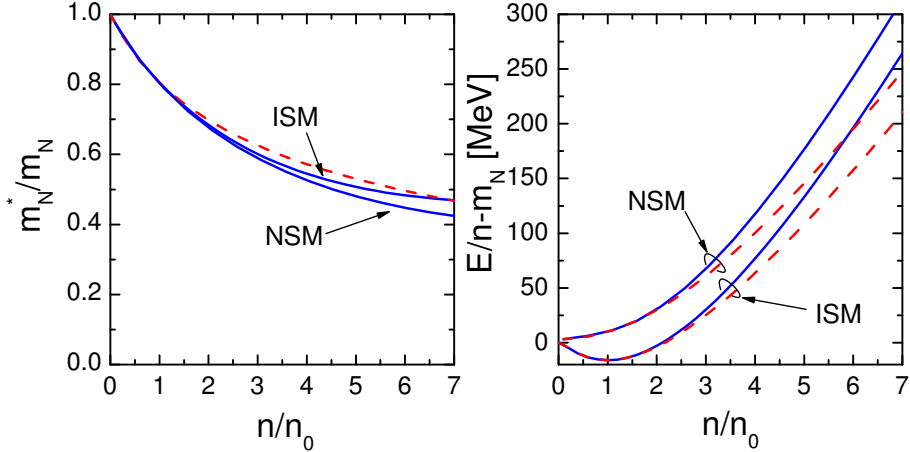}
\end{center}
\caption{The effective nucleon mass (left panel), and the energy per particle (right panel)
of the isospin symmetrical nuclear matter (ISM) and the neutron star
matter  (NSM)  within the MW(nu)  model with the scaling~(\ref{etnun}),
$z=0.65$, input-parameter set (\ref{param},\ref{MWnu-inp},\ref{param_sc_065}).
The dashed lines correspond to the  MW(u)  model ($z=0$) with the
input-parameter set  (\ref{param},\ref{MWnu-inp},\ref{param_sc_0}).}
\label{fig:eos_sc}
\end{figure}
In Fig.~\ref{fig:cbord_sc} we show a new  $c=0$ border on the
$(K,m_N^*)$-plane for the MW(nu) models using the input set (\ref{param}) and
the scaling functions (\ref{etnun}) for $z=0.65$\,. We draw the same
conclusion as in the case of MW(u) models.
In the framework of MW(nu) models (with $z=0.65$) keeping $c>0$ we
can't significantly decrease the effective nucleon mass $m^{*}_N (n_0 )$ not
increasing the compressibility beyond realistic values.

A very small value of the $c$-parameter in (\ref{param_sc_065}) that we have
obtained,
motivates a conjecture that the number of free parameters of the model can be still
reduced. If we did not care about  the global stability of the ground state
we  would put $c=0$ in (\ref{param_sc_065}). 
Taking care of this problem we could construct a model assuming $b=0$, but $c>0$.
Exploring $\eta_{\sigma}\neq 1$ we could put $b=c=0$.
In this paper we will not consider these possibilities of further modernizations of our models.

Fig.~\ref{fig:eos_sc} (left) demonstrates the density dependence
of the effective nucleon mass for the isospin symmetrical nuclear
matter (ISM) and for the NS matter (NSM)  for the given MW(nu) model
(solid lines: $z=0.65$, see (\ref{param},\ref{etnun},\ref{MWnu-inp},\ref{param_sc_065})),  and for
the  MW(u) model (dashed lines: $z=0$, (\ref{param},\ref{etnun},\ref{MWnu-inp},\ref{param_sc_0})).
Dashed curves for ISM and NSM cases cannot be distinguished. We
find that masses are slightly smaller for the MW(nu) model. We observe that
for $z=0.65$  the effective nucleon mass is monotonously
decreasing in the symmetrical nuclear matter  up to the density of
$7\,n_0$. The effective nucleon mass
becomes to increase  only at higher densities.  A dis-balance of
the universal scaling is just minor.
Right panel in
Fig.~\ref{fig:eos_sc} shows energies per particle. They are higher
in case of the MW(nu) model (for $n\gsim 2\,n_0$).

\begin{figure}
\begin{center}
\includegraphics[clip=true,width=\textwidth]{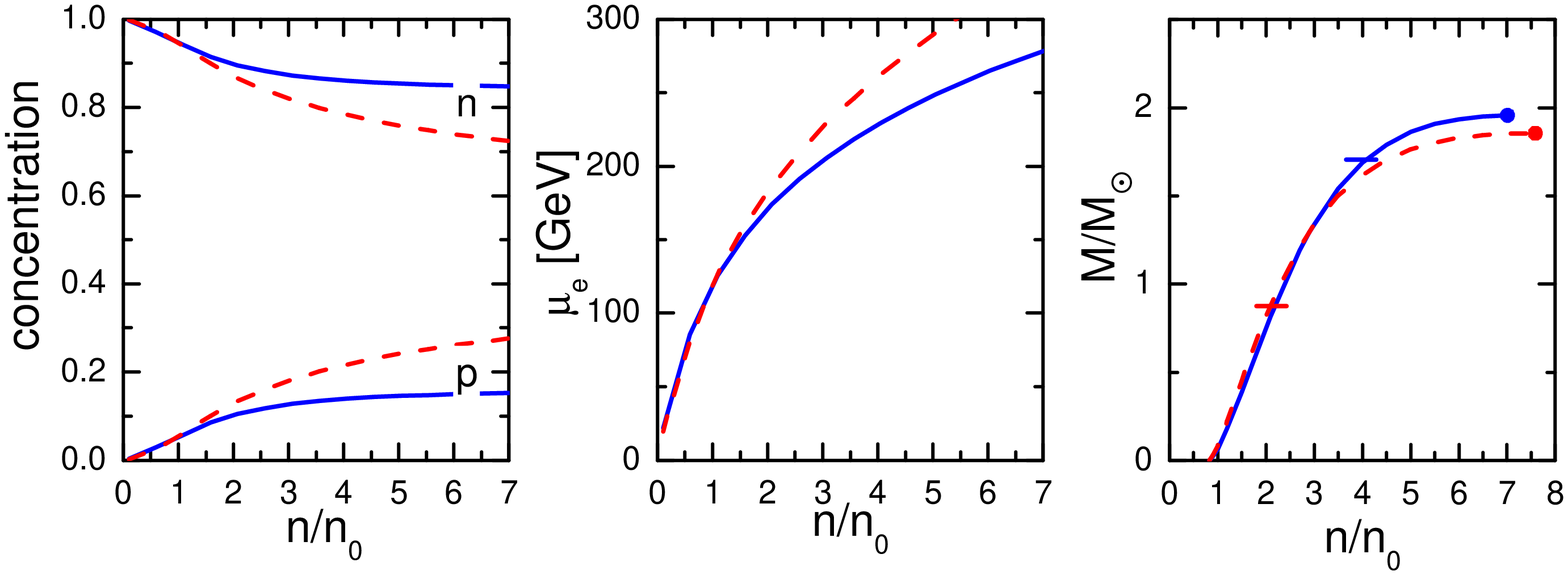}
\end{center}
\caption{The proton and neutron concentrations
(left panel), and  the electron chemical potential (middle panel) in the
neutron star  matter
for the very same models as in Fig.~\ref{fig:eos_sc}.
%%MW(nu)  model with the scaling~(\ref{etnun}), $z=0.65$
%%and  the parameter set (\ref{param_sc_065}) are shown by solid lines.
%%The dashed lines correspond to the MW(u) model  ($z=0$) with the
%%parameters (\ref{param_sc_0})\,.
The right panel shows the mass of the neutron star  as a
function of the central density. The DU thresholds are indicated by cuts.}
\label{fig:nspar_sc}
\end{figure}

In Fig.~\ref{fig:nspar_sc} (left) we show how the given MW(nu) model (the input-parameter
set (\ref{param},\ref{etnun},\ref{MWnu-inp},\ref{param_sc_065})), may help to cure a problem of
the low critical density for the DU process. For the MW(nu) model the DU process becomes
forbidden up to $4.0\,n_0$, whereas for the given MW(u) model (the
input-parameter set  (\ref{param},\ref{etnun},\ref{MWnu-inp},\ref{param_sc_0}))
we have
$n_{\rm crit}^{\rm DU}\simeq 2.08\,n_0$. The middle panel demonstrates the behavior of
the electron chemical potential. The right panel shows the NS mass as a function of
central density.
{\em{ With the MW(nu) model ($z=0.65$) the limiting NS mass is increased}} by about 5\%.
We see that due to a stiffening of the EoS we may now satisfy
the mentioned limit $M\gsim 2\,M_{\odot}$. If we used
another parameter choice, e.g., with a smaller effective nucleon mass at $n_0$, we could
construct an even stiffer EoS. The critical value of the NS mass for the
occurrence
of the DU process is $M(n_{\rm crit}^{\rm DU}\simeq 4.0~n_0 )\simeq 1.69~M_{\odot}$. This value
is essentially higher than the averaged value of the NS mass
measured in the NS binaries. Thus {\em{ we removed the problem with the low critical
density for the DU process.}}

\subsubsection{Fit to $A18+\delta
v+UIX^*$ and HHJ EoS}

Now let us show the efficiency of the MW(nu) models on another
example.
In ref.~\cite{HHJ} a parameterization of the nuclear EoS was
proposed (we called it the HHJ EoS), which is a good fit to the realistic EoS of the
Urbana-Argonne group~\cite{APR98} (A18+$\delta v$+UIX*) up to
$4\,n_0$. For larger densities the Urbana-Argonne EoS based on
non-relativistic interaction produces a supersonic velocity.
The HHJ EoS  treats this problem, smoothly incorporating  the causality constraint.
In this parameterization the energy density is given by
\be
&&E[n_p,n_n]=(n_p+n_n)\,\left[
e_{\rm B}\,u\,\frac{2.2-u}{1+0.2\,u}+a_{\rm
sym}\,u^{0.6}\,\frac{(n_p-n_n)^2}{(n_p+n_n)^2}
\right]\,,
\nonumber\\
&&u={(n_p+n_n)}/{n_0}\,,\,\,\, e_B=-15.8~{\rm MeV}\,\,\,\, a_{\rm
sym}=32~{\rm MeV}\,\,.
\label{HHJ}
\ee
In ref.~\cite{KV} it was demonstrated that
the HHJ EoS can be successfully fitted by the MW model, which uses
as an input set
\be
&&e_{\rm B}=-15.8~{\rm MeV}\,,\,\,
K=250~{\rm MeV}\,,\,\, m_N^*(n_0)=0.8\,m_N\,,\,\,
\nonumber\\
&&a_{\rm sym}=28~{\rm MeV}\,.\nonumber\\
\label{inp_param_MWU_hhj}
\ee
The corresponding parameter set
\be
&&C_\sigma^2=195.10\,,\quad C_\om^2= 90.911\,,\quad
C_\rho^2=77.276\,,
\nonumber\\
&& b=8.6964\times 10^{-3}\,,\quad
c=8.1411 \times 10^{-3}\,.
\label{param_MWU_hhj}
\ee
is slightly different from the original one used in ref.~\cite{KV},
where the nucleon mass was taken to be  938.918~MeV instead of 938~MeV, that
we adopted here.
The MW(u) model with such parameters gives approximately the same dependence of the
NS mass on the central density as the HHJ EoS~\cite{BGV04}.
However, as other RMF models, this MW(u)  model yields a low
threshold density for DU processes,  $n_{\rm crit}^{\rm DU}\simeq 2.6 n_0$
(corresponding to the NS mass $M\simeq 1.1 M_{\odot}$),
whereas  for  the HHJ EoS  $n_{\rm crit}^{\rm DU}\simeq 5.2 n_0$ ($M\simeq 1.8 M_{\odot}$).
This shortage can be corrected if we allow for a non-universal
scaling of the rho meson mass and the coupling constant, which
results in
\be
\eta_\sigma=\eta_\om=1\,,\quad
\eta_\rho(f)=\left(\frac{1+z\,f}{1+z\,f_0}\right)^2\,.
\label{etarho_hhj}
\ee
For $z=2.9$  the scaling ratio
$\Phi_{\rho}/\chi_{\rho}$
changes from unity at $n_0$ to $\simeq 1.4$ at $7n_0$. Thus a dis-balance of
the universal scaling that we have introduced is not too strong.
The threshold density for the DU processes is now shifted to
$n_{\rm crit}^{\rm DU}=5.2\,n_0$, in accordance with that for the HHJ EoS.

\begin{figure}
\begin{center}
\includegraphics[clip=true,height=6cm]{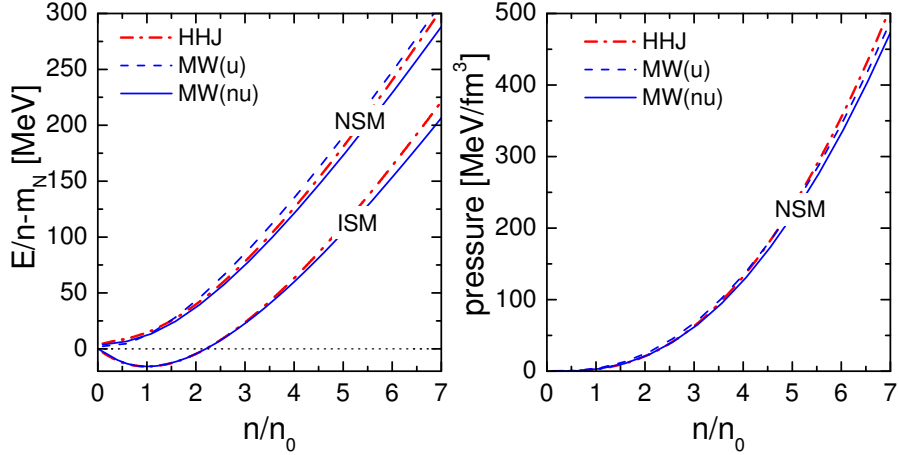}
\end{center}
\caption{Left panel: the energy per particle
of the isospin symmetrical  matter (ISM) and the
neutron star matter (NSM), as a function of the nucleon
density.
Right panel: the pressure of the neutron star matter.
Solid lines show calculations done within the MW(nu) model with the
scaling~(\ref{etarho_hhj}) and   $z=2.9$ with the input-parameter set
~(\ref{inp_param_MWU_hhj},\ref{param_MWU_hhj}).
Dashed lines correspond to the MW(u) model (z=0) for the very same input-parameter
set. Dash-dotted lines are for the HHJ EoS (\ref{HHJ}).}
\label{fig:eos_hhj}
\end{figure}

Fig.~\ref {fig:eos_hhj} (left panel) demonstrates the energy per
particle of the isospin symmetrical matter and of the neutron star
matter (NSM) calculated within the  MW(nu) model with the scaling
(\ref{etarho_hhj}) for $z=2.9$ and for the input-parameter set
~(\ref{inp_param_MWU_hhj},\ref{param_MWU_hhj}) in comparison with
the values calculated for the MW(u) model ($z=0$), with the same
input-parameter set,
(\ref{inp_param_MWU_hhj},\ref{param_MWU_hhj}), and for the HHJ
EoS. Dashed and solid curves are not distinguishable. The right
panel shows the pressure, as a function of the density for the
NSM. We see that in both cases (ISM and NSM) the thermodynamic
characteristics of the given  MW(nu) and MW(u) models are very
closed to the values calculated for the HHJ EoS.

\begin{figure}
\begin{center}
\includegraphics[clip=true,width=\textwidth]{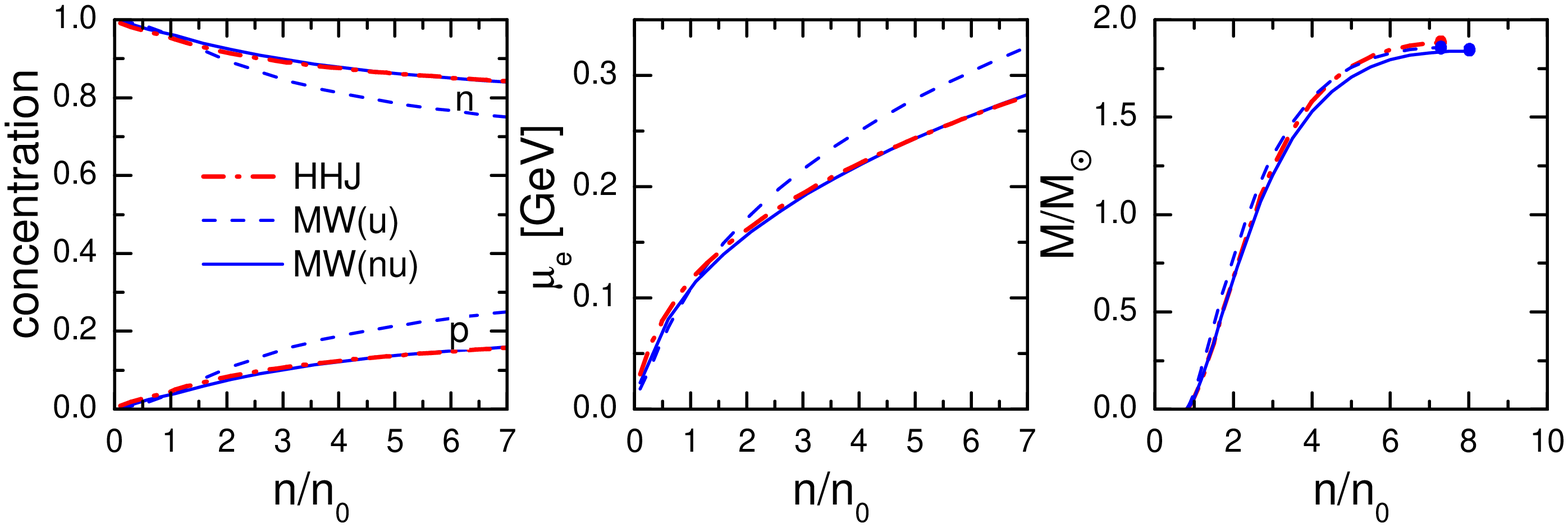}
\end{center}
\caption{The proton and neutron concentrations (left panel),
 the electron chemical potential in the neutron star
matter (middle panel) and
the mass of the neutron star  as a
function of the central density for the very same models as in Fig.~\ref{fig:eos_hhj}.
}
\label{fig:ns_mat_hhj}
\end{figure}

In Fig.~\ref {fig:ns_mat_hhj} (left and middle) we illustrate how the scaling
(\ref{etarho_hhj}) allows to change the composition of the NS
matter. We see that for the given MW(nu) model (input-parameter
set ~(\ref{inp_param_MWU_hhj},\ref{param_MWU_hhj})) the proton concentration and the
electron chemical potential are strongly reduced compared to the
corresponding MW(u) model (the same input-parameter set) and become equal to those for the HHJ
parameterization. Right panel shows NS masses as functions of the central
densities for the given three models. They are very close to each other.

Concluding this section, we have shown an example of the MW(nu)
model that allows to increase the limiting NS mass and to shift
the DU threshold to higher densities. We have also constructed another
MW(nu) model that can be used as a RMF model equivalent to the HHJ
parameterization, thus also demonstrating an appropriate fit to the $A18+\delta
v+UIX^*$
EoS.

\section{ Second
  order phase transition to $\rho^-$ condensate in NS}\label{sec:rho}

Now we turn to the discussion of the isospin asymmetrical matter
where, as we will show, mean fields of charged $\rho$ mesons may have
finite values. As we have mentioned, usually only one component,
$\rho_0^{(3)}$ is assumed to be non-zero. A new ansatz for the $\rho$
meson fields was proposed in ~\cite{v97}. Besides the
traditional $\rho_0^{(3)}\neq 0$ field it includes also spatial
components of  charged $\rho$ meson fields, $\rho_i^\pm
=(\rho_i^{(1)} \pm i\rho_i^{(2)})/\sqrt{2}\neq 0$\,, $i=1,2,3$\,.
As it was argued in ref.~\cite{v97}, the presence of the fields $\rho_i^{(3)}$,
$\rho_0^{(i)}$ results in an increase of the system energy. Therefore, these fields
can be put equal to zero. It was also found  that  the condition
$\big(\rho_i^+\rho_j^--\rho_i^-\rho_j^+\big)=0$ minimizes the
energy. This implies that the ratio $\rho_i^+/\rho_i^-$ is
constant, independent of the spatial index $i$. Then we may take
$\rho_i^-=a_i\,\rho_c$ and $\rho_i^+=a_i\, \rho_c^\dagger$, where
$\vec{a}=\{a_i\}$ is the spatial unit vector, and $\rho_c$ is a
complex amplitude of a charged $\rho$-meson field.

In terms of the $\rho_0^{(3)}$ and $\rho_c$ fields the corresponding contribution
to the density of the thermodynamic
potential
(effective energy)
(\ref{ERfun}) becomes
\be
\Omega_{\rho}[n_n,n_p;f,\rho_0^{(3)}, \rho_c]
&=&  \frac12 g_{\rho}\, {\chi}_\rho (n_p-n_n)\, \rho_0^{(3)}
- \frac12
\Big(\rho_0^{(3)}\Big)^2\, m_\rho^{2}\, \Phi_\rho^2
\nonumber\\
&-&\left[\Big(g_\rho{\chi'}_\rho \, \rho_0^{(3)}  -\mu_{ch}^\rho\Big)^2
-m_\rho^{2}\, \Phi_\rho^2\right]\, |\rho_c|^2\,.
\label{er}
\ee
Variation of the Lagrangian/(thermodynamic potential)  with respect to the fields
$\rho_0^{(3)}$ and $\rho_i^-$ yields equations of motion
\be
&&\left[ \Big( g_{\rho}\,{\chi}^{'}_\rho\,
\rho_0^{(3)}   -\mu_{ch}^\rho \Big)^2-m_\rho^2\,\Phi_\rho^2
\right]
\, \rho_c =0\,,
\nonumber\\&&
m_\rho^{2}\,\Phi_\rho^2\, \rho_0^{(3)} +
2\, g_\rho\,{\chi}^{'}_\rho\, \Big(g_\rho{\chi'}_\rho \,
\rho_0^{(3)} -\mu_{ch}^\rho\Big) |\rho_c|^2 \,  = \frac12 g_{\rho}\,
{\chi}_\rho\,
(n_p-n_n)\,.
\label{rhoem}
\ee
This system of equations has two solutions.
The first solution is the traditional one, cf. (\ref{solt}), (\ref{Er}),
\be
\rho_0^{(3)}= \frac12\, \frac{g_\rho}{m_\rho^2}\,
\frac{{\chi_{\rho}}}{\Phi_\rho^2}\, (n_p-n_n)\,,\quad \rho_c=0\,,
\label{solt1}
\ee
yielding the contribution to the density of the thermodynamic potential
\be
\Omega_{\rho}^{(1)}[n_n,n_p;f]=\frac{C_\rho^2 (n_n-n_p)^2}{8\,m^2_N\,
  \eta_\rho(f)}\,.
\ee
The second solution of  (\ref{rhoem}) (found previously in~\cite{v97}, neglecting
$\mu_{ch}^{\rho}$, and without scaling of couplings)
is as follows
\be
\rho_0^{(3)} &=& \frac{\mu_{ch}^\rho }{g_\rho{{\chi}'_\rho}}-
\frac{m_\rho}{g_\rho}\,\frac{\Phi_\rho}{{\chi}'_\rho}\,
\mathop{\rm sign}(n_n-n_p)\,,
\quad
|\rho_c|^2=\frac{|n_p-n_n|-n^\rho}{4\,m_\rho\, \eta^{1/2}_\rho\,{\chi}'_\rho}\,,
\\
n^\rho &=& 2\, \frac{m_\rho\, m_N^2}{C_\rho^2}
\,\frac{\eta^{1/2}_\rho\, \Phi_\rho^2}{{\chi}'_\rho}\,\left(1-\frac{\mu_{ch}^\rho}{m_\rho\,
    \Phi_\rho}\mathop{\rm sign}(n_n -n_p) \right)\,.\nonumber
\label{solc}
\ee
It exists only, if $|n_n-n_p|>n^{\rho}$\,.
The density of the thermodynamic potential corresponding to this solution
takes the form
\be
\Omega_{\rho}^{(2)}[n_n,n_p;f]=\Omega_{\rho}^{(1)}[n_n,n_p;f]-
\frac{C_\rho^2}{8\, m_N^2\, \eta_\rho}\, \Big(|n_n-n_p|-n^\rho \Big)^2\,.
\label{enrhoc}
\ee
Hence, for $n> n_{\rm c}^{\rho\,,\rm II}$, where the value $n_{\rm c}^{\rho\,,\rm
II}$ follows from the solution of the equation
\be\label{sec}
n^\rho =|n_n-n_p|\,,
\ee
the second solution (\ref{solc}) becomes  energetically favorable
compared to the traditional solution (\ref{solt1}).
The density $n_{\rm c}^{\rho\,,\rm II}$ which can be written as
\be\label{secc}
n_{\rm
c}^{\rho\,,\rm II}= 2\, \frac{m_\rho\, m_N^2}{C_\rho^2}
\,\frac{\eta^{1/2}_\rho\, \Phi_\rho^2}{(1-2\,Y_p^{(c)} ){\chi}'_\rho}\left(1-\frac{\mu_{ch}^\rho}{m_\rho\,
\Phi_\rho} \right)\,\,,\quad Y_p^{(c)} =\frac{n_p}{n_{\rm c}^{\rho\,,\rm II}}<\frac12\,,
\ee
is the critical density with respect to a second-order phase
transition. In principle a transition to a new solution
could be of a first order with
an abrupt change of the isospin composition and the density.
It would then happen at a smaller critical density $n=n_{\rm
c}^{\rho ,\rm I}<n_{\rm c}^{\rho\,,\rm II}$, cf. \cite{v97}. We studied this possibility
(see below)
and found out that in the presence of
electrons the first-order phase transition is not realized.
Thus we conclude that in the frameworks of
the RMF models that we exploit here and with taking into account  of
electrons the
charged $\rho$-meson condensation occurs by the second-order phase transition.
Note that the new solution (\ref{solc}) induces a spontaneous breaking of
the spatial symmetry of the system, since we fixed only one value $\rho_c$,
whereas the choice of
components $a_i$ is not fixed.

The charged $\rho$-meson chemical potential is found from the chemical
equilibrium conditions. Considering the reaction  $e+n \leftrightarrow \rho^- +n$
we find
\be
\mu_{ch}^\rho =\mu_e .
\ee
The number of protons, electrons, muons and
$\rho^-$ is governed by the local charge neutrality
condition
\be
n_p -n_e -n_\mu +n_{ch}^{\rho}=0, \quad n_{ch}^{\rho}=
2\,\Big(g_\rho{\chi'}_\rho \, \rho_0^{(3)}-\mu_{ch}^\rho\Big)
|\rho_c|^2 <0\,.
\ee
In the phase with $\rho_c\neq 0$ from  (\ref{Efun}), (\ref{solc})
we find
\be
&&\mu_e=\frac{1}{1+\chi_\rho(\bar f)/\chi'_\rho(\bar f)}\,\nonumber\\
&&\times\left(\sqrt{m_N^2\,\Phi_N^2(\bar f)+p_{{\rm F},n}^2}-
\sqrt{m_N^2\,\Phi_N^2(\bar f)+p_{{\rm F},p}^2}+m_\rho\,\frac{\Phi_\rho(\bar f)\chi_\rho(\bar f)}
{\chi'_\rho(\bar f)}\right)\,,
\\
&&n_p =\frac{(\mu_e^2-m_e^2)^{\frac32}}{3\,\pi^2}\theta (\mu_e -m_e) +
\frac{(\mu_e^2-m_\mu^2)^{\frac32}}{3\,\pi^2}\theta (\mu_e -m_\mu) +n_{ch}^{\rho}(\mu_e,n_p,n_n,\bar
f), \nonumber\\
&&n_{ch}^{\rho}=2\,
m_\rho\,\Phi_\rho(f)\, |\rho_c|^2\,. \nonumber
\label{bet-eq-rhoc}
\ee

The main uncertainty of our consideration is the unknown
modification of the non-Abelian $\rho$-meson self-interaction in
nuclear medium, which in our scheme is encoded in the scaling
parameter $\chi'_\rho$.
As we have mentioned, we see no reason for the scaling of
$\chi'_\rho$, thus taking $\chi'_\rho =1$. However
for the completeness of the consideration and in order to investigate the sensitivity
of our results
to the parameter variation
we consider also the case $\chi'_\rho\simeq
\chi_\rho\simeq \Phi_\rho$. Moreover note that some QCD motivated studies, cf.
\cite{bando83}, use smaller values for the $\rho$ meson coupling constants
$g_{\rho}=g_{\rho}^{'}\simeq m_{\rho}/F_{\pi}\simeq 5.8$,
$F_{\pi}\simeq 132~$MeV, than those follow from the fits in the RMF models. In this
respect the second choice $\chi'_\rho\simeq
\chi_\rho\simeq \Phi_\rho$ effectively demonstrates what could be if
$g_{\rho}$ were decreased.

\begin{figure}
\begin{center}
\includegraphics[clip=true,height=6cm]{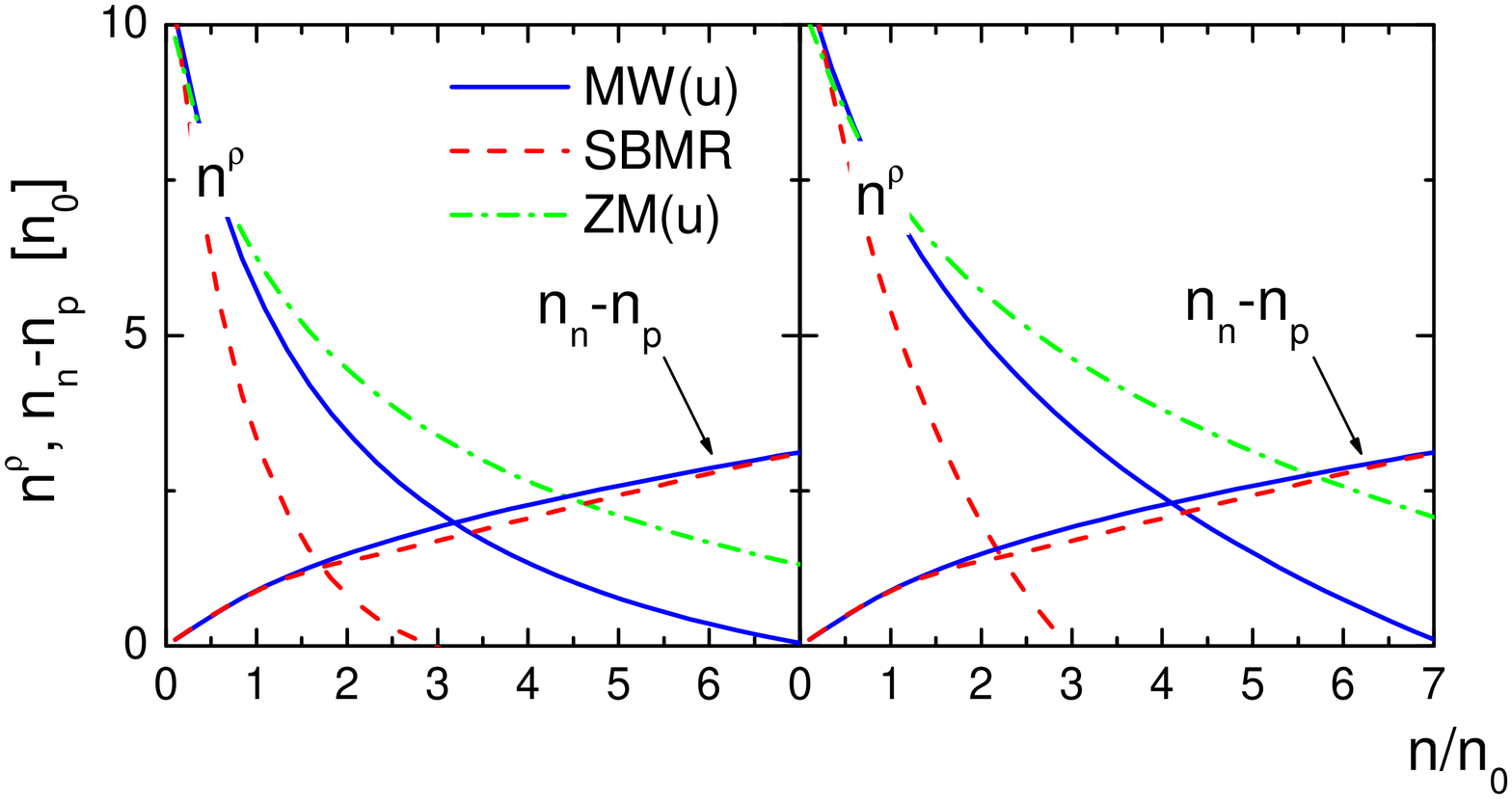}
\end{center}
\caption{Densities  $n^{\rho}$  from (\protect\ref{solc}) and
$n_n-n_p$ calculated for the neutron star matter within the MW(u)
(input-parameter set (\ref{param},\ref{MW-Km},\ref{parMW})),
ZM(u) (input-parameter set (\ref{param},\ref{parZM}))
and SBMR models.
Left panel shows the results for the choice ${\chi}'_\rho=1$ and
right panel, for ${\chi}'_\rho=\Phi_\rho$. Crossing points of curves $n^{\rho}$
with the curve $n_n-n_p$
correspond to
$n=n_{\rm c}^{\rho\,,\rm II}$, cf.
eq. (\ref{sec}).
}
\label{fig:ncritV}
\end{figure}

The quantity $n^{\rho}(n)$ given by (\ref{solc}) and $n_n-n_p$ are
depicted in Fig.~\ref{fig:ncritV} for three types of models MW(u)
(input-parameter set (\ref{param},\ref{MW-Km},\ref{parMW})), ZM(u)
(input-parameter set (\ref{param},\ref{parZM})), and SBMR discussed above. We use here $m_\rho=770$~MeV.
The actual value of the critical density, $n_{\rm c}^{\rho\,,\rm
II}$, for the second order phase transition to the charged
$\rho$-meson condensate state is given by the solution (\ref{secc}) of eq.
(\ref{sec}). We see that
the condensation is possible at $n_{\rm c}^{\rho\,,\rm II}=3.2~
n_0$ for the MW(u) model (would be $4.1\, n_0$ for the model with
the scaling of $\chi_\rho^{'}$), at $n_{\rm c}^{\rho\,,\rm
II}=4.5~n_0$ for the ZM(u) model ($5.6\, n_0$ for the model with
the scaling of $\chi_\rho^{'}$), and at $n_{\rm c}^{\rho\,,\rm
II}=1.7~n_0$ for the SBMR model ($2.2\, n_0$ for the model with
the scaling of $\chi_\rho^{'}$).

\begin{figure}
\begin{center}
\includegraphics[clip=true,height=6cm]{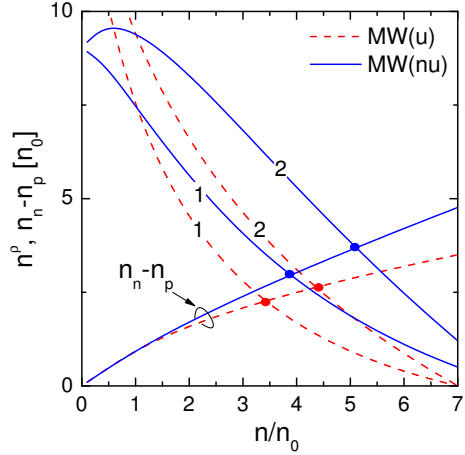}
\end{center}
\caption{Densities  $n^{\rho}$  from (\protect\ref{solc}) and
$n_n-n_p$ calculated for the neutron star matter within the
MW(nu)
model (\ref{etarho_hhj}), $z=2.9$ (the input-parameter set (\ref{inp_param_MWU_hhj});
~(\ref{param_MWU_hhj})) are shown by solid lines.
Dashed lines are for $z=0$ and the same choice of other parameters.
Fat dots  mark the critical
densities $n_{\rm c}^{\rho\,,\rm II}$. Curves 1 correspond to $\chi_{\rho}^{'}
=1$, whereas 2, to $\chi_{\rho}^{'}
=\Phi_{\rho}$.
}
\label{fig:ncritV1}
\end{figure}

In Fig.\ref{fig:ncritV1} we show the same, as in Fig.
\ref{fig:ncritV}, but for the MW(nu) model  with $z=2.9$ specified
by (\ref{inp_param_MWU_hhj},\ref{param_MWU_hhj},\ref{etarho_hhj}).
As the consequence of $\eta_{\rho}>1$, the MW(nu) model produces a
higher critical density for the charged $\rho$ meson condensation,
$n_c^{\rho,\rm II}\simeq 3.9\,n_0$ for $\chi_\rho^{'}=1$  and
$n_c^{\rho,\rm II}\simeq 5.1\,n_0$ for $\chi_\rho^{'}=\Phi_\rho $.
This is basically due to the smaller value of the asymmetry energy
in (\ref{inp_param_MWU_hhj}) and therefore, the smaller value of
the coupling constant $C_\rho$\,. We point out that in a most
realistic cases of the MW(u) and MW(nu) models the novel charged
$\rho$ meson condensation appears already for sufficiently low
densities $n\sim (3\div 4)~n_0$. As the price for the increase of
the threshold density for the DU process, the critical density for
the condensation increased in case of  this MW(nu) model.
The MW(nu) model (\ref{etnun}) for $z=0.65$, using the input-parameter set (\ref{param},\ref{MWnu-inp},\ref{param_sc_065})
yields $n_{c}^{\rho ,\rm II}\simeq 3.4\,n_0$
for $\chi_\rho^{'}=1$ and $n_{c}^{\rho ,\rm II}\simeq 4.6$ for
$\chi_\rho^{'}=\Phi_\rho$.

\begin{figure}
\begin{center}
\includegraphics[clip=true,height=5cm]{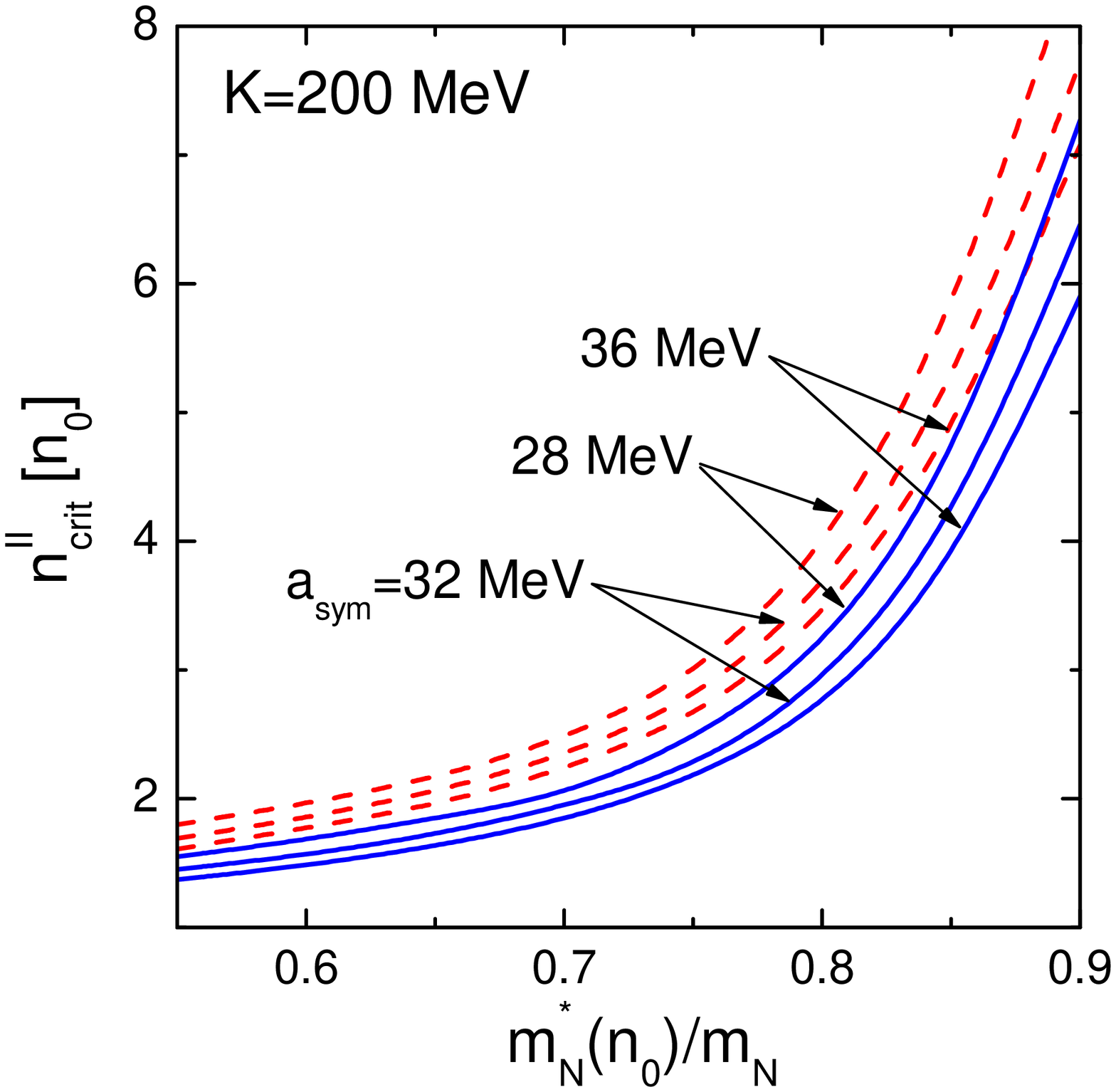}\quad
\includegraphics[clip=true,height=5cm]{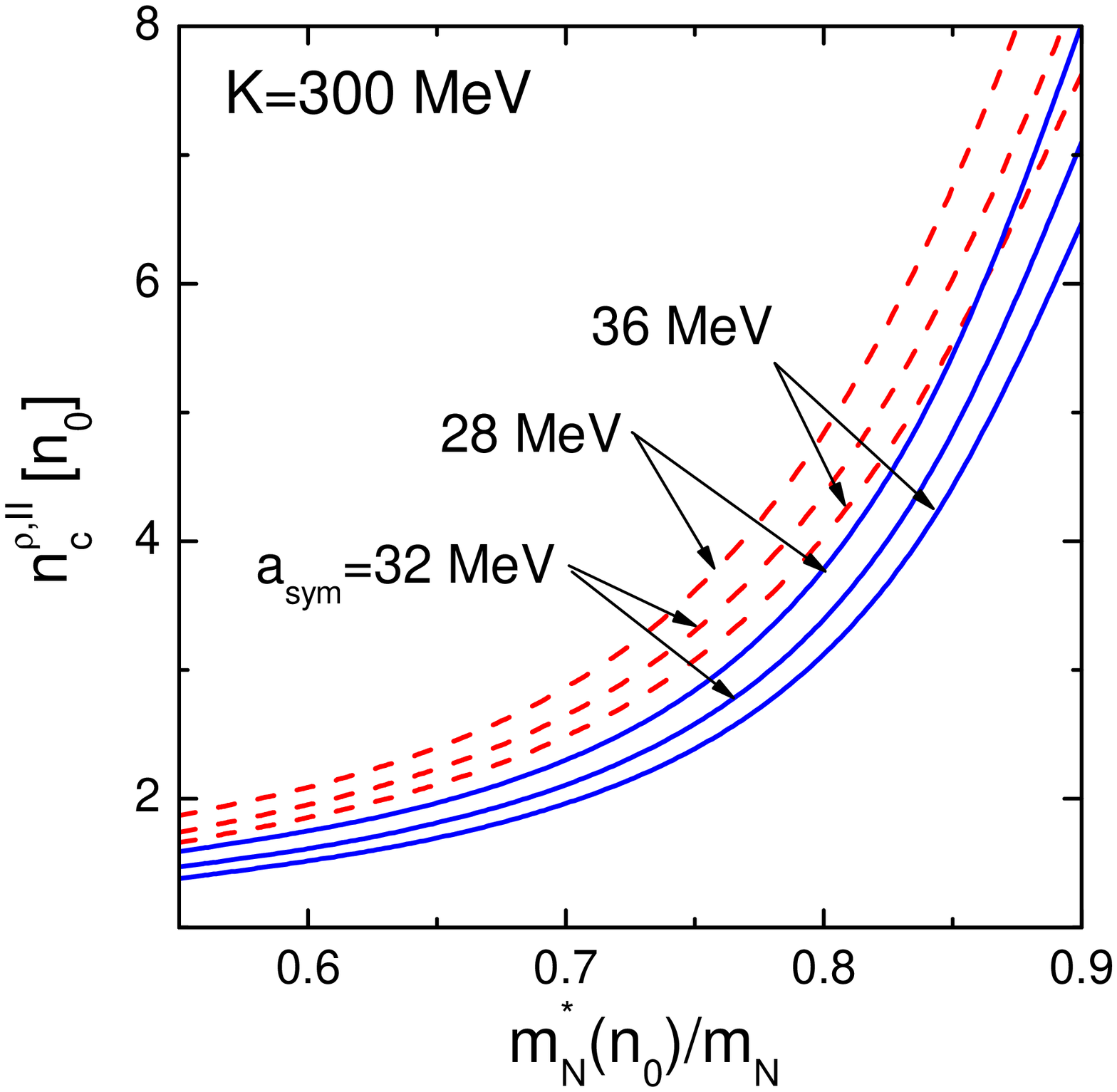}
\end{center}
\caption{
The critical density of the charged $\rho$ meson condensation, $n_{\rm
c}^{\rho,\,\rm II}$ as a function of the $m_N^{*}(n_0)$ used as an
input of the MW(u) model
for two values of the compressibility
modulus $K$ and three values of $a_{\rm sym}$. Other input values are taken
according to (\ref{param}).
The solid lines are calculated for the choice
${\chi}'_\rho=1$ and the dashed ones for
${\chi}'_\rho=\Phi_\rho$.
}
\label{fig:ncscan}
\end{figure}

The critical density, $n_{\rm c}^{\rho\,,\rm II}$ depends
on the particular values of the effective nucleon mass and compressibility
at saturation, which are used to constrain parameters of the
RMF model. We study these dependences on the example of the MW(u)
model, which allows for independent variation of the input
parameters $K$ and $m_N^*(n_0)/m_N$. We calculate the critical density
$n_{\rm c}^{\rho\,,\rm II}$,
as a function of the
input parameter $m_N^*(n_0)/m_N$ for $K=200$~MeV and 300~MeV for two
choices of the ${\chi}'_\rho$ scaling mentioned above.
Results are presented in Fig.~\ref{fig:ncscan}.
We see that $n_{\rm c}^{\rho\,,\rm II}$ is weakly dependent on $K$ but varies strongly
with the effective nucleon mass, especially for $m_N^*(n_0)/m_N\gsim 0.75$\,.
For $0.6\lsim m_N^*(n_0)/m_N\lsim 0.8$ the critical density varies within the interval
 $n_0\lsim n_{\rm c}^{\rho\,,\rm II}\lsim 3\, n_0$ for $\chi_\rho^{'}=1$.

\begin{figure}
\begin{center}
\includegraphics[clip=true,height=5cm]{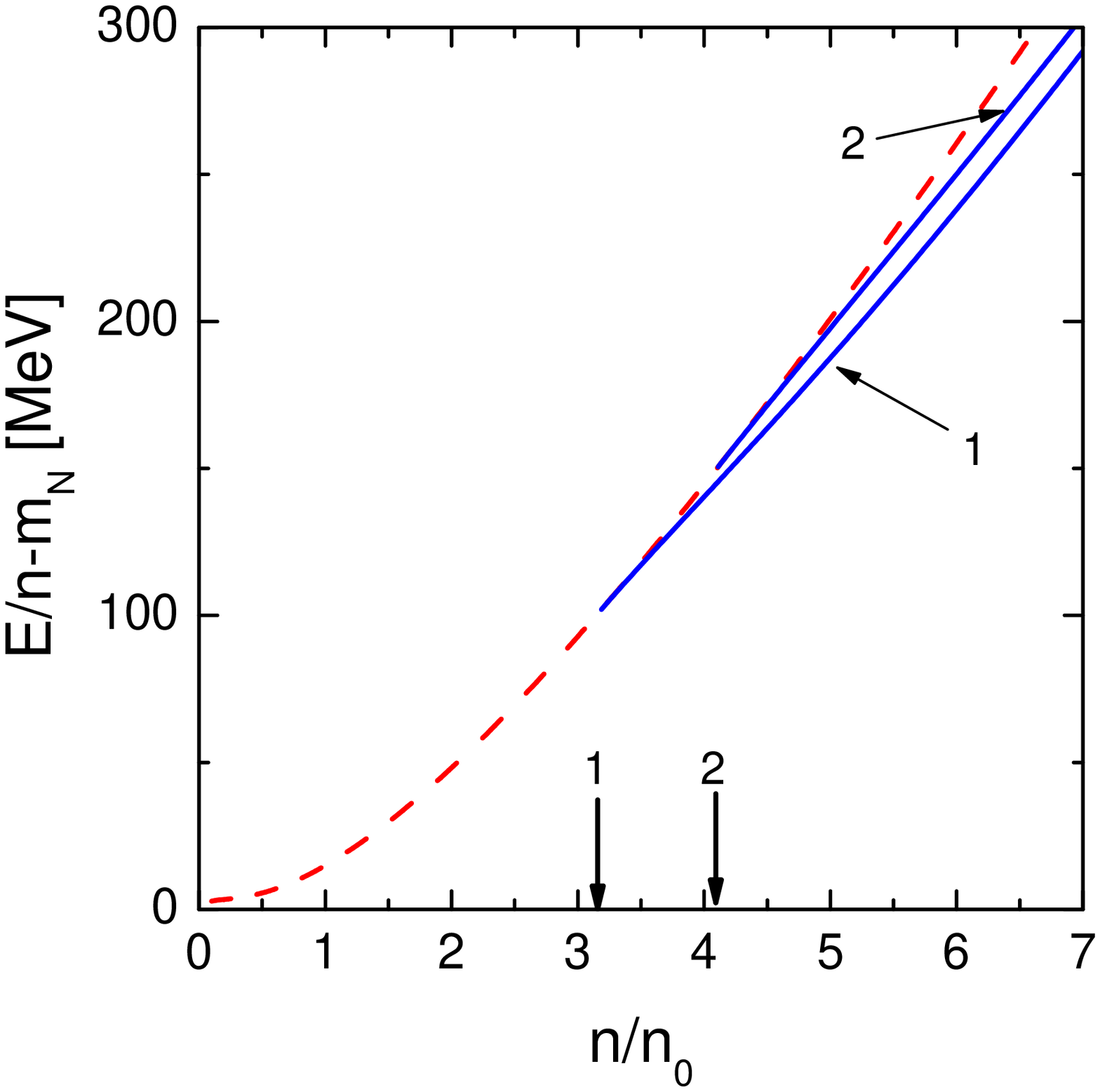}\quad
\includegraphics[clip=true,height=5cm]{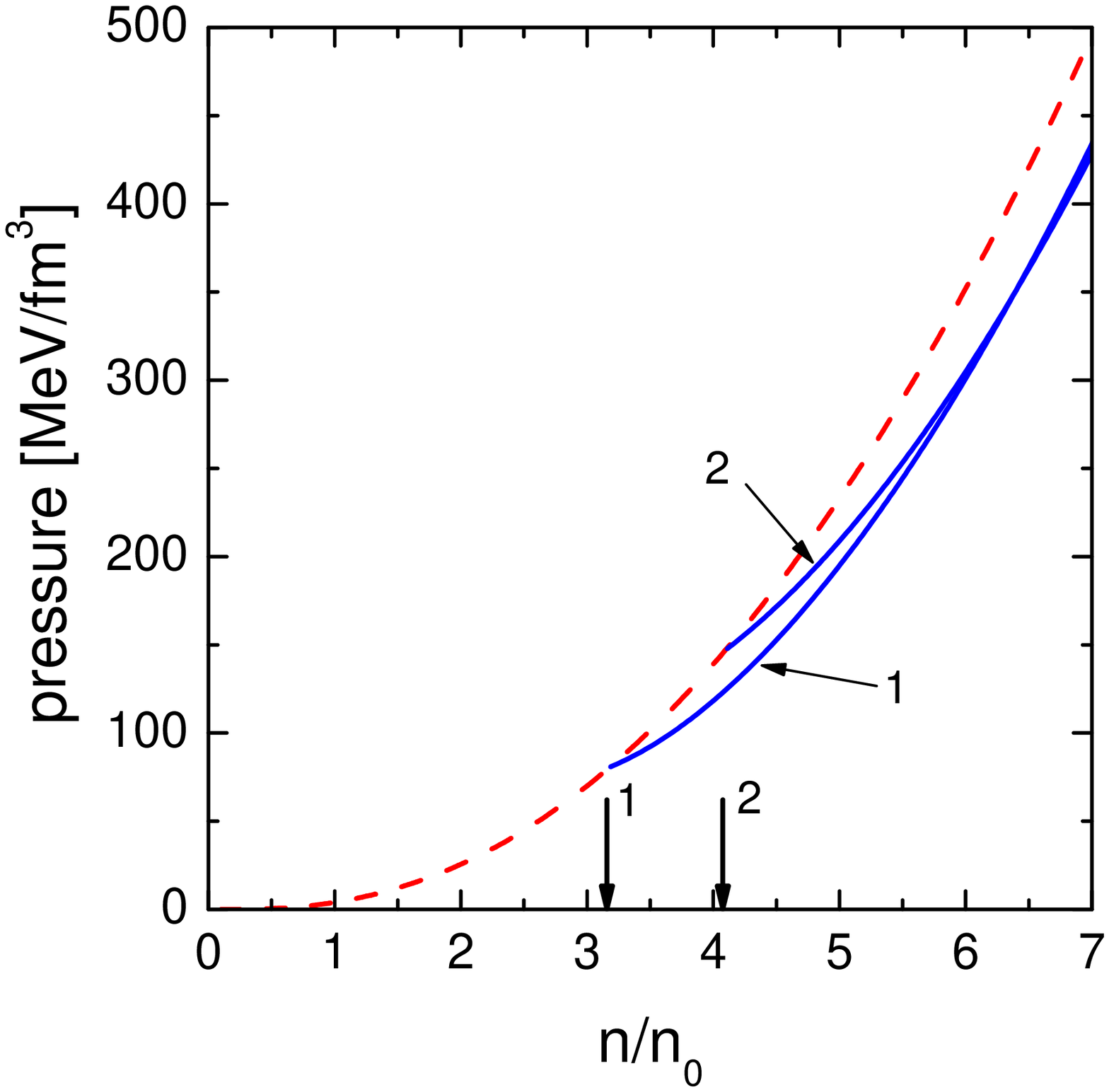}
\end{center}
\caption{The energy per particle (left panel) and
the pressure (right panel) of
the neutron star matter as a function of the nucleon density without (dash curves)
and with (solid curves) $\rho^-$ condensate
for  the MW(u) model (the  input-parameter set (\ref{param}),
(\ref{MW-Km}); (\ref{parMW})).
The lines labeled by "1" are calculated for the choice
${\chi}'_\rho=1$ and the lines labeled by "2" are for
${\chi}'_\rho=\Phi_\rho$.
}
\label{fig:enpress_Ior}
\end{figure}

Fig.~\ref{fig:enpress_Ior} demonstrates the energy per particle and
the pressure  of
the NS matter as a function of the nucleon density without (dash curves) and with
 $\rho^-$
condensate (solid curves)
for the case of the MW(u) model (the  input-parameter set
(\ref{param},\ref{MW-Km},\ref{parMW}))
cf. Figs.~\ref{fig:eos}, \ref{fig:eosN}. Curve 1 is
for $\chi_{\rho}^{'}=1$ and curve 2, for $\chi_{\rho}^{'}=\Phi_{\rho}$.
We observe no van~der~Waals behavior of the pressure typical for the first order phase transitions.
The EoS is softened in the presence of the second order phase transition to
the charged $\rho^-$ condensate but quantitatively the
softening effect is minor.

\begin{figure}
\begin{center}
\centerline{
\parbox{4.5cm}{\includegraphics[clip=true,width=4.5cm]{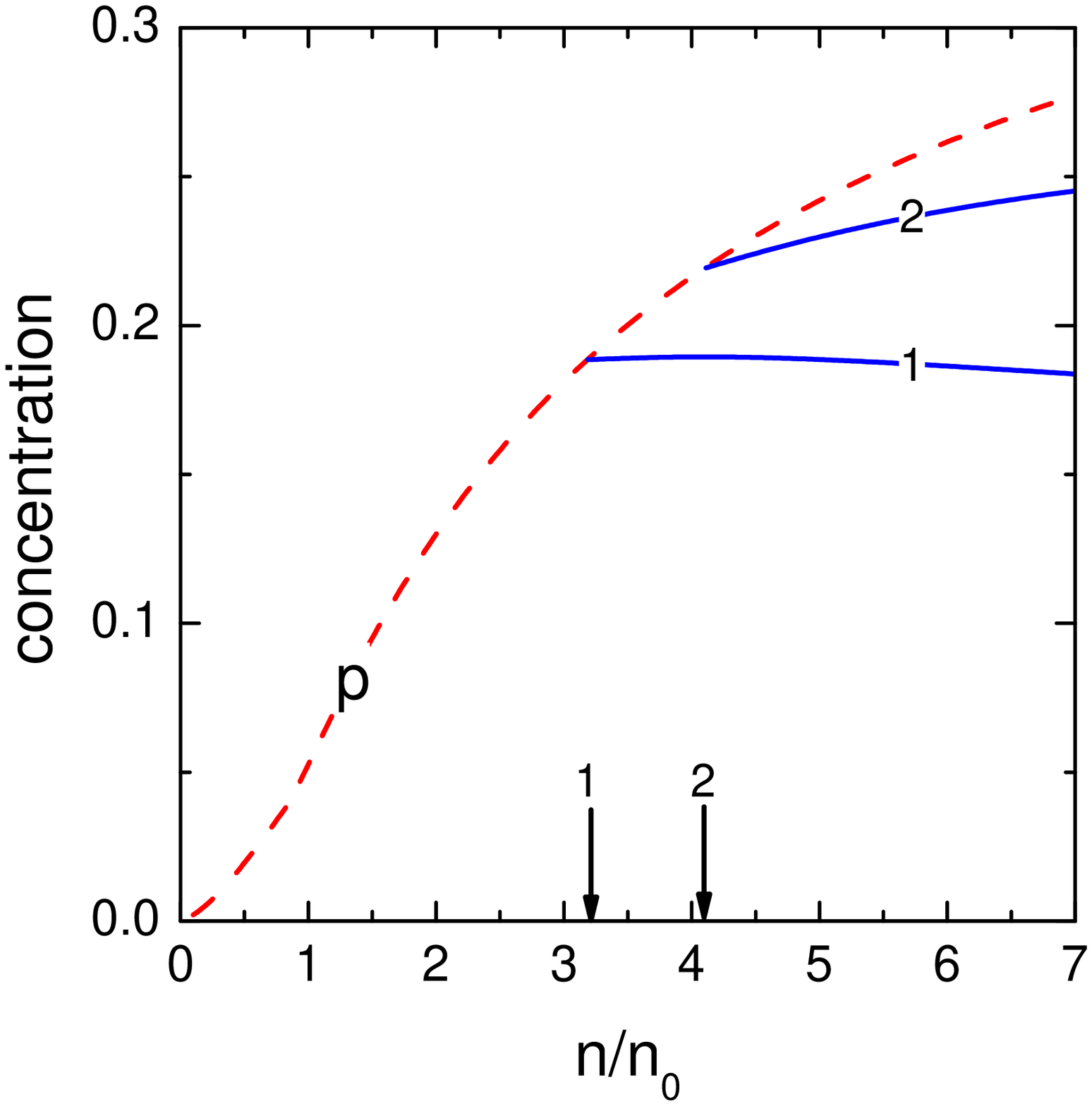}}\quad
\parbox{4.5cm}{\includegraphics[clip=true,width=4.5cm]{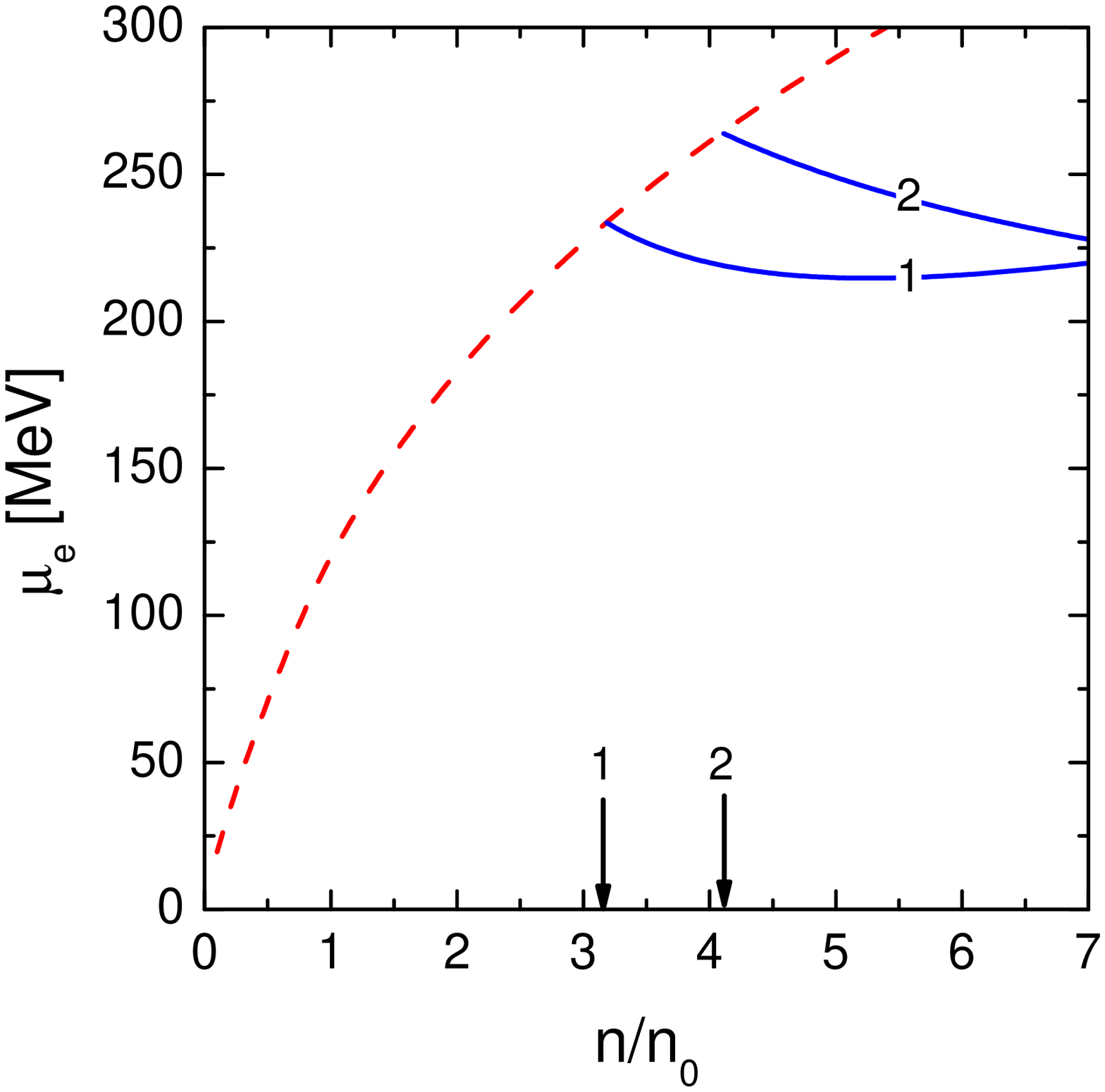}}\quad
\parbox{4.2cm}{\includegraphics[clip=true,width=4.2cm]{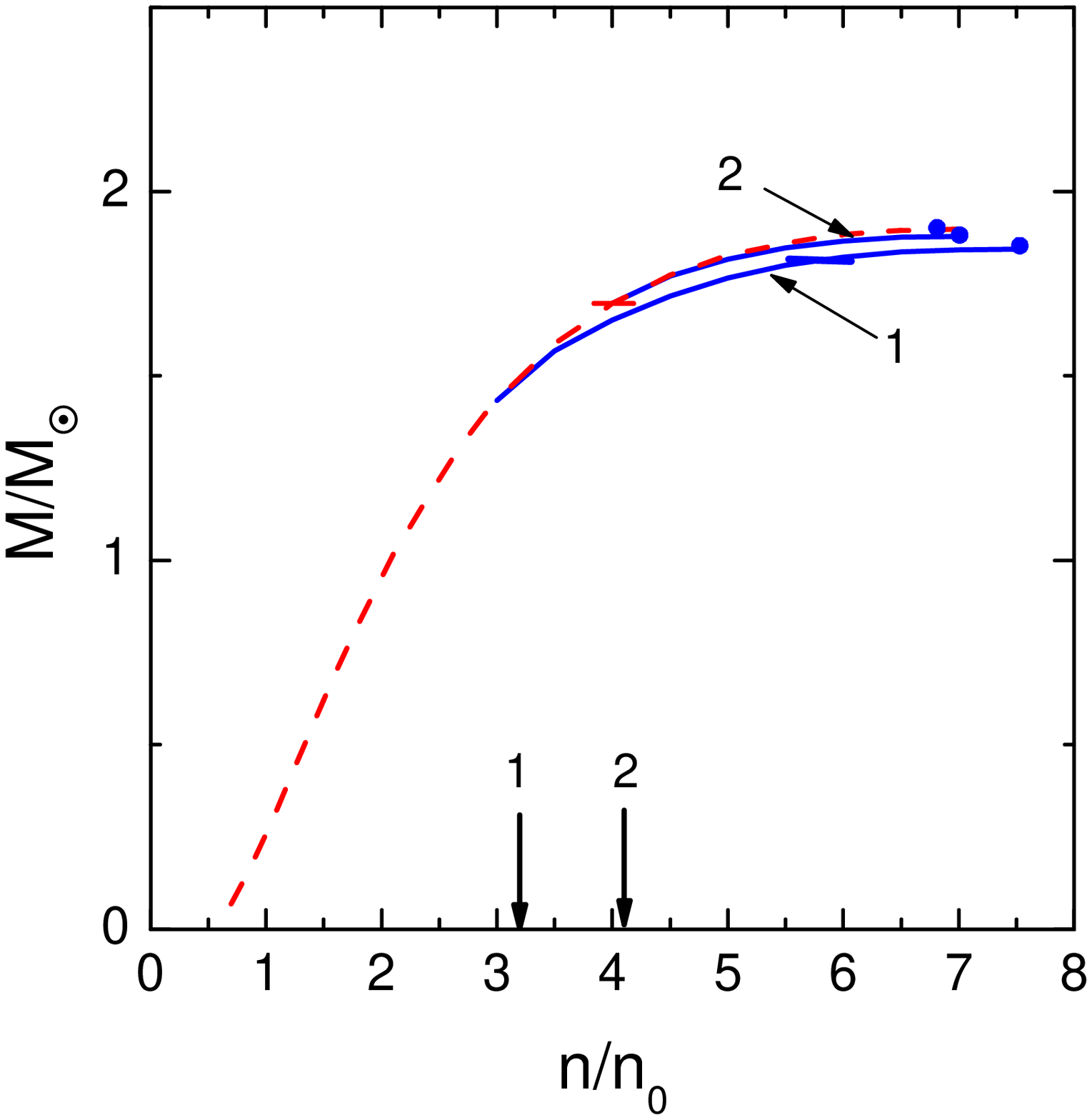}
}}
\end{center}
\caption{The proton concentration  (left panel) and
the electron chemical potential (middle panel) of
the neutron star matter as a function of the nucleon density without (dash curves)
and with (solid curves) $\rho^-$ condensate
for  the MW(u) model (the  input-parameter set (\ref{param},\ref{MW-Km},\ref{parMW})).
The right panel shows the mass of the neutron star as a function
of the central density.
The lines labeled by "1" are calculated for the choice
${\chi}'_\rho=1$ and the lines labeled by "2" are for
${\chi}'_\rho=\Phi_\rho$.
}
\label{fig:mue_Iore}
\end{figure}

In Fig.~\ref{fig:mue_Iore} we show the proton  fraction $Y_p =n_p /n$,
the electron chemical potential,  and the NS mass for
the NS matter as a function of the nucleon density without (dash curves) and with
(solid curves) $\rho^-$ condensate
for the case of the MW(u) model (the  input-parameter set
(\ref{param},(\ref{MW-Km},\ref{parMW})),
cf. Fig. \ref{fig:nspar}. Curves 1 are
for $\chi_{\rho}^{'}=1$ and curves 2 are for $\chi_{\rho}^{'}=\Phi_{\rho}$.
We see that a part of electrons is now replaced to the condensate $\rho^-$
mesons. Due to that the electron chemical potential decreases in presence of the
condensate compared to the case without the condensate. The proton fraction
becomes smaller in the presence of the $\rho^-$ condensate than would be in the absence
of the condensate. However, as we obtain {\em{ in the framework of  the MW(u) model, 
DU processes once started
at $2.07\,n_0$  stay operative
in the condensate phase up to rather high densities}} (up to $5.8~n_0$ for  case 1 and
up to the maximum central density for case 2).
The neutron star masses change very little with the appearance of the condensate.

\begin{figure}
\begin{center}
\includegraphics[clip=true,height=5cm]{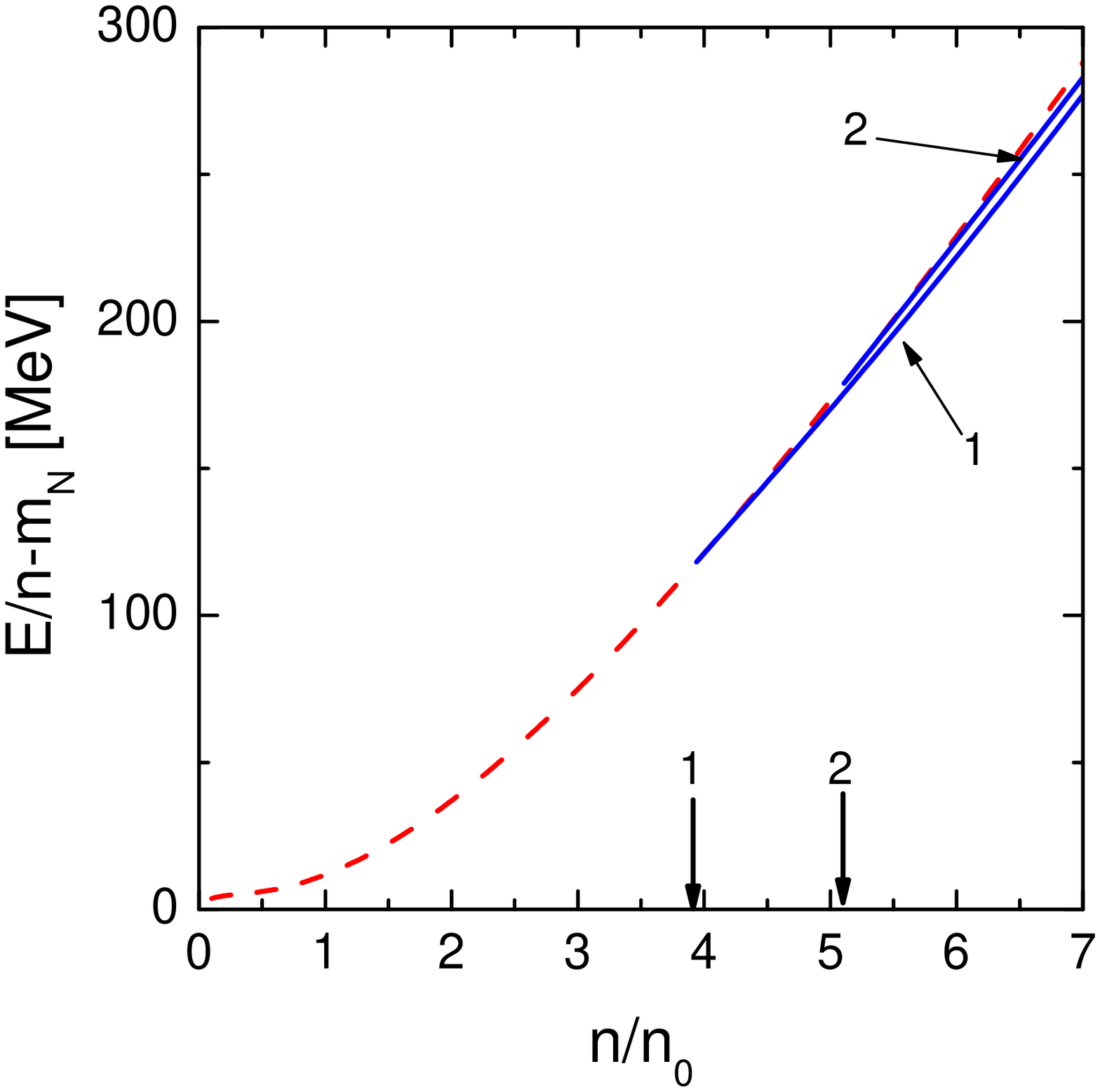}\quad
\includegraphics[clip=true,height=5cm]{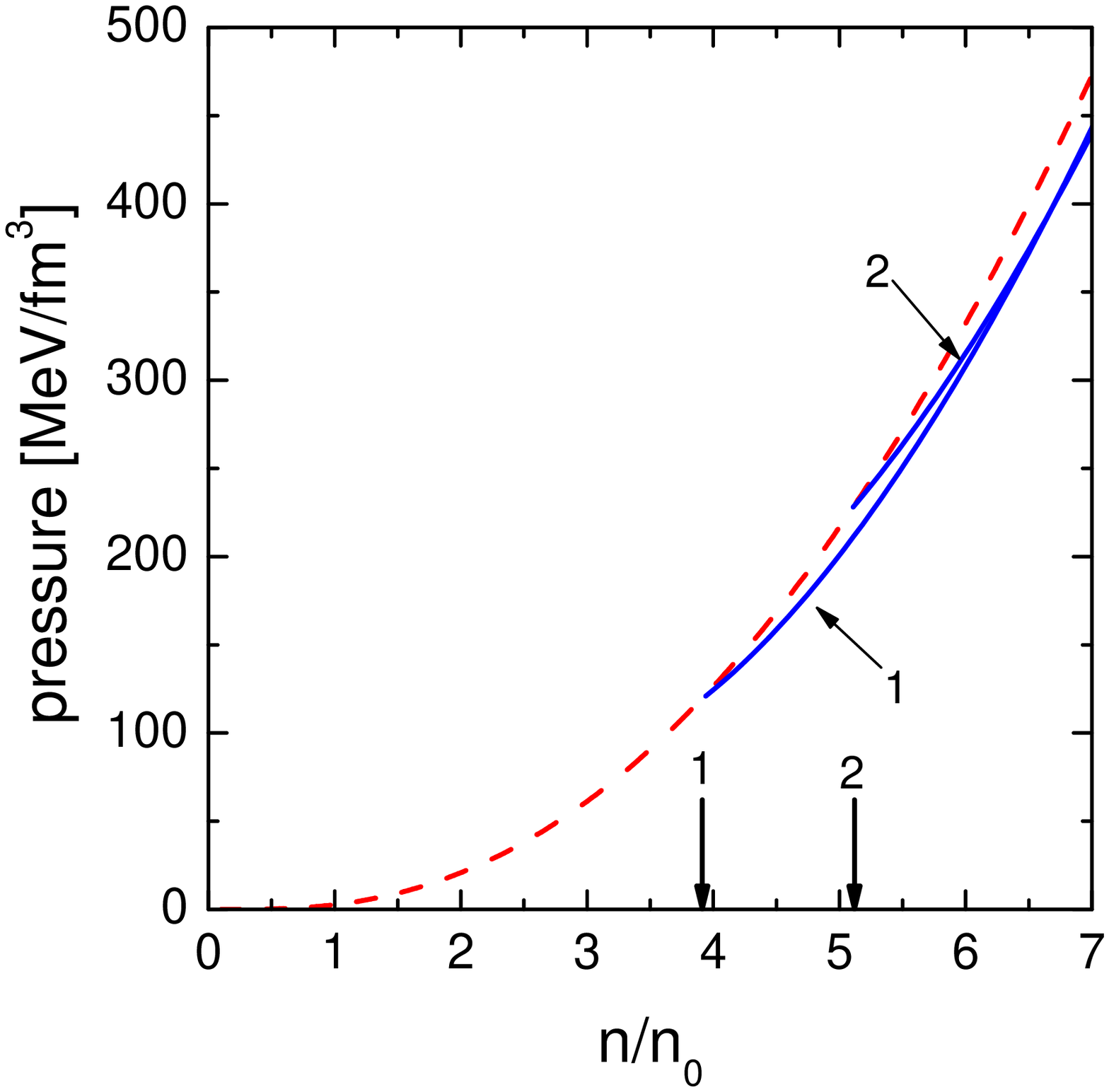}
\end{center}
\caption{The same as in Fig.~\ref{fig:enpress_Ior} but
for  the MW(nu) model with (\ref{etarho_hhj}), $z=2.9$,
the input-parameter set
~(\ref{inp_param_MWU_hhj},\ref{param_MWU_hhj}).
}
\label{fig:enpress_Ior-hhj}
\end{figure}

\begin{figure}
\begin{center}
\centerline{
\parbox{4.5cm}{\includegraphics[clip=true,width=4.5cm]{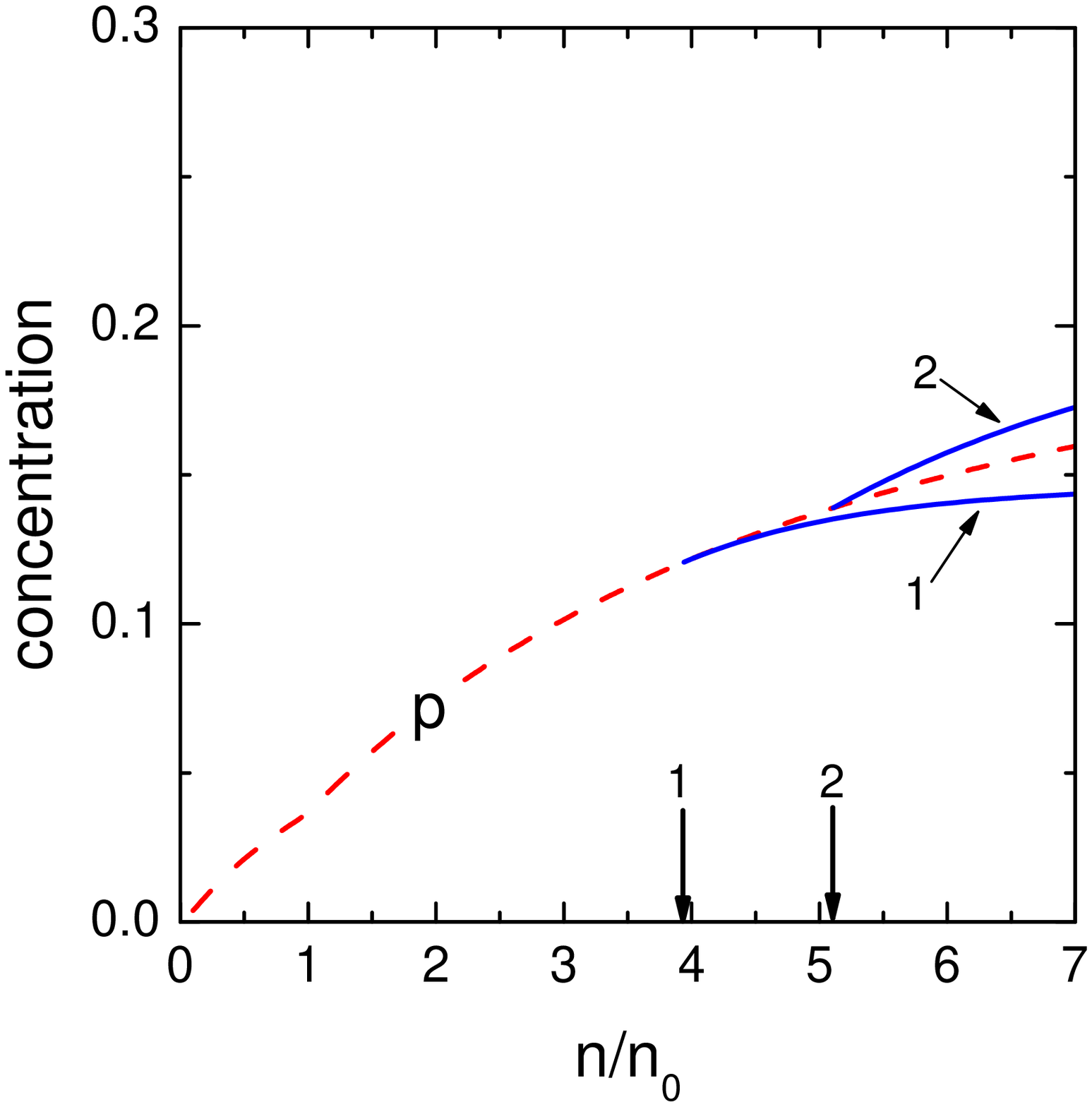}}\quad
\parbox{4.5cm}{\includegraphics[clip=true,width=4.5cm]{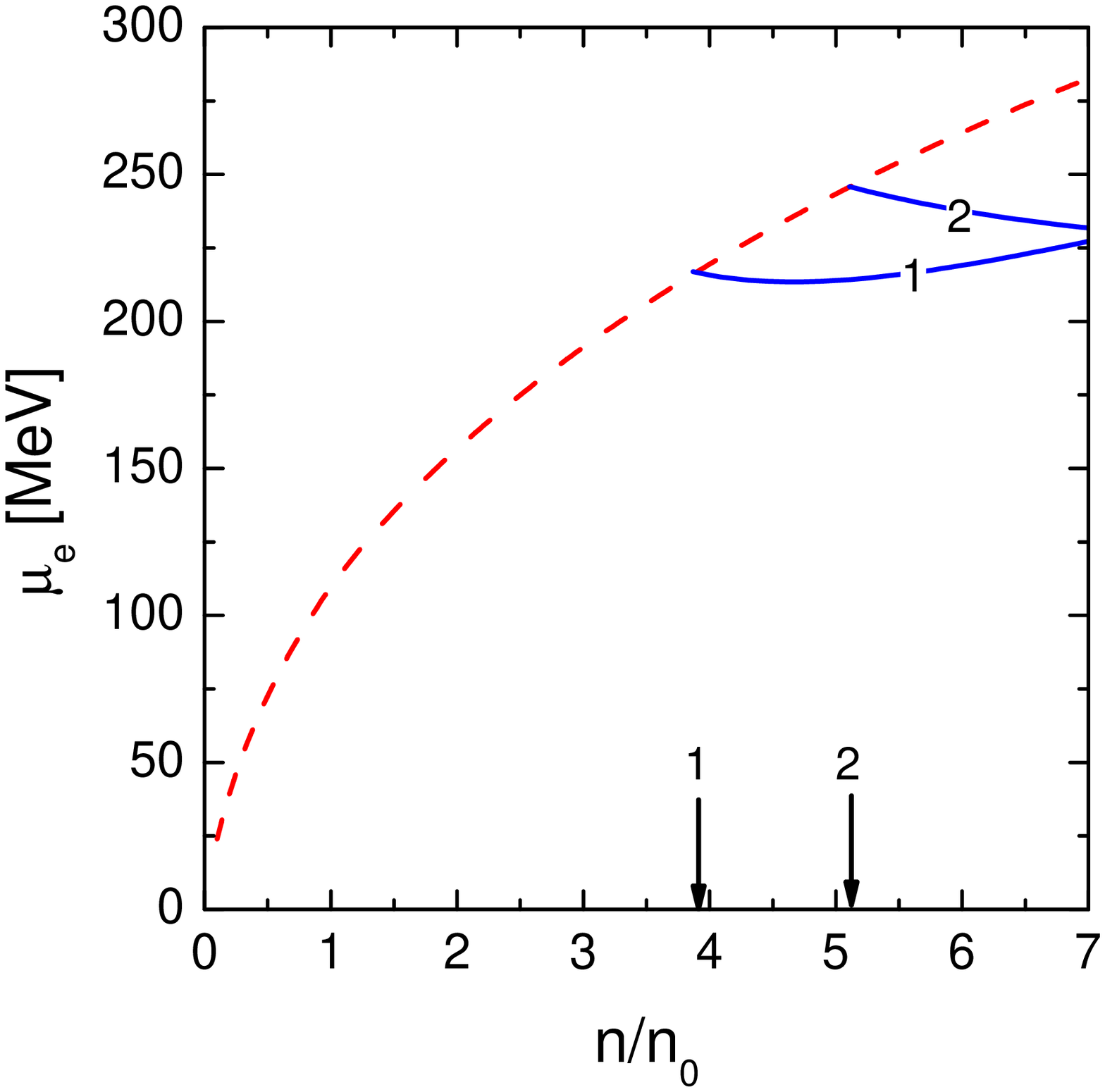}}\quad
\parbox{4.5cm}{\includegraphics[clip=true,width=4.5cm]{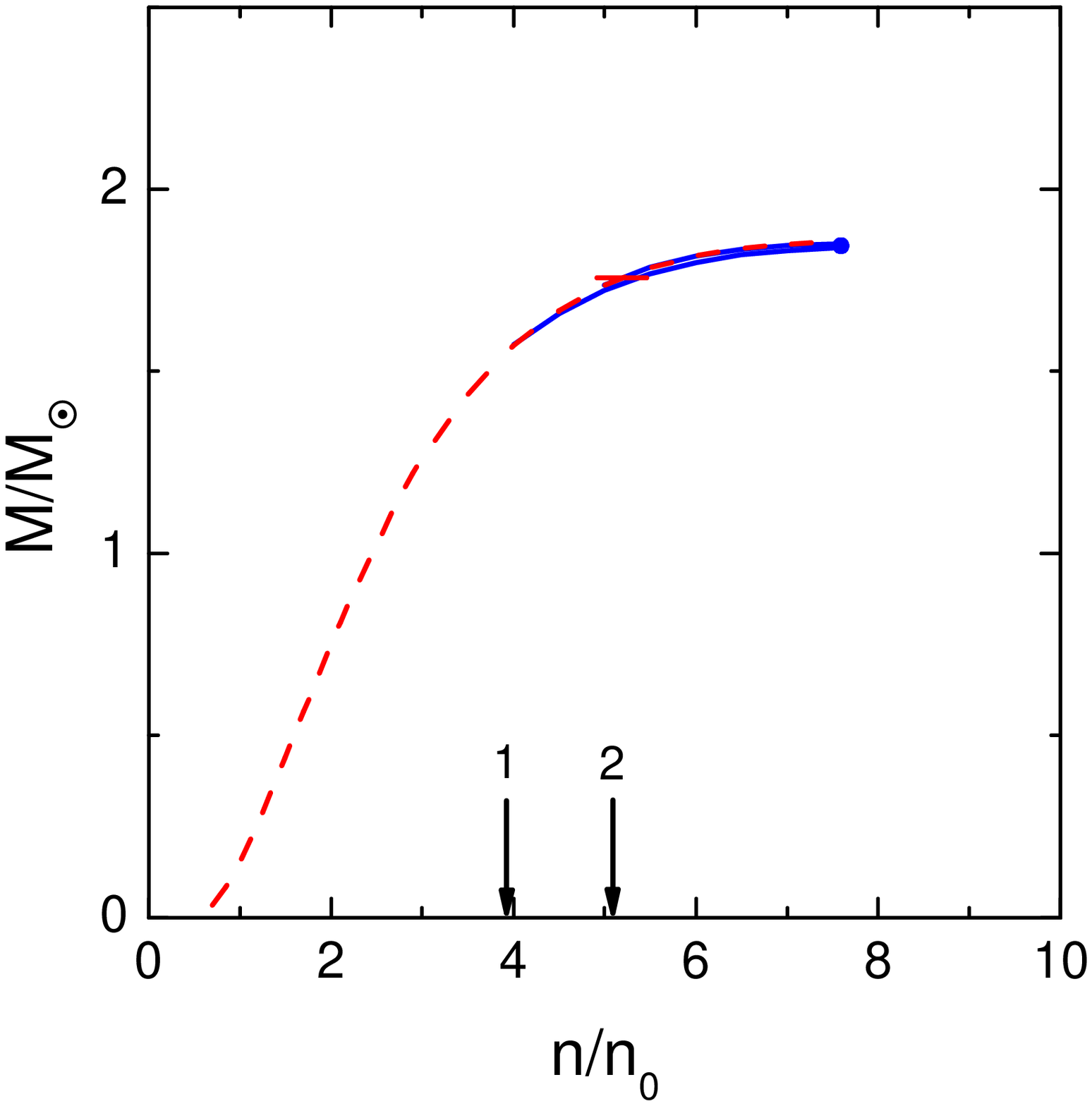}
}}
\end{center}
\caption{The same as in Fig.~\ref{fig:mue_Iore} but
for  the MW(nu) model with (\ref{etarho_hhj}), $z=2.9$, the input-parameter set
~(\ref{inp_param_MWU_hhj},\ref{param_MWU_hhj}).
}
\label{fig:mue_Iore-hhj}
\end{figure}

Figure \ref{fig:enpress_Ior-hhj} shows the same as Figure
\ref{fig:enpress_Ior} but for the MW(nu) model with
(\ref{etarho_hhj}), $z=2.9$, the input-parameter set
~(\ref{inp_param_MWU_hhj},\ref{param_MWU_hhj}).
Again we find no van~der~Waals behavior of
the pressure and only weak softening of the EoS. The proton
concentration, electron chemical potential and the NS masses for
this model are shown in Fig.~\ref{fig:mue_Iore-hhj}. The only
qualitative difference is seen in the behavior of the proton
fraction for the case $\chi_{\rho}^{'}=\Phi_{\rho}$. It increases
with increase of the density in this particular case. However the triangle
inequality for the DU processes is not fulfilled since a part of the
negative charge is taken by the $\rho^-$.
{\em{ Thus in the case of the MW(nu) model, $z=2.9$,  
DU processes do not occur in the phase with the condensate at all.}}

\begin{figure}
\begin{center}
\includegraphics[clip=true,height=5cm]{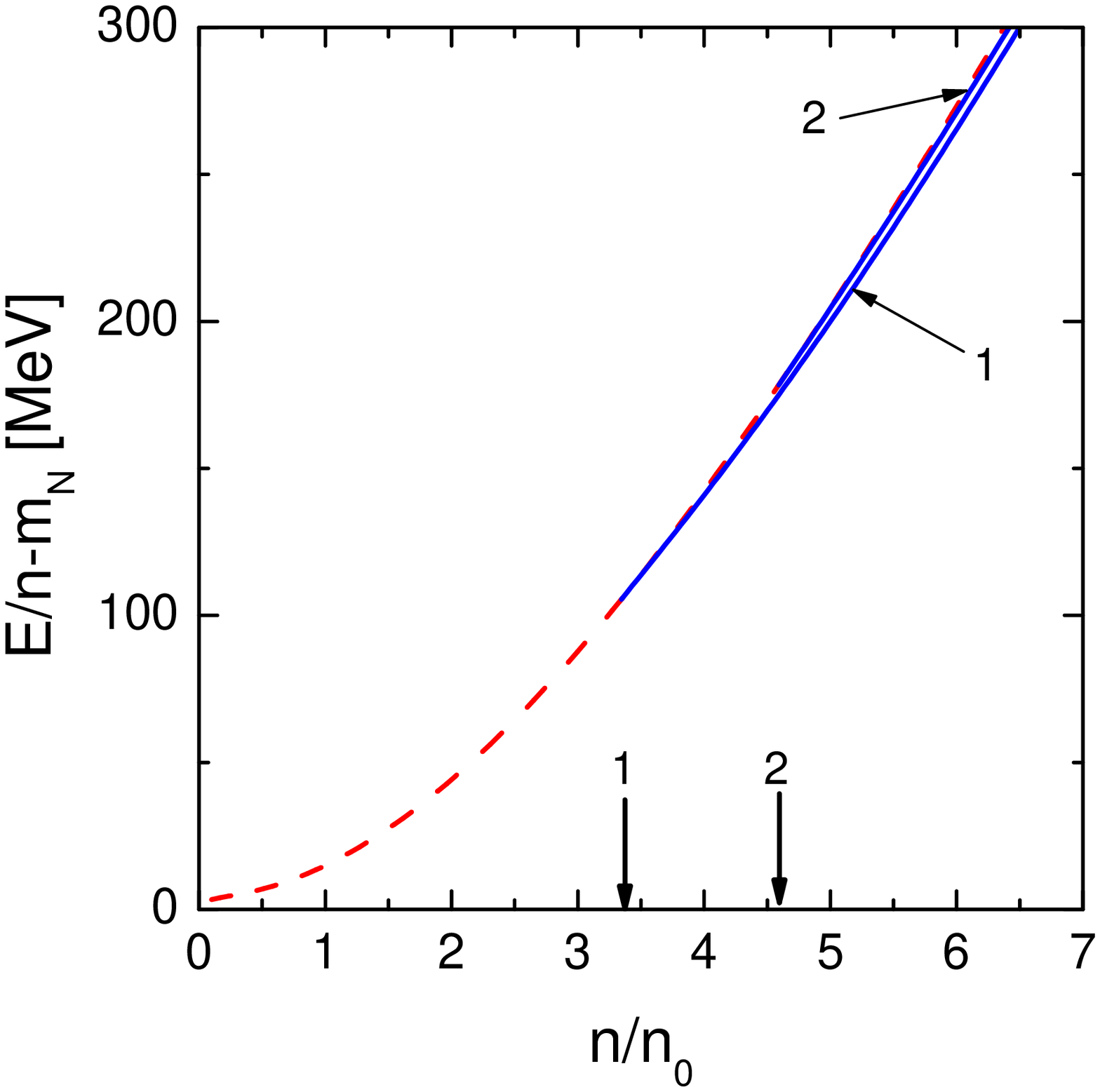}\quad
\includegraphics[clip=true,height=5cm]{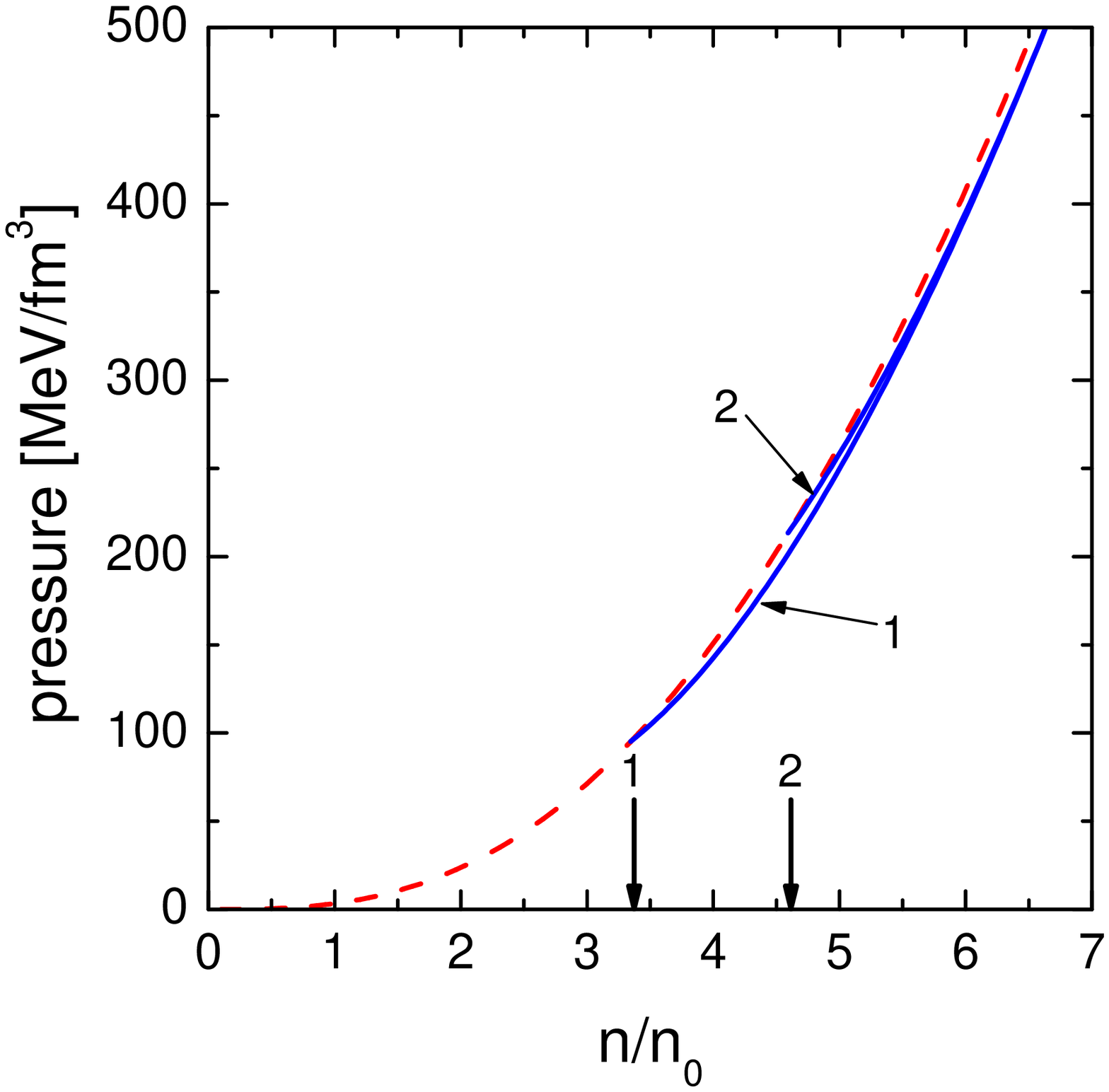}
\end{center}
\caption{The same as in Fig.~\ref{fig:enpress_Ior} but
for  the MW(nu) model (\ref{etnun}), $z=0.65$, the input-parameter set
(\ref{param},\ref{MWnu-inp},\ref{param_sc_065}).
}
\label{fig:enpress_Ior-mt}
\end{figure}

\begin{figure}
\begin{center}
\centerline{
\parbox{4.5cm}{\includegraphics[clip=true,width=4.5cm]{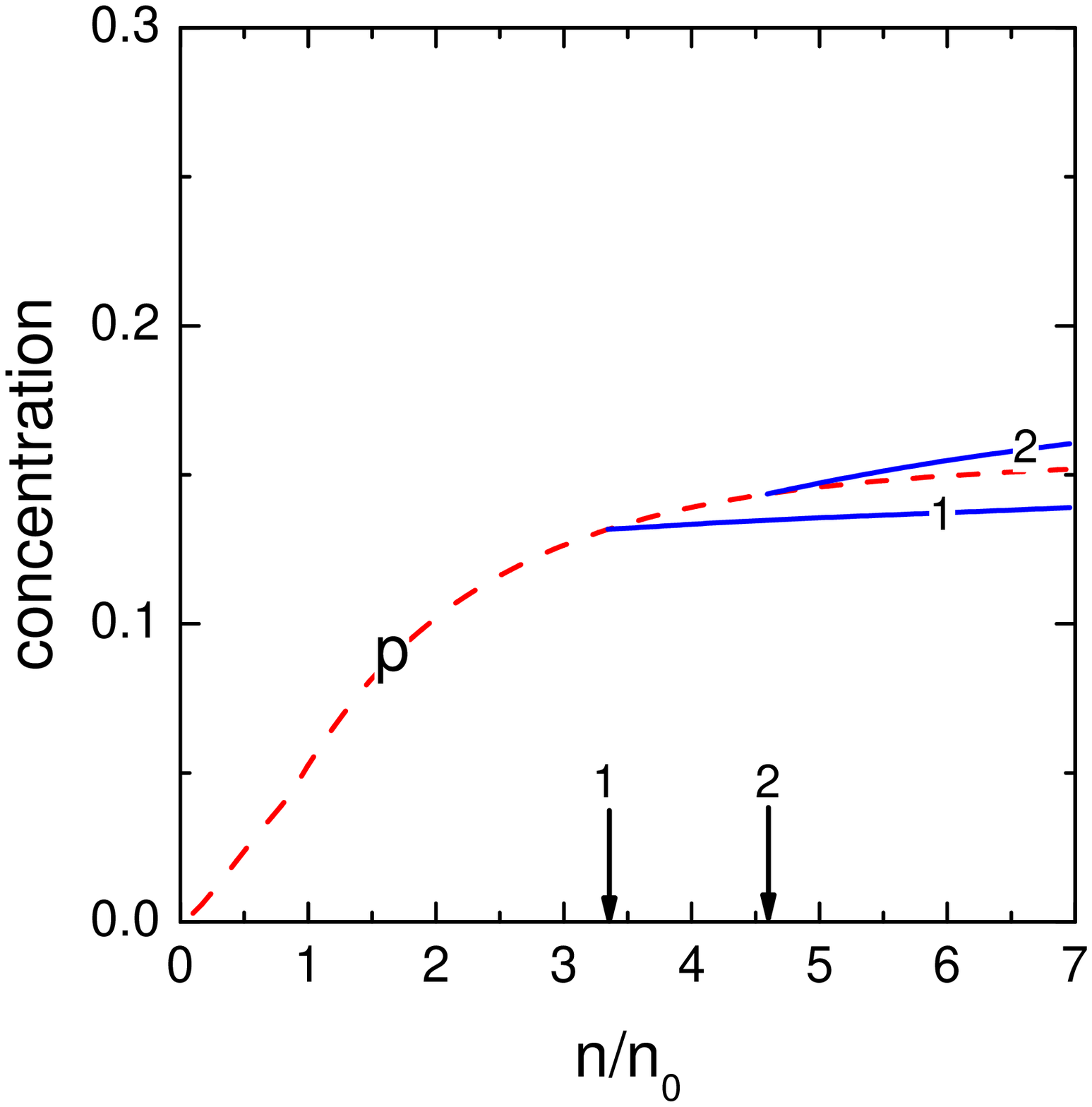}}\quad
\parbox{4.5cm}{\includegraphics[clip=true,width=4.5cm]{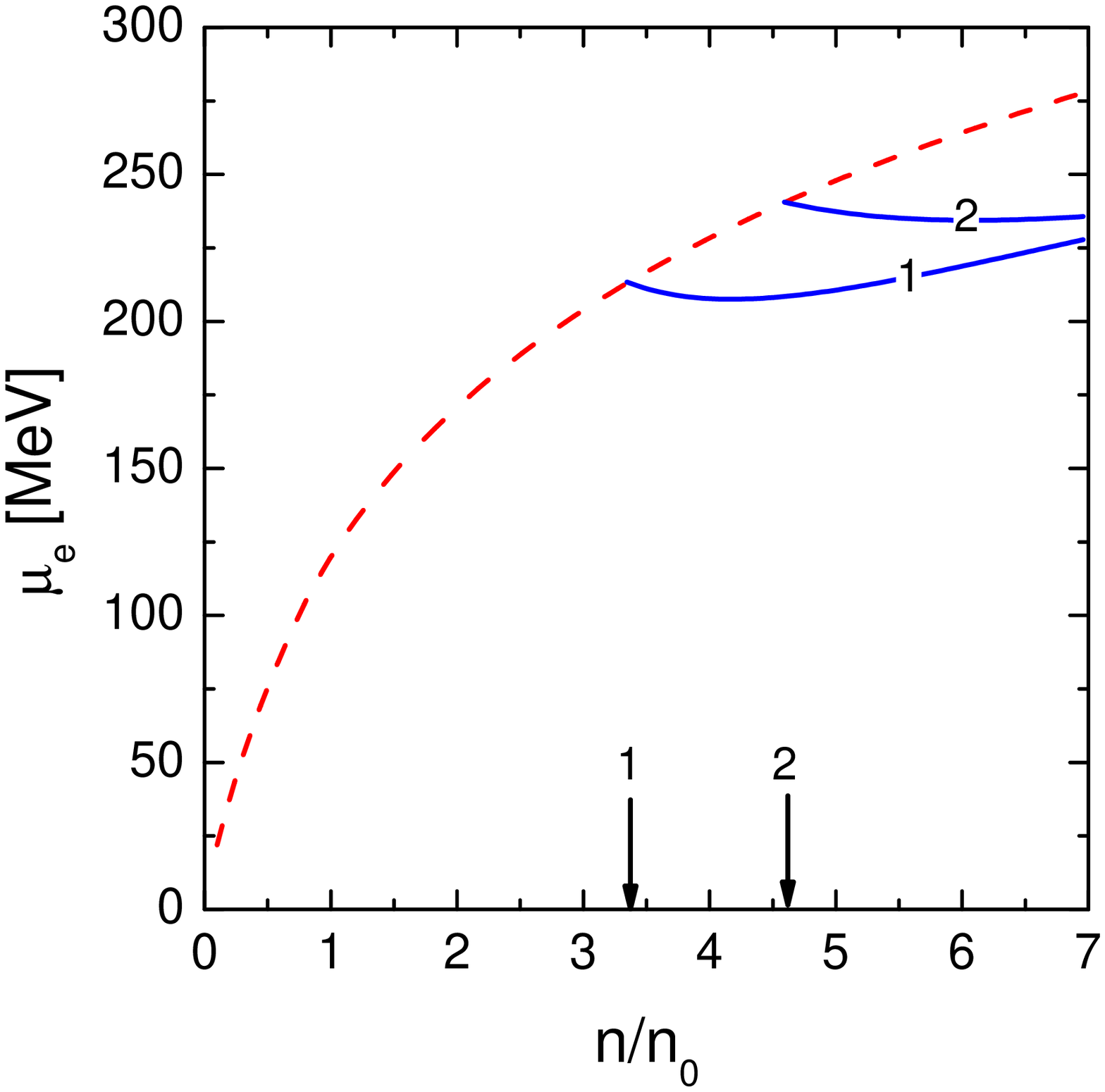}}\quad
\parbox{4.5cm}{\includegraphics[clip=true,width=4.5cm]{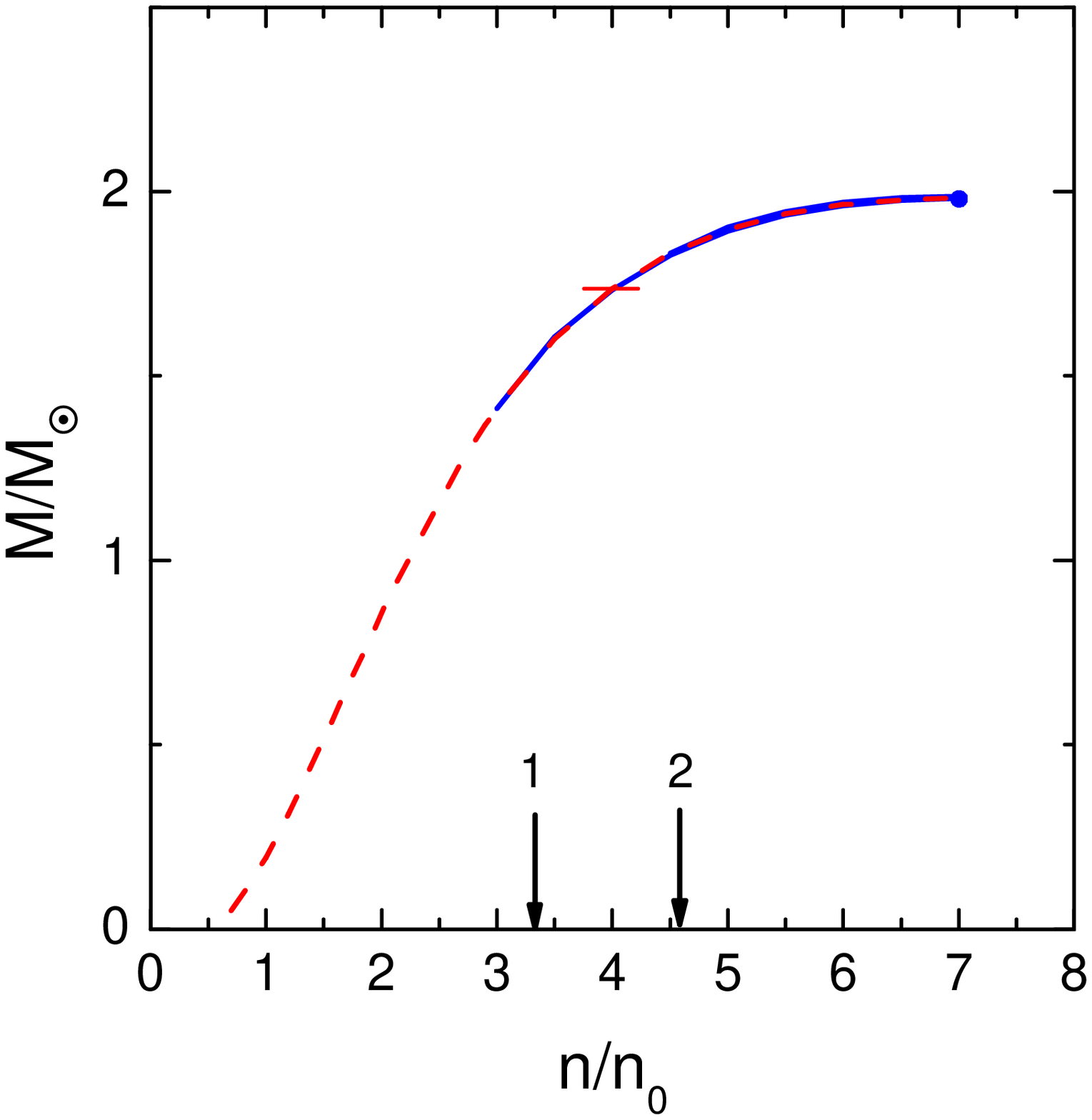}
}}
\end{center}
\caption{The same as in Fig.~\ref{fig:mue_Iore} but
for  the MW(nu) model (\ref{etnun}), $z=0.65$, the input-parameter set
(\ref{param},\ref{MWnu-inp},\ref{param_sc_065}).
}
\label{fig:mue_Iore-mt}
\end{figure}

Fig.~\ref{fig:enpress_Ior-mt} shows the same as
Fig.~\ref{fig:enpress_Ior-hhj} but for the MW(nu) model
(\ref{etnun}), $z=0.65$, the input-parameter set
(\ref{param},\ref{MWnu-inp},\ref{param_sc_065}).
Fig.~\ref{fig:mue_Iore-mt} shows the same as Fig.
\ref{fig:mue_Iore-hhj} for the given model. Qualitative behavior
of the quantities for the given MW(nu) model (\ref{etnun}) with
$z=0.65$ is the same as for the MW(nu) model (\ref{etarho_hhj})
with $z=2.9$. {\em{Also in the case  of the MW(nu) model, $z=0.65$,
DU processes are
non-operative in the condensate phase for both cases
$\chi_{\rho}^{'}=1$ and $\chi_{\rho}^{'}=\Phi_{\rho}$.}}

Concluding this section we found  that a new phase of the dense
nuclear matter  with a charged $\rho$ meson
mean field might exist in the NS interiors already for not too high densities.
For the most realistic cases we found $n_{\rm c}^{\rho\,,\rm II}\lsim 4~n_0$.

\section{Neutrino emissivity in the DU processes on $\rho^-$
  condensate}\label{sec:rhocool}

As we have demonstrated, the standard DU processes $n\rightarrow
pe\bar{\nu}$, $pe\rightarrow
n\nu$ are forbidden in cases of the MW(nu) models
(\ref{etnun},\ref{param_sc_065}) and (\ref{etarho_hhj},\ref{param_MWU_hhj})
up to sufficiently high density ($n>4~n_0$ and  $n>5.2~n_0$ for the case $\chi_{\rho}^{'}=1$).
Therefore we  consider these models as most realistic ones.
However a new channel of the DU-like processes on the condensate charged
$\rho^-$ meson, e.g.,
$n\rho^{-}_c \rightarrow ne\bar{\nu}$ becomes operative for $n>n_c^{\rho , \rm
  II}$. Since the charged $\rho^-$
condensate arises in the s-wave, the calculation of its contribution to the
emissivity can be done similar to that
for the kaon condensation processes,
except additional requirement of the chiral symmetry
used in ref.~\cite{BKPP}. Note that ref. \cite{BKPP} did not take into account
nucleon-nucleon correlations in vertices that yields an additional
suppression factor $\lsim 10^{-1}$ for all condensate
processes, cf. \cite{VS84,MSTV,V01}. Recovering the missing factor we easily estimate
the neutrino emissivity
\be
\epsilon_{\nu}^{\rho^-_c}\sim 10^{26}\, g_{\rho
N}^2\,\chi_{\rho}^2\,
\left(\frac{m_N^*}{m_N}\right)^2 \, \left(\frac{\mu_e}{m_N}\right)\,T_9^6
\frac{n^{\rho}_{ch}}
{m_{\rho}\,\Phi_{\rho}\,m_{\pi}^2} \, \Gamma^4\,,  \quad \frac{\rm erg}{\rm cm^3
  \cdot sec},
\ee
$m_{\pi}=140~$MeV,
$T_9 =T/10^9 \mbox{K}$,
$\Gamma^4 =\Gamma_s^2 \Gamma_{w-s}^2$ is the nucleon-nucleon correlation
factor,
\be
&&\Gamma_s =\widetilde{\Gamma}(f^{'} , \varepsilon \simeq \mu_e , k\simeq
p_{\rm Fn}),\quad \widetilde{\Gamma}(x,\varepsilon ,k)=[1-4C_0 x
A_{np}(\varepsilon ,k)]^{-1},\nonumber\\
&&\Gamma_{w-s}^2 =\widetilde{\Gamma}^2(f^{'} ,\varepsilon =q\simeq p_{\rm Fe}),
\ee
$C_0 \simeq 0.77 m_{\pi}^{-2}$ relates to  the density of states at the Fermi
surface for $n=n_0$,
$f^{'}\simeq 0.5\div 1$
is the Landau-Migdal parameter in isospin
channel, cf. \cite{M83,MSTV,ST98}, $A_{np}$ is the
proton -- neutron hole loop (without spin degeneracy factor), see \cite{V01}.
%%%$A_{np}(\varepsilon \ll kp_{\rm Fn}/(m_N \Phi_N))\simeq -m_N \Phi_N p_{\rm Fn} /(2\pi^2)$,
%%%$A_{np}(\varepsilon\simeq q)\simeq -n_n/(2\mu_e)$.
As we have mentioned, the presence of the correlation factor is crucially important. It may suppress the rate
by a factor $\sim 10^{-1}\div 10^{-2}$. Another suppression factor arises from a small
concentration $n^{\rho}_{ch}$ in the new phase, $n^{\rho}_{ch}<n_p$.

In presence of the nucleon superfluidity the emissivity of the
DU-like processes is additionally suppressed by a  factor
$\mbox{min}[\xi_{nn},\xi_{pp}]$, $\xi_{ii}\sim e^{-\Delta_{ii}/T}$
for $T<T_{ci}$, where $\Delta_{ii}$ is the pairing gap in the
corresponding nn or pp channel. According to \cite{SF,BGV04} the
$^3P_2$ $nn$-pairing gap should be very small, less than $10~$keV.
If so, $\mbox{min}[\xi_{nn},\xi_{pp}]\simeq \xi_{pp}$. Final rough
estimation of the emissivity is
\be
\epsilon_{\nu}^{\rho^-_c}\sim
(10^{24}\div 10^{25}) \, T_9^6 \, \mbox{min}[\xi_{nn},\xi_{pp}]
\quad \frac{\rm erg}{\rm cm^3 \cdot sec}.
\ee
This rate is typically either smaller or of the  order of that expected for the
charged pion condensate. However  it is much smaller than
that for the standard DU processes, if the latter were permitted.
On the other hand, for typical densities and temperatures
$T\sim 0.3\div 3\cdot 10^8$~K
the emissivity on the charged $\rho^-$
condensate   is somewhat larger than
the emissivity  of the competing medium-modified Urca processes,
as they have been computed in \cite{SVSWW,BGV04}. 

As has been shown in ref.~\cite{BGV04} within HHJ EoS, the
critical density of the charged pion condensate should not be less
than $(2.5\div 2.7) n_0$. Otherwise even rather low mass NS would
cool too fast in disagreement with the data. A similar  statement
can be done in case of the charged $\rho^-$ condensation. Thus we see that
the values $n_c^{\rho ,\rm II}\sim (3\div 4)~n_0$ that we have
obtained within our MW(u) and MW(nu) models do not contradict to
the analysis of the NS cooling \cite{BGV04}.

\section{Conclusion}\label{sec:concl}

In this paper we studied generalized relativistic mean field
models that include in-medium changes of meson masses and
coupling constants. We demonstrated equivalence of a certain class of
the models. Practically we showed that the energy density for
homogeneous matter depends only on three independent combinations
of mass and coupling scaling functions.
Often this equivalence is not realized  and one considers
some models as different, whereas in reality they are  identical
for the description of homogeneous matter on the mean field level.

We showed that a naive modification of solely boson masses
would lead to severe drawbacks, like discontinuous changes of mean
fields and hadron masses with the density. To get rid off these
problems the mass scaling should be essentially compensated  by
the scaling of the corresponding coupling constants.

For linear RMF models, i.e., without a non-Abelian  $\rho$ interaction and a non-linear
$\sigma$ potential $U$, we demonstrated equivalence of the models
with universal scaling $g_i^*\,m_i/g_i\,m_i^*\simeq 1$\,,
$i=\sigma, \om,\rho$, and the models without any scaling of those
masses and couplings.
For non-linear RMF models ($U\neq 0$) the above equivalence would
require a certain relation between corresponding non-linear
potentials.

We have demonstrated the efficiency of the generalized
relativistic mean field models based on the modified Walecka (MW) model with a
non-universal scaling of meson masses and couplings (MW(nu)).
It allows to enlarge the threshold density for the direct Urca
process and produces a stiffer equation of state without any 
changes of the EoS near the saturation density. These possibilities are
favored  by "the nuclear medium cooling scenario" for the neutron
star cooling, cf. \cite{BGV04} and by recent measurements of heavy
neutron stars, cf. \cite{S04}.

Our generalized mean field models with a non-universal scaling of meson masses
and couplings might be useful in discussion of possibilities of different
phase transitions in neutron star interiors, since they allow to compensate a
softening of the equation of state due to a phase transition by a stiffening
due to a stronger decrease of masses compared to a decrease of couplings.
Especially the model might be helpful in discussion of hybrid stars.
With the standard hadron models the quark matter either begins to appear
already at rather low neutron star mass, or does not appear at all, cf. \cite{BGV1}.
In the former case the limiting mass of the star is found to be rather low
(typically $\lsim 1.8~M_{\odot}$).
If one used a model assuming a smaller nucleon mass at the saturation, one
could get a still  higher value of the limiting neutron star mass.

We also demonstrated an example of the relativistic mean field
model that fits well the modern Urbana-Argonne equation of state
(more precisely, the fit \cite{HHJ} that cures the causality
problem), yielding the same threshold density for the direct Urca
process.

The MW(nu) models can be useful for the description of properties
of finite nuclei. Here equivalence is absent due to
the spatial inhomogeneity of the field profiles. In standard MW
models one is forced to decrease the $\sigma$ meson mass in order
to appropriately fit the proton density profile and the binding
energy as a function of the atomic number, cf. \cite{VYTkaon}. In
the framework of the MW(nu) models there is still a room to fit
the scaling laws for effective masses and couplings.

It would be interesting to apply the model to the description of
heavy-ion collisions. Early naive studies of the pion yield in
heavy-ion collisions required extremely hard EoS, with the
compressibility $K\sim 1000$~MeV. Further it was demonstrated, cf.
\cite{MSTV,V83,KV96} and references therein, that the problem might be
naturally solved within the standard modified Walecka model (MW) when
one includes contributions from  soft pion modes. These
modes yield very small contribution to thermodynamical
characteristics of the cold nuclear matter but produce a large
contribution for the hot nuclear matter. The physical
reason is obvious: soft modes are easily excited with the
temperature increase. Due to a decrease of $\sigma$, $\omega$, $\rho$
and nucleon masses related modes become to be softer. Thus our mean field
models may give an appropriate basis to go beyond the mean field level
considering thermal excitations.

Recent studies of the collective isospin flow \cite{THIM}
extracted the density
dependence of the symmetry energy but did not introduce any
astrophysical constraints and a possible soft mode contribution. It is therefore a challenge to fit the
corresponding data using the constructed above models with
inclusion of thermal contribution from soft modes.

If one introduces a rho-meson field as a non-Abelian gauge boson
(cf.~\cite{bando83}) in a relativistic mean-field model which supports the decreasing
$\rho$-meson mass, one should take care of the possibility of the
charged rho meson condensation in neutron star matter.
Following~\cite{v97}, we found that at density $n>n_{\rm
c}^{\rho\,,\rm II}$ the charged $\rho$-meson mean field appears.
We demonstrated that in the framework of our relativistic mean
field models the novel phase  arises by a second order phase
transition. The critical density $n_{\rm c}^{\rho\,,\rm II}$ is
rather low in the neutron star matter (typically $\sim 3\div
4\,n_0$), if one allows for a decrease of the rho meson mass.
We also included the corresponding decrease of the effective
nucleon-rho coupling, as well as the scaling of other meson masses
and couplings. If one used a model assuming a smaller nucleon mass
at the saturation, one could get a still  smaller value of the
critical density for the charged $\rho$ condensation. Charged
$\rho$ condensation slightly softens the equation of state. In the
neutron-enriched matter the charge of the $\rho$-meson condensate
is negative. We evaluated the emissivity of the direct Urca-like
processes on the condensate $\rho^-$ meson. The resulting rate
behaves similar to that for the $\pi^-$ condensate. The analysis
of the cooling data indicates that the value $n_{\rm c}^{\rho\,,\rm II}$
should be not too small, $n_{\rm c}^{\rho\,,\rm II}\gsim 2.5\div
2.7~n_0$.

Our non-Abelian $\rho$ meson field is in some sense analogous to
the gluon field. Dropping of the effective $\rho$ meson mass
enforces a similarity. Effective nucleon and other meson masses also
decrease with increase of the density. Decrease of the
corresponding effective couplings motivates  a future  asymptotic
freedom in the quark-gluon phase. All these features  might be in
favor of a smooth transition from the hadron  degrees of freedom
to the quark-gluon degrees of freedom. Up to now there is no appropriate
description of a cross-over temperature region seen from the lattice results
\cite{Fodor}.
Formulated above
models might be helpful in construction of such a description.

\noindent
{\bfseries{Acknowledgments}}\\
We would like to thank A.D. Jackson for stimulating discussions.
We also thank Yu. B. Ivanov for the discussion of the results.
D.N.V. highly appreciates hospitality and support of GSI
Darmstadt. The work has been
supported in part by DFG (project 436 Rus 113/558/0-2), and by
RFBR  grant NNIO-03-02-04008.

\appendix
\section{Thermodynamic potential of RMF model in presence of the electric field}
\label{sec:rhonse}

So far we have disregarded the electromagnetic interactions. But
only including the coordinate dependent electric field one may
properly discuss structures, that may occur at first order phase
transitions in charged systems. Indeed, a charged structure may
exist, only if there is an  inhomogeneous (not constant) electric
potential. Although we did not find a first order phase transition
in the framework of our RMF model EoS, this result might be still
model dependent. Here we demonstrate how one can include
electromagnetic interactions into the scheme. We consider the NS
matter consisting of the neutrons, protons, electrons and muons.
We incorporate mean meson fields including charged $\rho$ meson
field, $\rho_\mu^-=a_\mu\,\rho_c$, $\rho_\mu^-=a_\mu\,\rho_c^\dag$ with
$a_\mu=(0,\vec{a})$  and the electric field.

The interaction with the static electric field $eA^\mu
=(-V,\vec{0})$ can be introduced according to the minimal coupling prescription via the
gauge  replacement
\be
\prt_\mu \Psi_N &\to& \left(\prt_\mu+\frac{i}{2} e A_\mu\,
  (1+\tau_3)\right)\Psi_N\,,\quad
\prt_\mu\Psi_l \to (\prt_\mu -i eA_\mu)\Psi_l\,,
\nonumber\\
\vec{\rho}_{\mu\nu}&\to& \vec{\rho}_{\mu\nu}-e\,
A_\mu\,[\vec{n}_3\times \vec{\rho\,}_\nu]+e\,
A_\nu\,[\vec{n}_3\times \vec{\rho\,}_\mu]\,,
\label{minc}
\ee
where $(\vec{n}_3)^a=\delta^{a3}$ is the unit vector in the isospin
space, $a=1,2,3$\,.
The frequencies of the charged fields
acquire the shift $ \varepsilon\rightarrow \mu -V$ for negatively charged fields ($e$,
$\mu^-$ and $\rho^-$ in our case) and
$\varepsilon \rightarrow \mu +V$ for positively charged fields (for proton).

The density of the total thermodynamic potential is as follows:
\be\label{omAp}
\Omega &=&\Omega_N +\Omega_M +\Omega_V +\Omega_e +\Omega_\mu ,\\
\Omega_N &=&\left(\intop_0^{p_{{\rm F},n}}+\intop_0^{p_{{\rm F},p}}\right)
\frac{\rmd p p^2}{\pi^2}\,\sqrt{m_N^2\, \Phi_N^2(f)+p^2}-n_n \nu_n -n_p \nu_p ,\\
&&\nu_p =\mu_{n}-\mu_e +V-g_\om \chi_\om \om_0 -\frac{1}{2}g_\rho \chi_\rho \rho_0^{(3)},
\nonumber\\
&&\nu_n =\mu_{n}-g_\om \chi_\om \om_0 +\frac{1}{2}g_\rho \chi_\rho \rho_0^{(3)},
\nonumber\\
\Omega_M &=&\frac{1}{2}(\nabla \sigma )^2 +\frac{1}{2}m_\sigma^2\Phi_\sigma^2 +U(\sigma)\nonumber\\
&-&\frac{1}{2}(\nabla \om )^2 -\frac{1}{2}m_\om^2\Phi_\om^2 -
\frac{1}{2}(\nabla \rho_0^{(3)} )^2 -\frac{1}{2}m_\rho^2\Phi_\rho^2
+\Omega_{ch}^\rho ,\\
\Omega_{ch}^\rho &=&|\nabla\rho_c |^2+\left[m_\rho^2\Phi_\rho^2-(\mu_{ch}^\rho
  -V-g_\rho \chi_\rho^{'}\rho_0^{(3)} )^2\right]|\rho_c |^2 , \\
\Omega_V &=&-\frac{1}{8\pi e^2}(\nabla V)^2 ,\\
\Omega_{l} &=&\intop_0^{p_{{\rm F},l}}
\frac{\rmd p p^2}{\pi^2}\,\sqrt{m_l^2\, +p^2}-n_l (\mu_l -V) \quad l=e,\,\mu\,,
\ee
for $|\mu_e -V|>m_\mu$.

With the help of the variation
\be\label{eqBL}
\frac{\delta}{\delta n_i} \left(\int \Omega d\vec{r}\,\right) =0
\quad i =n,\,\,p,\,\,e,\mu
\ee
and
\be
\frac{\delta}{\delta \xi_i} \left(\int \Omega d\vec{r}\,\right) =0,
\quad  \xi_i =\sigma ,\,\,\om,
\,\,\rho_0^{(3)} ,\,\,\rho_c, \,\,V
\ee
we recover necessary equations of motion. In the body of the paper we used the same
equations but for fields independent of coordinates.

Chemical potentials of neutrons, protons, electrons and muons are found from
equations of motion (\ref{eqBL}):
\be\label{mun}
\mu_n =\sqrt{m_N^2\, \Phi_N^2(f)+p^2_{\rm F,n}}+g_{\om}\chi_\om \om_0
-\frac{1}{2}g_{\rho}\chi_{\rho}\rho_0^{(3)} ,
\ee
\be\label{mup}
\mu_p +V=\sqrt{m_N^2\, \Phi_N^2(f)+p^2_{\rm F,p}}+g_{\om}\chi_\om \om_0
+\frac{1}{2}g_{\rho}\chi_{\rho}\rho_0^{(3)} ,
\ee
\be
\mu_l -V =\sqrt{m_l^2 +p^2_{\rm F,l}}\,.
\ee
Chemical potentials of proton, neutron, electron,
muon and charged $\rho$ are related to each other following equilibrium
conditions in
reactions $n\leftrightarrow p+e$, $n\leftrightarrow
p+\mu^-$,$n\leftrightarrow
p+\rho^-$. Thus $\mu_n =\mu_p+\mu_e$, $\mu_e =\mu_\mu
=\mu_{ch}^\rho$.

The local charge neutrality condition that we used in the paper body is now replaced by the Poisson
equation for the electric potential
\be\label{Poisson}
\Delta V =-4\pi (n_p -n_e -n_\mu +n^{\rho}_{ch}),
\ee
where
\be
n_{e,\mu} = \frac{[(V-\mu_{e,\mu})^2 -m_{e,\mu}^2]^{3/2}}{3\pi^2}\theta
(|V-\mu_{e,\mu}|-m_{e,\mu})
\ee
for electrons and  muons and
\be
n^{\rho}_{ch}=\Big(g_{\rho}\,{\chi}'_\rho\,\rho_0^{(3)}+V-\mu_{ch}^{\rho}\Big)\,
|\rho_c|^2 <0
\ee
for the charged $\rho^-$ meson condensate. The local charge
neutrality condition appears as the solution of the Poisson
equation for the charged structure of the size much larger than
the maximum value among Debye screening lengths of different
species. For the case of Maxwell construction, solving
(\ref{Poisson}) one may describe the boundary layer between
phases. With these findings one may also solve the problem of
possible occurrence of the mixed phase structures. This discussion
is in full analogy to the framework  of ref.~\cite{VYTkaon}
describing mixed phase at the first order  phase transition to the
s-wave negative kaon condensate.

%------------------------------------------------------------%


\begin{thebibliography}{99}
\bibitem{G00}
N.K.~Glendenning, {\it Compact Stars: Nuclear Physics,
Particle Physics, and General Relativity}, 2nd edition,
Springer (N.Y.) (2000);
F.~Weber, {\it
Pulsars as Astrophysical Laboratories for
Nuclear and Particle Physics }, IoP Publishing, 1999.
\bibitem{APR98}
A. Akmal, V.R. Pandharipande and D.G. Ravenhall, Phys. Rev. {\bf C58} (1998) 1804.
\bibitem{BB77}
J. Boguta and A.R. Bodmer, Nucl. Phys. A {\bf 292} (1977) 413.
\bibitem{Boguta81}
J. Boguta, Phys. Lett. B {\bf 106} (1981) 255.
\bibitem{MSTV} A.B. Migdal, E.E. Saperstein, M.A. Troitsky, D.N. Voskresensky,
  Phys. Rep. {\bf 192} (1990) 179.
\bibitem{LP}
J.M. Lattimer, C.J. Pethick, M. Prakash, P. Haensel, Phys. Rev. Lett. {\bf 66}
(1991) 2701.
\bibitem{YAKDU}
D.G. Yakovlev, O.Y. Gnedin, A.D. Kaminker, K.P. Levenfish, A.Y.
Potekhin, Adv.~Space~Res. {\bf 33} (2004) 523
\bibitem{TTTTT02}
S. Tsuruta, M.A. Teter, T. Takatsuka, T. Tatsumi, R. Tamagaki,
Ap. J., {\bf 571} (2002) L143.
\bibitem{T04}
S. Tsuruta, 2004, in: Proceedings of IAU Symposium
``Young Neutron Stars and their Environments'', F. Camilo, B.M. Gaensler
(Eds.), vol. 218,  [arXiv:astro-ph/0401245].
\bibitem{BGV04}
D. Blaschke, H. Grigorian, and D.N. Voskresensky, Astron.\ Astrophys. {\bf 424},
979 [arXiv:astro-ph/0403170].
\bibitem{KV}
E.E. Kolomeitsev and  D.N. Voskresensky, Phys. Rev., {\bf C68} (2003) 015803.
\bibitem{S04}
I.H. Stairs, Science, {\bf 304} (2004) 547.
\bibitem{TC99}
S.E. Thorsett, and D. Chakrabarty, Ap. J. {\bf 512} (1999) 288.
\bibitem{Q03}
H. Quaintrell et al.,Astron.\ Astrophys., {\bf 401} (2003) 303.
\bibitem{Miller}
W. Zhang, T.E. Strohmayer and J. H. Swank, Ap. J. {\bf 482} (1997) L167;
M.C. Miller, F.K. Lamb and D. Psaltis, Ap. J. {\bf 508} (1998) 791.
\bibitem{HHJ}
H. Heiselberg and  M. Hjorth-Jensen, eprint:~astro-ph/9904214;
Phys. Rep. {\bf 328} (2000) 237.
\bibitem{NK}
D. Kaplan and A. Nelson, Phys. Lett., {\bf B175} (1986) 57.
\bibitem{chiral}V.~Koch,
Int.\ J.\ Mod.\ Phys.\ E {\bf 6} (1997) 203.
\bibitem{BR}
G.E.~Brown and  M.~Rho,
Phys.\ Rev.\ Lett.\ {\bf  66} (1991) 2720;
Phys.\ Rep.\ {\bf 396} (2004) 1.
\bibitem{SBMR97}
C.~Song, G.E.~Brown, D.-P.~Min and M.~Rho, Phys. Rev., C  {\bf 56}
2244.
\bibitem{consist}
A.A.~Shanenko, E.P.~Yukalova and V.I.~Yukalov,
Nuovo Cim.\ A {\bf 106} (1993) 1269;
C.~Song, D.P.~Min and M.~Rho,
Phys.\ Lett.\ B {\bf 424} (1998) 226;
T.S.~Biro, A.A.~Shanenko and V.D.~Toneev,
Phys.\ Atom.\ Nucl.\  {\bf 66} (2003) 982.
\bibitem{manka}
I.N.~Mishustin, J.~Bondorf, and M.~Rho, Nucl. Phys. A {\bf 555}
(1993) 215;\\
R.~Ma\'nka and I.~Bednarek,
J.\ Phys.\ G: Nucl.\ Part.\ Phys.\ {\bf 27} (2001) 1975.
\bibitem{THIM}
K. Tuchitani et al., arXiv:nucl-th/0407004.
\bibitem{dan}
P. Danielewicz, R. Lacey and W.G. Lynch, Science {\bf 298} (2002) 1592.
\bibitem{Gai}
T. Gaitanos, M. Di Toro, G. Ferini, M. Colonna and H.H. Wolter,  arXiv:nucl-th/0402041.
\bibitem{v97}
D.N.~Voskresensky, Phys.\ Lett.\, B {\bf 392} (1997) 262.
\bibitem{bando83}
M.~Bando, T.~Kugo, S.~Vehara, K.~Yamawaki and T.~Yamagida,
Phys.\ Rev.\ Lett.\ {\bf 54} (1983) 1215;
M.~Bando, T.~Kugo, K.~Yamawaki, Phys. Rep. {\bf 164} (1988) 217.
\bibitem{migdal} A.B.~Migdal, JETP Lett. {\bf 28} (1978) 35.
\bibitem{linde} A.D.~Linde, JETP Lett. {\bf 27} (1978) 441.
\bibitem{bodmer}
A.R.~Bodmer, Nucl.\ Phys.\ A {\bf 526} (1991) 703.
\bibitem{furnstahl}
R.J.~Furnstahl, H.B.~Tang and B.D.~Serot,
Phys.\ Rev.\ C {\bf 52} (1995) 1368;
R.J.~Furnstahl, B.D.~Serot and H.B.~Tang,
Nucl.\ Phys.\ A {\bf 598} (1996) 539;
Nucl.\ Phys.\ A {\bf 615} (1997) 441
[Erratum-ibid.\ A {\bf 640} (1998) 505].
\bibitem{ASK}
B.K. Agrawal, S. Shlomo and V. Kim Au, Phys. Rev. {\bf C68} (2003) 031304.
\bibitem{asym} P.E.~Haustein, At.\ Data Nucl.\ Data Tables {\bf
39} (1988) 185.
\bibitem{walecka}
J.D.~Walecka, Ann.\ Phys.\ (N.Y.)  {\bf 83} (1974) 491.
\bibitem{lklb98} C.H.~Lee, T.T.S.~Kuo, G.Q.~Li and G.E.~Brown,
Phys.\ Rev.\ C {\bf 57} (1998) 3488.
\bibitem{zm}
J.~Zimanyi and S.A.~Moszkowski, Phys.\ Rev.\ C {\bf 42} (1990) 1416.
\bibitem{rmdb99} R.~Rapp, R.~Machleidt, J.W.~Durso, G.E.~Brown,
  Phys.\ Rev.\ Lett.\ {\bf 82} (1999) 1827.
\bibitem{LBLK97}
G.Q. Li, G.E. Brown, C.H. Lee and C.M. Ko,
arXiv:nucl-th/9702023.
\bibitem{Rho-vm} M.~Rho, Prog.\ Theor.\ Phys.\ Suppl.\  {\bf 149} (2003)
142 [arXiv:hep-ph/0301008].
\bibitem{Rho} M. Rho, arXiv:nucl-th/0212076.
%%%\bibitem{KS82}
%%%V.A. Khodel and E.E. Saperstein, Phys. Rep. {\bf 92} (1982) 183.
\bibitem{G92}
N.K.~Glendenning, Phys. Rev. {\bf D46} (1992) 1274.
\bibitem{GS}
N.K.~Glendenning and J.~Schaffner-Bielich, Phys. Rev. {\bf C60}
(1999) 025803;
M.~Christiansen and N.K.~Glendenning, eprint:~astro-ph/0008207.
\bibitem{VYT}
D.N. Voskresensky, M. Yasuhira, T. Tatsumi, Phys. Lett. {\bf B541} (2002) 93; Nucl. Phys. {\bf A723}
(2003) 291.
\bibitem{VYTkaon}
T. Maruyama, T. Tatsumi, D.N. Voskresensky,
T. Tanigawa, S. Chiba,
arXive:nucl-th/0311076.
\bibitem{BGV1}
D. Blaschke, H. Grigorian, D.N. Voskresensky,
arXiv:astro-ph/0403171.
\bibitem{BKPP}
G.E. Brown, K. Kubodera, D. Page, P. Pizzochero, Phys. Rev. {\bf D37} (1988)
2042.
\bibitem{VS84}
D.N. Voskresensky, A.V. Senatorov, JETP Lett.
{\bf 40} (1984) 1212; Sov. Phys. JETP
{\bf 63} (1986) 885.
\bibitem{V01}
 D.N.~Voskresensky, in book: "Physics of Neutron Star
Interiors",
Lecture Notes in Physics, Eds. D. Blaschke, N.K. Glendenning, A. Sedrakian,
Springer, Heidelberg (2001), p. 467-502.
\bibitem{M83}
A.B. Migdal.''Theory of finite Fermi systems and properties of nuclei'' (in
Russian), Nauka, Moscow, 1983.
\bibitem{ST98}
E.E. Saperstein, S.V. Tolokonnikov, JETP Lett., {\bf 68} (1998) 553.
\bibitem{SF}
A. Schwenk, B. Friman, Phys. Rev. Lett., {\bf 92} (2004) 082501.
\bibitem{SVSWW}
C. Schaab, D. Voskresensky, A.D. Sedrakian, F.
Weber, M.K. Weigel, Astron. Astrophys. {\bf 321} (1997) 591.
\bibitem{V83} D.N. Voskresensky, Nucl. Phys. {\bf A555} (1993) 293.
\bibitem{KV96} E.E.~Kolomeitsev, B.~K\"ampfer and D.~N.~Voskresensky,
Int.\ J.\ Mod.\ Phys.\ E {\bf 5} (1996) 313.
\bibitem{Fodor}
Z. Fodor, and S.D. Katz, JHEP {\bf{0203}} (2002) 014;
F. Csikor, G.I. Egri, Z. Fodor, S.D. Katz, K.K. Szabo, and
A.I. Tothet, hep-lat/0301027;  F. Karsch, and E. Laermann,
hep-lat/0305025.
\end{thebibliography}
\end{document}